\documentclass[12pt]{article}

\usepackage{amsmath,amssymb,latexsym,theorem,epsfig,amscd}
\usepackage[all]{xy}
\CompileMatrices

\newcommand{\bbz}{{\mathbb Z}}

\newcommand{\cp}[1]{{\mathbb P}^{#1}}
\newcommand{\op}[1]{\operatorname{#1}}
\newcommand{\ocp}[1][]{{\mathcal O}_{\cp{1}}{#1}}

\newcommand{\ses}[3]{\xymatrix{
0 \ar[r] & {#1} \ar[r] & {#2} \ar[r] & {#3} \ar[r] & 0 \\ }}
\newcommand{\les}[8]{\xymatrix{       &      & ...  \ar[r]  &  {#1}    \ar@{->} `r[d] `[l] `^dl[dlll]  `^dr/14pt[dll]    [dll] \\
&  {#2} \ar[r] & {#3} \ar[r] & {#4}  \ar `r/10pt[d] `[l]  `^dl[dlll]  `^dr/14pt[dll]   [dll] \\ 
& {#5} \ar[r]  & {#6} \ar[r] & {#7}  \ar `r/10pt[d] `[l]  `^dl[dlll]  `^dr/14pt[dll]   [dll] \\
&  {#8} \ar[r] & ... & & }}

\newcommand{\ts}[4]
{\xymatrix{
0 \ar[r] & {#1} \ar[r] & {#2} \ar[r] & {#3} \ar[r] & 0 \\  
         &            & {#4} \ar[u] \ar@{.>}[ul]^{\alpha} \ar@{.>}[ur]_{\beta}  &            & \\
         &            &  0  \ar[u] &            &
}}

{\theorembodyfont{\rmfamily} }
{\theorembodyfont{\rmfamily} }

\numberwithin{equation}{section}

\begin{document}

\begin{titlepage}

\vspace{-5cm}

\title{
   \hfill{\normalsize hep-th/0212221}\\ \vspace{-0.3cm}
   \hfill{\normalsize UPR-1015-T} \\[1em]
   {\LARGE Torus-Fibered Calabi-Yau Threefolds with Non-Trivial Fundamental Group}
\\
[1em] }
\author{
     Burt A. Ovrut$^1$, Tony Pantev$^2$ and
Ren\'e Reinbacher$^1$ \\[0.5em]
   {\normalsize $^1$Department of Physics, University of Pennsylvania} \\[-0.4em] {\normalsize Philadelphia, PA 19104--6396}\\
   {\normalsize $^2$Department of Mathematics, University of Pennsylvania} \\[-0.4em]
   {\normalsize Philadelphia, PA 19104--6395, USA}\\ }
\date{}
\maketitle
\begin{abstract}
\noindent
We construct smooth Calabi-Yau threefolds $Z$, torus-fibered over a $dP_{9}$ base, with fundamental group $\pi_1={\mathbb Z}_{2} \times {\mathbb Z}_{2}$. To do this, the structure of rational elliptic surfaces is studied and it is shown that a restricted subset of such surfaces admit at least a ${\mathbb Z}_{2} \times {\mathbb Z}_{2}$ group of automorphisms. One then constructs Calabi-Yau threefolds $X$ as the fiber product of two such
$dP_{9}$ surfaces, demonstrating that the involutions on the surfaces lift to a freely acting ${\mathbb Z}_{2} \times {\mathbb Z}_{2}$  group of automorphisms  on $X$. The threefolds $Z$ are then obtained as the quotient $ Z=X/({\mathbb Z}_{2} \times {\mathbb Z}_{2})$.  These Calabi-Yau spaces  $Z$ admit stable, holomorphic $SU(4)$ vector bundles  which, in conjunction with ${\mathbb Z}_{2} \times {\mathbb Z}_{2}$  Wilson lines, lead to standard-like models of particle physics with naturally suppressed nucleon decay.

\end{abstract}

\thispagestyle{empty}

\end{titlepage}

\section{Introduction}
In several papers \cite{hw1,hw2}, Ho\v rava and Witten showed that the strongly coupled heterotic string, $M$-theory compactified on an $S^{1}/{\mathbb Z}_{2}$ orbifold, is described by two end-of-the-world fixed ten-planes, each with an $E_{8}$ worldvolume supermultiplet, separated by the eleventh dimension. Furthermore, it was demonstrated in \cite{w1} that, when compactified on a Calabi-Yau threefold, this theory could, in the low momentum limit, describe realistic grand unified theories (GUTs) in four-dimensions. In the phenomenologically relevant case when the Calabi-Yau radius is much smaller than the size of the orbifold, the process of compactification proceeds through a five-dimensional effective theory, which was first computed in \cite{losw1,losw2}. This five-dimensional theory, called heterotic $M$-theory, is comprised of two end-of-the-world three branes, one our observable universe and the other a hidden sector, as well as a variable number of wrapped five-branes inside the five-dimensional geometrically warped bulk space. Thus, heterotic $M$-theory emerges as a fundamental paradigm for ``brane universe'' scenarios within the context of strongly coupled heterotic strings. The residual gauge supermultiplets on the end-of-the-world branes need no longer be $E_{8}$. Instead, non-trivial gauge instantons on the Calabi-Yau threefold can break these to smaller, and more physically relevant, subgroups. In \cite{low1,low2,dlow}, it was shown how to construct large classes of gauge instantons on elliptically fibered Calabi-Yau threefolds with trivial homotopy that break $E_{8}$ to phenomenologically interesting GUT theories with $E_{6}$, $SO(10)$ and $SU(5)$ gauge groups. These instantons,  generically, have structure groups and connections that do not correspond to the ``standard'' $SU(3)$ embedding.

A deficiency of this approach is that, apparently, it is impossible to break $E_{8}$ to the standard model gauge group $SU(3) \times SU(2) \times U(1)$, or even a standard model-like variation of this group, on a Calabi-Yau  threefold with trivial fundamental group. The reason is straightforward. The most obvious mechanism to break the above GUT groups to the standard model is through extensions of the gauge instantons by a non-trivial flat bundle, that is, Wilson lines. However, these do not exist on a manifold where the fundamental group, $\pi_{1}$, is the identity. This deficiency was overcome in \cite{schoenCY,dopw-i,dopw-ii,dopw-iii,dopw-iv}, where torus-fibered Calabi-Yau threefolds with $\pi_{1}={\mathbb Z}_{2}$ were constructed. Furthermore, in \cite{dopw-i,dopw-ii,dopw-iii,dopw-iv}, stable gauge instantons with $SU(5)$ structure group, leading to $SU(5)$ GUT theories at low energy, were produced for the first time. When these instantons are extended by ${\mathbb Z}_{2}$ Wilson lines, the low energy gauge group is exactly $SU(3) \times SU(2) \times U(1)$. Therefore, within this context, a standard-like model of particle physics can be achieved.

As emphasized long ago by Witten \cite{w2}, a potential problem of $SU(5)$ models
with ${\mathbb Z}_{2}$ Wilson lines is too rapid nucleon decay precipitated 
by dimension 4 operators. Although this need not be the case, there is no 
obvious symmetry that prevents it. One solution of this problem is to proceed
not through $SU(5)$ but, rather, through an $SO(10)$ GUT group. When broken by
Wilson lines to a standard-like model, the low energy gauge group often
contains a $U(1)_{B-L}$ symmetry that eliminates the dimension 4 operators. It
would seem worthwhile, therefore, to extend the results of \cite{dopw-i,dopw-ii,dopw-iii,dopw-iv} to $SO(10)$ GUT
theories broken by Wilson lines. However, since $SO(10)$ is a larger group
than $SU(5)$, it is necessary to have a larger group of Wilson lines, at least
${\mathbb Z}_{2} \times {\mathbb Z}_{2}$, to achieve a standard-like model. To
do this, one must first construct torus-fibered Calabi-Yau
threefolds with fundamental group ${\mathbb Z}_{2} \times {\mathbb Z}_{2}$ or
larger. Once this is done, gauge instantons with structure group $SU(4)$,
which lead to unbroken $SO(10)$ at low energy, must be produced on these
manifolds. If this can be achieved, then Wilson lines will break this theory
to standard-like models at low energy with naturally suppressed nucleon decay.
In this paper, we accomplish the first of these tasks, namely, we construct
torus-fibered Calabi-Yau threefolds with $\pi_{1}={\mathbb Z}_{2}
\times {\mathbb Z}_{2}$.

To accomplish this, we proceed as follows. As stated above, our context will
be Calabi-Yau threefolds, $X$, fibered over a base surface ${B^{'}}$. As discussed elsewhere \cite{dlow}, the requirement that $X$ be Calabi-Yau restricts the base surfaces to be either a Hirzebruch surface ${\mathbb F}_{r}$ where $r$ is a non-negative integer, an  Enriques surface, ${\cal{E}}$, a del Pezzo surface, $dP_{i}$ for $i=1,\dots,8$, or a rational  elliptic surface $dP_{9}$. Of these, perhaps the most interesting one  is $dP_{9}$, which is itself elliptically fibered over the base complex projective line ${\mathbb P}^{1}$. The fibered structure of $dP_{9}$ naturally admits involutions. Since we are interested in fundamental groups containing ${\mathbb Z}_{2}$ factors, it seems reasonable to choose ${B^{'}}=dP_{9}$ for our construction.

With this in mind, in Section 2 we discuss various properties of rational
elliptic surfaces $B$ that are essential in this paper. These include their
structure as a fibration, the Kodaira classification of the fibers in the
discriminant and a detailed construction of their Weierstrass model. 
We recall, in the process, that the complex dimension of the moduli space of
generic rational elliptic surfaces is eight.
Section 3 is devoted to a complete description of all allowed involutions 
$\tau_{B}$ on a rational elliptic surface. We show that these come in three
different types. The first is of the form $\tau_{B}=t_{\xi} \circ (-1)_{B}$, 
where $\xi$ is any section of $B$ and $(-1)_{B}$ is the natural inversion 
on the fibers, while the second is $\tau_{B}=t_\xi$, where $\xi$ intersecting 
any fiber is a point of order two. These two involution types share the
property that their projection onto the base ${\mathbb P}^{1}$ is trivial. The
third type of involution has the structure $\tau_{B}=t_{\xi} \circ \alpha_{B}$,
where $\alpha_{B}$ is a zero section preserving involution mapping fibers to
fibers and $\xi$ satisfies an appropriate constraint. The projection of this
third type of involution onto the base ${\mathbb P}^{1}$ is non-trivial. We
indicate why the first type of involution is unlikely to induce a fixed
point free involution on the full Calabi-Yau threefold $X$ and, hence, do
not consider it further. The second type of involution can induce a fixed
point free action on $X$, but can only exist in restricted cases 
when a global section of order two is defined. Hence, this type of involution 
is of secondary importance and we discuss it only in the appropriate context. 
However, the third type of involution is of fundamental importance. These
involutions, when $\xi$ is not the zero section, can, indeed, lead to a freely
acting involution on $X$. Furthermore, they occur on generic rational elliptic
surfaces. Hence, the next few sections are dedicated to analyzing 
this third type of involution. This analysis follows the work of \cite{dopw-i}.

To begin with, we must show under what conditions an involution
$\alpha_{B}$ exists. We examine this question in detail in Section 4, proving
that such involutions exist in a five-parameter subspace of the eight
parameter moduli space of rational elliptic surfaces. This is accomplished by
showing that any quotient space $B/\alpha_{B}$ must be the double cover of the
product surface $Q={\mathbb P}^{1} \times {\mathbb P}^{1}$ branched over a
curve $M=T \cup r$, where $T$ and $r$ have bidegree (1,4) and (1,0)
respectively. The existence of such a branched surface then indicates 
the existence of an involution $\alpha_{B}$ on $B$. Counting the moduli space of
these rational elliptic surfaces then reduces, in essence, to counting the moduli
of the curves $T$ and $r$.

Having identified the five parameter space of rational elliptic surfaces $B$
that admit an involution $\alpha_{B}$, we then discuss, in Section 5, the
allowed set of involutions of the form $\tau_{B}=t_{\xi} \circ \alpha_{B}$. This
depends on the space of constrained sections $\xi$. For a generic
surface $B$ in this five-parameter space, such sections form a
rank four lattice. Hence, there are many such involutions. However, we
show that no two of these involutions commute and, therefore, that the maximal group
of involutions on $B$ is ${\mathbb Z}_{2}$. It was demonstrated in \cite{dopw-i,dopw-ii,dopw-iii} that the homology ring of these surfaces does not contain a ${\mathbb Z}_{2}$ invariant fiber class, a pre-requisite for constructing standard-like models. Therefore, in the remainder of Section 5, we further restrict the moduli space of rational elliptic surfaces with the intent of introducing such invariant
fiber classes. This is accomplished by restricting the curve $T$ of the 
branch locus in space $Q$ to be of the form $T=t \cup s$, where curves $t$
and $s$ have bidegree (1,3) and (0,1) respectively. We show that the
surfaces $B$ constructed in this manner form a four parameter subspace of the
five parameter space of surfaces admitting involutions $\alpha_{B}$. The
restriction on the moduli space of these surfaces also greatly reduces the
number of constrained sections $\xi$. For a generic four parameter surface,
these are reduced to the zero section, which is not of interest, 
plus three pairs of explicit sections. There are now only six involutions 
of the type $\tau=t_{\xi} \circ \alpha_{B}$ but, again, no two of these 
involutions commute and, hence, the maximal group of involutions remains ${\mathbb
Z}_{2}$. However, we have achieved our goal. We show explicitly that 
the restriction to this four-parameter family of surfaces $B$ introduces one
extra fiber class which, as will be shown in \cite{opr}, is invariant 
under the involution group. As
demonstrated in \cite{dopw-i,dopw-ii,dopw-iii}, the existence of this class allows one to construct standard-like models within the context of $SU(5)$ gauge instantons with ${\mathbb Z}_{2}$ Wilson lines. 

However, since we are interested in $SU(4)$ gauge instantons with larger groups
of involutions, we must place even further restrictions on the parameter space
of rational elliptic surfaces $B$. This is accomplished in Section 6. In this
section, we further restrict the curve $T$ of the branch locus in space $Q$ to
be of the form $T={t}^{'} \cup i \cup s$, where curves 
${t}^{'}$, $i$ and $s$ have bidegree $(1,2), (0,1)$ and $(0,1)$
respectively. We show that surfaces $B$ constructed in this manner form a
three-parameter subspace of the four-parameter space discussed above. The
restriction on the moduli space of these surfaces further reduces the number
of constrained sections $\xi$. Specifically, we show that, in the generic
case, these are now reduced to a single section of order 2, which we call
$e_{6}$, plus two pairs of explicit sections. The existence of the section of
order two allows two involutions of the form $\tau_{B}=t_{\xi} \circ \alpha_{B}$
to commute. We find that the maximal group of commuting involutions is now
enlarged to ${\mathbb Z}_{2} \times {\mathbb Z}_{2}$, as we desired. However,
a fundamental problem remains. In analogy with the situation in \cite{dopw-i,dopw-ii,dopw-iii}, it is
essential to find a fiber class that is invariant under this enlarged group of
involutions. Despite the fact that several new classes are introduced in this
restricted moduli space, none of them is found to be ${\mathbb Z}_{2} \times
{\mathbb Z}_{2}$ invariant. To find such a class, we must make one further
restriction.

Hence, for the remainder of Section 6 we further restrict the curve $T$ to be
of the form $T=t^{''} \cup j \cup i \cup s$, where
$t^{''}$ has
bidegree $(1,1)$ and the remaining curves bidegree $(0,1)$. We show that the
surfaces $B$ constructed in this manner form a two-parameter subspace of the
three-parameter moduli space above. As in previous cases, the number of
restricted sections $\xi$ is also reduced, now consisting of two different 
sections of order 2, the section $e_{6}$ discussed above plus a new section
which we denote by $e_{4}$, and one remaining pair of explicit sections. In
addition, we show that there is one extra section of order two, distinct
from $e_{6}$ and $e_{4}$, that intersects this pair at a point and satisfies the
constraint on $\xi$. This last section is not generic and appears because of
the strong restriction on the surfaces $B$. The appearance of these extra
sections of order two now allows three involutions of the form  
$\tau_{B}=t_{\xi} \circ \alpha_{B}$ to commute. We find that the maximal group of 
commuting involutions is now enlarged to ${\mathbb Z}_{2} \times {\mathbb
Z}_{2} \times {\mathbb Z}_{2}$. In addition, there is now an extra fiber class
that is invariant under a ${\mathbb Z}_{2} \times {\mathbb Z}_{2}$ subgroup of
the involution group. This, as we will show elsewhere \cite{dopr-i,dopr-ii}, is sufficient to
allow one to construct standard-like models in the $SU(4)$, ${\mathbb Z}_{2} 
\times {\mathbb Z}_{2}$ Wilson line context. Hence, although it is embedded in
a larger involution group, it is actually this ${\mathbb Z}_{2} \times 
{\mathbb Z}_{2}$ subgroup that plays an active role.

In Section 7, we construct Calabi-Yau threefolds $X$ that are
elliptically fibered over a base surface $dP_{9}$ and that admit a freely acting
 ${\mathbb Z}_{2} \times {\mathbb Z}_{2}$ group of involutions. We begin by
demonstrating that all Calabi-Yau threefolds elliptically fibered over ${B^{'}}=dP_{9}$
can be written as a fiber product $X= B \times_{{\mathbb P}^{1}} {{B}^{'}}$.
We then compute the Chern classes of the tangent bundle of $X$ explicitly,
verifying that $c_{1}(T_{X})=0$ and that $X$ admits a global, holomorphic
three-form, two requisite properties for $X$ to be a Calabi-Yau threefold. The
factorization of $X$ into a fiber product of rational elliptic surfaces
allows us to lift involutions from these surfaces to the threefold. We
show in Section 7 that any Calabi-Yau threefold $X=B
\times_{{\mathbb P}^{1}}{{B^{'}}}$, with both $B$ and ${B}^{'}$
restricted to the two-parameter region of moduli space discussed in Section 6,
admits a freely acting ${\mathbb Z}_{2} \times {\mathbb Z}_{2}$ as its
maximally commuting set of involutions. Actually, one can lift the full
${\mathbb Z}_{2} \times {\mathbb Z}_{2} \times {\mathbb Z}_{2}$ group of
involutions to $X$, but only the ${\mathbb Z}_{2} \times {\mathbb Z}_{2}$
subgroup acts freely. In addition, the invariant fiber classes on $B$ and
${{B}^{'}}$ induce a ${\mathbb Z}_{2} \times {\mathbb Z}_{2}$ invariant
class on $X$. We now complete the program addressed in this paper by modding
out the freely acting involution group on $X$ to produce the threefold
$Z=X/({\mathbb Z}_{2} \times {\mathbb Z}_{2})$. It is then shown that $Z$ has
vanishing first Chern class and a global, holomorphic three-form and, hence,
is a Calabi-Yau space. Unlike $X$, which is elliptically fibered, we show that
$Z$ admits no global section and, therefore, is only torus-fibered. Clearly,
$Z$ has the non-trivial homotopy group $\pi_{1}(Z)={\mathbb Z}_{2} \times {\mathbb Z}_{2}$, as well as an extra class which descends from the invariant class on $X$. This was the goal of this paper.

This work is the first in a series of papers \cite{opr,dopr-i,dopr-ii,dopr-iii} which will construct
standard-like models of particle physics with naturally suppressed nucleon
decay based on structure group $SU(4)$ gauge instantons and Wilson lines. The
${\mathbb Z}_{2} \times {\mathbb Z}_{2}$ transformation properties of the
homology ring of $Z$ will be described in \cite{opr}. In \cite{dopr-i} we will construct
stable, holomorphic vector bundles with structure group $SU(4)$ on $Z$ and
show that, with the addition of Wilson lines, these can be adjusted to give 
three family standard-like models. Finally, we will give a more detailed
discussion of the underlying mathematics, much of which has been suppressed in
the previous publications, in \cite{dopr-ii}.

\section{Rational Elliptic Surfaces $B$} \label{RES}

In this section, we derive some of the basic properties of rational elliptic surfaces. We will give two different geometric descriptions of such surfaces and formulate their Weierstrass models.

Let $\cp{2}$ be a complex projective space of dimension two and denote a point in  $\cp{2}$ by $(z_{0}:z_{1}:z_{2})$ where $ z_{i}$ are projective coordinates. Specify by $l\cong \cp{1}$ a divisor of $\cp{2}$ given by a linear equation in the coordinates $z_i$. Then these  coordinates can be considered as global sections in $ H^{0}(\cp{2},{\mathcal O}_{{\mathbb P}^{2}}(l))$, where ${\mathcal O}_{{\mathbb P}^{2}}(l)$ is the line bundle on $\cp{2}$ with a unique section vanishing along the given line $l$. In fact, any curve $C\subset \cp{2}$  can be given as the zero set of a section of a line bundle on $\cp{2}$. Each curve characterizes a line bundle uniquely, which is denoted by ${\mathcal O}_{{\mathbb P}^{2}}(C)$. Since on $\cp{2}$ all line bundles with global holomorphic sections are isomorphic to ${\mathcal O}_{{\mathbb P}^{2}}(Nl)$ for positive $N$, it follows that any curve $C$ can be specified by  a homogeneous polynomial of degree $N$ in the projective coordinates $(z_{0}:z_{1}:z_{2})$. Here, we are interested in the case when $C$ is a genus one Riemann surface, that is, a torus $T^2$. In this case, it is not difficult to show that $ N=3$. We conclude that a given torus $T^2$ can be ``cut out'' of $\cp{2}$ by specifying a unique homogeneous polynomial $f$ of degree three in the complex projective coordinates.

With this in mind, we consider the product $\cp{1}\times\cp{2}$ with the natural projections $\pi_i : \cp{1}\times\cp{2} \to \cp{i}$ for $i=1,2$. Denote a point in $\cp{1}$ by $(t_{0}:t_{1})$, where $ t_{i} \in H^{0}(\cp{1},{\mathcal O}_{{\mathbb P}^{1}}(1))$ for $i=0,1$ are projective coordinates on $\cp{1}$, and a point in $\cp{2}$ by $(z_{0}:z_{1}:z_{2})$, as above. Let 
\begin{equation} \label{B}
B \subset \cp{1}\times\cp{2}
\end{equation}
be an effective divisor with associated line bundle ${\mathcal O}_{{\mathbb P}^{1}\times{\mathbb P}^{2}}(B)$. By definition, this line bundle admits a unique section $f_t$ whose zero locus is precisely $B$. Extending the previous discussion, we see that $f_t$ must be a homogeneous polynomial of degree $M$ in the projective coordinates $(t_{0}:t_{1})$ and of  degree $N$ in the projective coordinates $(z_{0}:z_{1}:z_{2})$, where $M,N$ are positive integers. We define the space $B$ by explicitly giving the section $f_t$, which is chosen to be the degree $(1,3)$ polynomial 
\begin{equation} \label{f_t}
f_t=t_0 f_0 + t_1f_1,
\end{equation}
where $f_0,f_1$ are two homogeneous polynomials of degree $3$ in $(z_{0}:z_{1}:z_{2})$.

What are the structure and properties of this specific space $B$ ? To begin with, since $B$ is a divisor in $\cp{1}\times\cp{2}$
\begin{equation}
dim_{\mathbb C}B=2,
\end{equation}
so $B$ is a complex surface. Now note that $B$ inherits two natural projections
\begin{equation}
\beta_i : B \to \cp{i}
\end{equation}
for $i=1,2$ where
\begin{equation}
\beta_i=\pi_i|_B.
\end{equation}

First consider $\beta_2: B \to \cp{2}$. For any given point $z=(z_{0}:z_{1}:z_{2}) \in \cp{2}$, the fiber $\beta_2^{-1}(z)$ is given by the solution of the equation 
\begin{equation} \label{f_t1}
t_0 f_0(z) + t_1f_1(z)=0.
\end{equation}

\begin{figure}[!ht]
\begin{center}
\input{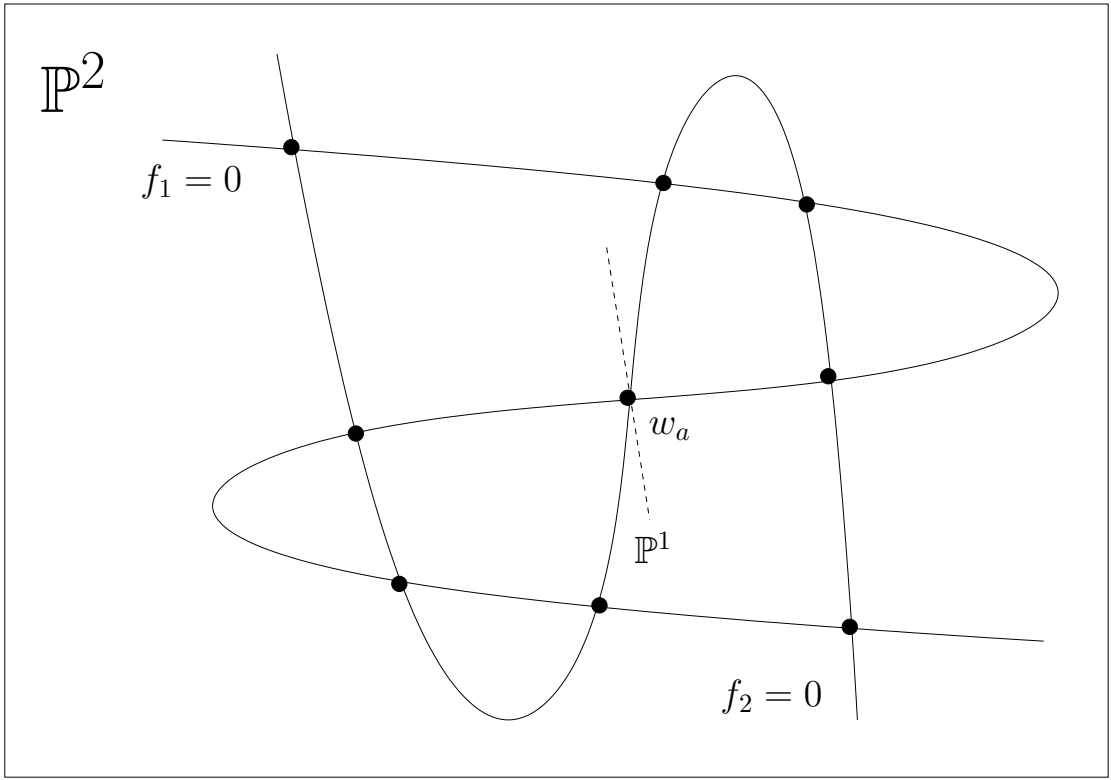} 
\end{center}
\caption{The curves $f_1=0$ and $f_2=0$ intersecting at nine points. }
\label{fig1} 
\end{figure}

\noindent
For a general point $z \in \cp{2}$, at least one of $f_0(z), f_1(z)$ does not vanish and, hence, the solution of (\ref{f_t1}) is a unique point in $\cp{1}$. However, as illustrated in Figure~\ref{fig1}, there are nine distinct points, call them $\omega_a$ with $a=1,...,9$, where $f_0(\omega_a)= f_1(\omega_a)=0$. It is clear that at each of these points the solution of equation  (\ref{f_t1}) is the complete complex projective space $\cp{1}$. This leads to the first characterization of the surface $B$, namely, that $B$ is  $\cp{2}$ blown-up by $\cp{1}$ at each of the nine points satisfying $f_0= f_1=0$. Specifically, 
\[
\beta_2 : B \to \cp{2}
\]
is such that the pre-image $\beta^{-1}(\omega_a)\cong \cp{1}$ for $ a=1,...,9$. For further details on the blow-up of surfaces, we refer the reader to \cite{gh}. $B$ is related to del Pezzo surfaces and, hence,  is denoted by
\begin{equation} 
B=d{\mathbb P}_9.
\end{equation}
Under a mild general position requirement, each subset of eight points $\omega_a$ determines the ninth. It follows that the surface $B$ depends only on $2\cdot 8- dim \mathbb{P}GL(3,\mathbb{C})=8$ parameters, where $ \mathbb{P}GL(3,\mathbb{C})$  is the automorphism group of $\cp{2}$. That is,
\begin{equation} \label{dim-8}
dim\mathcal{M}(B)=8,
\end{equation}
where $\mathcal{M}(B)$ is the moduli space of generic surfaces $B$.

Now consider the remaining natural projection $\beta_1:B\to\cp{1}$. For any given point $t=(t_{0}:t_{1}) \in \cp{1}$, the fiber $\beta_1^{-1}(t)$ is given by the solution of the equation
\begin{equation} \label{f_t2}
t_0 f_0(z) + t_1f_1(z)=0.
\end{equation}
Since at least one of $t_0, t_1$ is non-zero, equation (\ref{f_t2}) is simply the zero locus of some homogenous polynomial of degree three in the projective coordinates $(z_{0}:z_{1}:z_{2})$ of $\cp{2}$. It follows from the above discussion that this cuts out a locus $T^2 \subset \cp{2}$. That is
\begin{equation}
\beta_1^{-1}(t)\cong T^2.
\end{equation}
This leads to the second characterization of surface $B$. That is, with respect to the natural projection
\begin{equation}\label{beta}
\beta : B \to \cp{1},
\end{equation}
where here and henceforth we denote $\beta_1$ by $\beta$, $B$ is a torus fibration over $\cp{1}$.

 In fact, $B$ is an elliptic fibration, as we now demonstrate. First, note that every fiber $\beta^{-1}(t)$ intersects each of the nine blown-up $\cp{1}$ curves, since $f_0(\omega_a)= f_1(\omega_a)=0$ for $a=1,...,9$ solves equation (\ref{f_t2}) for any value of $t$. It follows that each of these curves is a global section of the fibration (\ref{beta}). We denote these nine sections by $e_i$, $i=1,...,9$ where each $e_i\cong \cp{1}$. Now choose any one  of these nine sections. For concreteness, we will always choose $e_9$. The intersection of $e_9$ with any fiber $\beta^{-1}(t)\cong T^2$ marks a point on $T^2$. This can be chosen as the identity element which turns $T^2$ into an Abelian group. Since this choice is canonically given at each fiber, (\ref{beta}) becomes an elliptic fibration with zero section
\begin{equation}\label{11}
e\equiv e_9.
\end{equation}
Note, that the space of global sections of (\ref{beta}) form a Abelian group as well, with $e$ as the identity element. Since $B$ is elliptically fibered, and it is constructed as the blow-up of $\cp{2}$, it is  referred to as a rational elliptic surface.

An important topological invariant of surface $B$ is the Euler characteristic
\begin{equation}\label{12}
\chi{(B)}=\sum_{i=0}^{4}(-1)^i rank (H_{i}(B,\bbz)).
\end{equation}
We first evaluate $\chi(B)$ using the characterization of $B$ as a blow-up, $d{\mathbb P}_9$, of $\cp{2}$ at nine separated points. Since each blow-up adds a linearly independent class to $H_{2}(B,\bbz)$, we see 
\begin{equation}
\chi{(B)}=\chi(\cp{2})+9.
\end{equation}
Then, using the fact that $\chi(\cp{2})=3$, we conclude that 
\begin{equation}\label{B12}
\chi{(B)}=12.
\end{equation}
Let us now evaluate $\chi(B)$ using the second characterization of $B$ as the elliptic fibration (\ref{beta}). If $B$ were to be a simple product $\cp{1}\times T^2$ then, using the relation $\chi{(\cp{1}\times T^2)}=\chi{(\cp{1}})\chi{(T^2)}$ and the fact that $\chi(T^2)=0$, we would conclude that $\chi(B)=0$, which is incorrect. We would arrive at the same conclusion if $B$ was a $T^2$ fiber bundle over $\cp{1}$. It follows that there must be a set of points $D$ in the base $\cp{1}$ of (\ref{beta}) such that, over every point $d \in D $, the fiber $\beta^{-1}(d)$ degenerates and is no longer a smooth $T^2$. Then $\chi(\beta^{-1}(d))\neq 0$ and the above inconsistency can disappear. The set $D \subset \cp{1}$ is a divisor of $\cp{1}$ called the discriminant of the fibration.

To proceed, we need to specify $\chi(\beta^{-1}(d))$. There are many possible degenerations of the torus, each with a different topological structure and Euler characteristic. Such degenerations have been classified by Kodaira \cite{kodaira-casIII}. In this paper, we will be concerned with only two of these. The first, and the simplest, is called an $I_1$ fiber and is simply the degeneration of a single $A$-type one-cycle on $T^2$ to a point. This can be resolved topologically into a sphere by removing the singular point and adding two points at the ends of the tube and, hence, 
\begin{equation}\label{I_1}
\chi{(I_1)}=1.
\end{equation}
The second is called an $I_2$ fiber and consists of the union of two $\cp{1}$'s intersecting transversely in two distinct points. Therefore, its Euler characteristic is
\begin{equation}\label{18}
\chi{(I_2)}=2.
\end{equation}
In our setup, the discriminant $D$ will always be the union $D=D_1 \cup D_2$, where any point in $D_i$ is associated with the fiber $I_i$ for $i=1,2$. Then
\begin{equation}
\chi(B)=\chi(\cp{1}\setminus D)\chi(T^2)+\sum_{i=1}^2 \chi(D_i)\chi(I_i).
\end{equation}
Using (\ref{B12}), (\ref{I_1}), (\ref{18}) and the fact that $\chi(T^2)=0$, this expression becomes
\begin{equation}
\chi(D_1)+2\chi(D_2)=12.
\end{equation}
This has seven solutions of the form
\begin{equation}\label{21}
\chi(D_1)=2n,\;\;\;\;\; \chi(D_2)=m,\;\;\;\;\; n+m=6,
\end{equation}
where $n$ and $m$ are non-negative integers. The generic solution is given by $n=6, m=0$. In this case, $\chi(D_1)=12$  and $D_2=\emptyset$ indicating that $D_1$ consists of twelve isolated points. That is, in this case, the fibration (\ref{beta}) has twelve separated $I_1$ fibers. Now consider the solution given by $n=5, m=1$. Here, $\chi(D_1)=10$ and $\chi(D_2)=1$, indicating that $D_1$ consists of ten isolated points and $D_2$ has one point. Hence, the fibration (\ref{beta}) has ten separated $I_1$ fibers and one $I_2$ fiber. This second solution has a simple interpretation. Remember from the above discussion that  $\chi(I_1)=1$. Selecting any pair of separated $I_1$ fibers and pushing them together will produce a singular fiber $f_{sing}$ with $\chi(f_{sing})=2$. However, there are  two different Kodaira fibers with Euler characteristic two, $I_2$ and the cusp, which is of type $II$. In this paper, as we show in detail below, the degeneration is of the first type, that is, $f_{sing}=I_2$. This reduces the number of $I_1$ fibers from twelve to ten and introduces one $I_2$ fiber, precisely the second solution. Hence, the second solution is simply a special region  in the moduli space of the generic solution of (\ref{21}). Clearly, the same is true of the remaining five solutions. They are all obtained from the generic solution by combining pairs of $I_1$ fibers. For example, the $n=0, m=6$ solution has $D_1=\emptyset$ and $\chi(D_2)=6$, indicating that the fibration (\ref{beta}) has six separated $I_2$ fibers. This is obtained from the generic solution by choosing six pairs of $I_1$ fibers and moving in the moduli space until each pair merges into an $I_2$ fiber. These special points in moduli space will play  a substantial role in the subsequent discussion. 

It is reasonably clear from the previous discussion, and can be verified, that $H_{2}(B,\bbz)$ has a basis of classes given by $\beta_2^{-1}(l)$, which we will also denote as $l$, and $e_i, i=1,...,9$ with the intersection numbers 
\begin{equation}\label{intersection}
l^{2}=1,\;\;\; e_i \cdot e_j=- \delta_{ij},\;\;\; l\cdot e_i=0.
\end{equation}
Note, for example, that the class of a generic fiber, denoted by $f$, can be written as 
\begin{equation}\label{fibre}
f = 3\ell - \sum_{i=1}^{9}e_{i}.
\end{equation}
To see this, recall that, prior to blowing up, the fiber is given by a curve in the class $3l$. However, on the blow-up $B=d{\mathbb P}_9$ of $\cp{2}$, the pre-image of this class will contain the nine exceptional curves $e_i, i=1,...,9$. But, as we have demonstrated, these curves are sections of (\ref{beta}). Therefore, to obtain the pure fiber class one must subtract off each of the nine exceptional curves, yielding expression (\ref{fibre}). The fibers are referred to as the proper transforms of the cubics in $\cp{2}$ passing through the points $\omega_a$.

An important bundle on $B$ is the canonical bundle $K_B=\wedge^2T^{*}B$. Using the relation  
\begin{equation}
K_B=\beta_2^{*} K_{\cp{2}}\otimes \sum_{i=1}^9 {\mathcal O}_{B}(e_i),
\end{equation}
and the fact that $K_{\cp{2}}={\mathcal O}_{\cp{2}}(-3l)$, it follows that 
\begin{equation}
K_B={\mathcal O}_{B}(-3l + \sum_{i=1}^9 e_i).
\end{equation}
Hence, the canonical class, which we will also denote by $K_B$, is
\begin{equation}\label{26}
K_B=-3l + \sum_{i=1}^9 e_i=-f.
\end{equation}

The elliptic fibration (\ref{beta}) can be given a Weierstrass representation as follows. First consider the projective bundle
\begin{equation}
p : P \to \cp{1},
\end{equation}
where
\begin{equation}\label{P}
P={\mathbb P}({\mathcal O}_{\cp{1}}\oplus{\mathcal L}^2\oplus{\mathcal {L}}^3)
\end{equation}
and
\begin{equation}\label{Cn}
{\mathcal L}={N}^{*}_{e/B}={\mathcal O}_e(-e)\cong{\mathcal O}_{\cp{1}}(1)
\end{equation}
is the conormal bundle of the zero section $e=e_9$. We have used the fact that the normal bundle of the divisor $e$ is given by the restriction of $\mathcal{O}_{B}(e)$ to $e$, that is $N_{e/B}=\mathcal{O}_{e}(e)$. Using the intersection numbers (\ref{intersection}), it follows that ${N}^{*}_{e/B}\cong {\mathcal O}_{\cp{1}}(1)$. Also note that ${\mathcal O}_{P}(1)\otimes p^{*}{\mathcal L}^m$ for any integer $m$ is a line bundle over $P$. Then there are three canonical sections
\begin{equation}\label{xyz}
x\in H^0(P,{\mathcal O}_{P}(1)\otimes p^{*}{\mathcal L}^2),\;\;\;\;
y\in H^0(P,{\mathcal O}_{P}(1)\otimes p^{*}{\mathcal L}^3),\;\;\;\; 
z\in H^0(P,{\mathcal O}_{P}(1)).
\end{equation}
Using the fact that ${\mathcal O}_{P}(1)|_{p^{-1}{(b)}\cong \cp{2}} \cong {\mathcal O}_{\cp{2}}(l)$ for any $b \in \cp{1}$, we see that $x,y,z$ restricted to any $\cp{2}$ fiber are coordinates of $\cp{2}$. In fact, they are precisely the $(z_{0}:z_{1}:z_{2})$ projective coordinates defined previously. Now consider the line bundle ${\mathcal O}_{P}(3)\otimes p^{*}{\mathcal L}^6$ on $P$. The key point, which we will not prove here, is that the Weierstrass model of $B$, which we denote by $W_B$, satisfies
\begin{equation}\label{W_B}
{\mathcal O}_{P}(W_B)={\mathcal O}_{P}(3)\otimes p^{*}{\mathcal L}^6.
\end{equation}
Furthermore, there exist a map $\nu:B\to W_B$. Since for generic $B$ the map $\nu$ is an isomorphism, we will, where further specification is not required,  denote $W_B$ simply as $B$. Then (\ref{W_B}) implies that $B=d{\mathbb P}_9$ is an effective divisor in $P$. That is, $B$ can be described as a submanifold in the total space of the projective bundle $P$ over $\cp{1}$. Setting $\beta=p|_B$, we recover the fibration (\ref{beta}). Relation (\ref{W_B}) is very useful in that it tells us that $B$ can be cut out of $P$ as the zero locus of some unique section 
\begin{equation}\label{s}
s \in H^0(P,{\mathcal O}_{P}(3)\otimes p^{*}{\mathcal L}^6).
\end{equation}
What is the section $s$ ? It follows from the explicit ${\mathcal O}_{P}(3)$ factor in (\ref{s}) that is must be a homogeneous polynomial of degree three in the sections $x,y,z$. There are ten possible terms, which we group as
\[
  x^3,\; y^2z \:\:\:\:\:\:\;\;\;\; x^2y,\; xy^2,\; y^3 \:\:\:\:\:\:\;\;\;\; x^2z,\; xyz,\; xz^2,\; yz^2,\; z^3.
\]
Each term in the first group has the structure of (\ref{s}) and is allowed. Now consider, for example, $x^2y$ in the second group. Clearly
\begin{equation}
x^2y  \in H^0(P,{\mathcal O}_{P}(3)\otimes p^{*}{\mathcal L}^7),
\end{equation}
which must be multiplied by a coefficient which is a section of $p^{*}{\mathcal L}^{-1}$ if it is to match (\ref{s}). But $p^{*}{\mathcal L}^{-1}$ admits no global holomorphic sections and, hence, $x^2y$ as well as all other terms in the second group are disallowed. Similarly, each term in the third group has the form of (\ref{s}) if it is multiplied by a coefficient which is a section of $p^{*}{\mathcal L}^m$ for $m>0$. Since $p^{*}{\mathcal L}^m$ for $m > 0$ admit global sections, all terms in the third group are allowed. The construction of $s$ is simplified somewhat be the realization that three of the terms in the last group, traditionally taken to be $x^2z,\; xyz$ and $yz^2$, can be set to zero by the appropriate choice of $x,y$ and $z$. It follows that
\begin{equation}\label{34}
s=y^{2}z - x^{3} - g_{2}xz^{2} -
g_{3}z^{3},
\end{equation}
where
\begin{equation}\label{35}
g_{2} \in H^{0}(P,p^{*}{\mathcal O}_{{\mathbb P}^{1}}(4)),\;\;\; g_{3}
\in H^{0}(P,p^{*}{\mathcal O}_{{\mathbb P}^{1}}(6))
\end{equation}
are determined by the explicit choice of $B$. The surface $W_B$, that is, $B$ is then given by the zero locus of $s$
\begin{equation}\label{We}
y^{2}z = x^{3} + g_{2}xz^{2} +
g_{3}z^{3},
\end{equation}
which is the Weierstrass representation of the fibration (\ref{beta}).

\section{Involutions on B}\label{INV}

In this section, we will give a complete classification of all possible involutions  on rational elliptic surfaces and describe some of their features. Let $\tau_{B} : B \to B $ be any involution on $B$, that is, $\tau^2_B=id$. Clearly $\tau_{B}$ is an automorphism of $B$ and, hence, 
\begin{equation}
\tau_{B}^{*}K_{B} \cong
K_{B}.
\end{equation}
However, it was shown previously in (\ref{26}) that $K_B={\mathcal O}_{B}(-f)$. If follows that $\tau_B$ preserves the fiber class $f$ of the fibration $\beta: B \to \cp{1}$. Note that this does not mean that fibers need to be invariant under $\tau_B$, only that $\tau_B$ maps fibers to fibers. We conclude that any involution $\tau_{B} : B \to B $ induces an involution $\tau_{\cp{1}} : \cp{1} \to  \cp{1}$ on the base satisfying $\tau_{\cp{1}}^2=id$ and 
\begin{equation}\label{2.2}
\tau_{\cp{1}}\circ \beta = \beta \circ \tau_B.
\end{equation}

\bigskip
\noindent
The possible involutions $\tau_{\cp{1}}$ on $\cp{1}$ are easy to classify. One can show that for an appropriate choice of the projective coordinates $(t_{0}:t_{1})$ on $\cp{1}$, $\tau_{\cp{1}}$ acts either as 
\begin{equation}
t_0 \to t_0,\;\;\; t_1 \to t_1,
\end{equation}
that is, $\tau_{\cp{1}}=id$ or non-trivially as 
\begin{equation}\label{i2}
t_0 \to t_0,\;\;\; t_1 \to -t_1,
\end{equation}
which has two fixed points at
\begin{equation}\label{noway}
0\equiv (1:0),\;\;\;\;\infty\equiv (0:1).
\end{equation}
That $0$ is a fixed point is obvious. To see that $\infty$ is indeed fixed, recall that projective coordinates are only determined up to multiplication by non-zero complex numbers. Therefore, after acting with $\tau_{\cp{1}}$ we are allowed to re-scale by $-1$ without changing the point. Hence, any involution on $B$ acts on the base $\cp{1}$ either as the identity or as a non-trivial involution with two fixed points. Note also that a non-trivial involution on $\cp{1}$ is uniquely determined by its fixed points.

Let us first discuss the possible involutions $\tau_B$ on $B$ whose induced involutions on $\cp{1}$ are $\tau_{\cp{1}}=id$. Such involutions $\tau_B$ clearly act along fibers, leaving each fiber stable, that is, invariant but not necessarily point-wise fixed.

One such involution, which we denote by $(-1)_B$, is immediately apparent and defined as follows. Recall from (\ref{beta}) and (\ref{11}) that $B$ is an elliptic fibration with the zero section  chosen to be $e=e_9$. The zero section $e$ marks a fixed origin, also called $e$, on each $T^2$ fiber. Choose one such fiber. Marking the origin automatically defines an  Abelian group on the points of $T^2$ with $e$ as the zero element. We define $(-1)_B |_{T^2}: T^2 \to T^2$ by 
\begin{equation}
(-1)_B |_{T^2}(a)=-a,
\end{equation}
where $a$ is any element of the Abelian group  $T^2$ and $-a$ its inverse. Clearly, $(-1)_B |_{T^2}^2=id$ and, hence, $(-1)_B|_{T^2}$ is an involution. Note that $(-1)_B|_{T^2}$ has four fixed points, the trivial fixed point $e$ as well as three non-trivial fixed points $e{'},e{''}$ and $e{'''}$. $(-1)_B$ is defined as the mapping on $B$ whose restriction to each fiber $T^2$ is $(-1)_B|_{T^2}$. Clearly $(-1)^2_{B}=id$ and, hence, $(-1)_B$ is an involution on $B$. Furthermore, since $(-1)_B$ leaves each fiber stable, it follows that the associated $\tau_{\cp{1}}=id$. Note that $(-1)_B$ is a natural involution arising on any elliptic fibration. 

Are there other involutions $\tau_B$ such that $\tau_{\cp{1}}=id$ ? The answer is affirmative. Most of them arise in a specific way from $(-1)_B$. To construct them, we must first introduce the  notion of a translational automorphism on $B$. Let $\Gamma(B)$ be the space of global sections of the fibration (\ref{beta}). $\Gamma(B)$ is non-empty since we know it contains $e_1,...,e_9$ defined above. Furthermore, as mentioned previously, it has an Abelian group structure. Let $\xi \in \Gamma(B)$. Then $\xi$ marks a point, also called $\xi$, on each fiber $T^2$. Define the mapping $t_{\xi}|_{T^2} : T^2 \to T^2$ by
\begin{equation}
t_{\xi}|_{T^2}(a)=a+\xi,
\end{equation}
where $a$ is any point on $T^2$. Clearly, this map is a translation along the fiber $T^2$. Furthermore, unless $\xi=e,e{'},e{''}$ or $e{'''}$, $t_{\xi}|_{T^2}$ is not an involution. Finally, we can define $t_\xi$ on $B$ as the map whose restriction to each fiber $T^2$ is $t_{\xi}|_{T^2}$. Clearly $t_\xi$ is a translational automorphism of $B$. There will be such an automorphism associated with every section of the fibration (\ref{beta}). Now consider
\begin{equation}\label{in1}
\tau_B=t_{\xi}\circ (-1)_B.
\end{equation}
Since 
\begin{equation}
 (-1)_B \circ t_{\xi}=t_{-\xi}\circ (-1)_B,
\end{equation}
it is clear that $\tau_B^2=id$ and, hence,  $\tau_B$ is an involution on $B$. By construction, $\tau_B$ leaves every fiber stable and, thus, the associated involution is $\tau_{\cp{1}}=id$. On a generic surface, all involutions $\tau_B$ which act trivially as $\tau_{\cp{1}}=id$ on $\cp{1}$ are of the form (\ref{in1}). Note that only $(-1)_B$ preserves the zero section. 

As we show later, for specific choices of $B$, there are other involutions whose induced action on $\cp{1}$ is trivial. They are of the form
\begin{equation}\label{in1a}
\tau_B=t_{\xi},
\end{equation}
where $\xi\neq e$ is a section of $B$ which intersects every fiber at a point which is invariant under $(-1)_B$, that is, at $e^{'},e^{''}$ or $e^{'''}$. For generic $B$, such sections do not exist. However, they may exist as sections of restricted rational elliptic surfaces.

Let us now consider involutions $\tau_B$ on $B$ whose induced involution on $\cp{1}$ is non-trivial, that is, of type (\ref{i2}). Here, we simply present the result which classifies all such involutions, referring the reader to \cite{dopw-i} for an explicit proof. Let $\beta : B \to \cp{1}$ and $\tau_B$ be an  involution of $B$ acting non-trivially on $\cp{1}$. Then, there exists an involution $\alpha_B : B \to B$ which maps fibers of $B$ to fibers with the property that
\begin{equation}\label{fe}
\alpha_B(e)=(e),
\end{equation}
where $e$ is the zero section of $B$, and a section $\xi \in\Gamma(B)$ satisfying
\begin{equation}\label{xi}
\alpha_B(\xi)=(-1)_B(\xi),
\end{equation}
such that
\begin{equation}\label{tau}
\tau_B=t_\xi\circ\alpha_B.
\end{equation}
First note that $\tau_B$ is indeed an involution. To see this, use condition (\ref{xi}) to show that
\begin{equation}\label{47}
t_\xi\circ\alpha_B=\alpha_B\circ t_{-\xi}.
\end{equation}
It follows immediately that $\tau_B^2=id$ and, hence, $\tau_B$ is an involution on $B$.

What are the properties of $\alpha_B$ and $\xi$ ? Note that (\ref{2.2}) and (\ref{tau}) imply
\begin{equation}
\beta \circ \alpha_B=\tau_{\cp{1}}\circ\beta
\end{equation}
and, therefore, the action of $\alpha_B$ on $\cp{1}$ is determined by $\tau_{\cp{1}}$. As mentioned above, each non-trivial involution $\tau_{\cp{1}}$ on $\cp{1}$ has two fixed points, which we call $0$ and $\infty$. Therefore, any involution $\alpha_B$ and, hence, any involution of the form (\ref{tau}) leave two fibers stable, the fiber $f_0=\beta^{-1}(0)$ and the fiber $f_\infty=\beta^{-1}(\infty)$. Next consider a section $\xi \in \Gamma(B)$ satisfying (\ref{xi}). This equation, indeed, puts some restrictions on $\xi$, but to compute these it is necessary to have more information about the action of $\alpha_B$. We will return to this important point at the end of this section.

To understand the involutions with non-trivial $\tau_{\cp{1}}$ better, we will now analyze how $\alpha_B$ acts on the Weierstrass representation $W_B$ of $B$ in $P$. Recall from (\ref{We}) that $W_B$ is given by the equation
\begin{equation}\label{We1}
y^{2}z = x^{3} + g_{2}xz^{2} +
g_{3}z^{3},
\end{equation}
where, using the identity $H^0(P, p^{*}{\mathcal L}^n)=H^0(\cp{1},{\mathcal O}_{\cp{1}}\otimes {\mathcal L}^n)$ for any integer $n$, (\ref{35}) becomes
\begin{equation}\label{50}
g_{2} \in H^{0}(\cp{1},{\mathcal O}_{{\mathbb P}^{1}}(4)),\;\;\;\; g_{3}
\in H^{0}(\cp{1},{\mathcal O}_{{\mathbb P}^{1}}(6)).
\end{equation}
Furthermore, it is easy to see that the zero section $e$ is given by the solution of (\ref{We1}) for
\begin{equation}\label{51}
x=z=0.
\end{equation}
It is straightforward to find the action of the natural involution $(-1)_B$ on $W_B$. It is 
\begin{equation}\label{52}
(-1)_{W_B} : (x:y:z) \to (x:-y:z).
\end{equation}
Clearly this action is an involution, which leaves each fiber stable and the zero section invariant. That is, it represents $(-1)_B$ on $W_B$. Recall that on a generic surface $B$, there are no sections whose intersection point with every fiber is a non-trivial point of order two. However, there exists a tri-section, that is, a divisor on $B$ which intersects each fiber in the three points  $e{'},e{''}$ and $e{'''}$. It can be found by solving (\ref{We1}) with
\begin{equation}\label{g}
y=0.
\end{equation}

Let us now find  the action of $\alpha_B$ on $W_B$. First consider the bundle $\ocp{(1)}$. The non-trivial involution $\tau_{\cp{1}}:\cp{1}\to \cp{1}$ allows one to construct a second bundle, the pull-back $\tau_{\cp{1}}^{*}\ocp{(1)}$, along with a natural bundle map
\begin{equation}\label{t}
t_{\cp{1}}: \tau_{\cp{1}}^{*}{\mathcal O}_{\cp{1}}(1) \to {\mathcal O}_{\cp{1}}(1).
\end{equation}
Note that, in general, $\tau_{\cp{1}}^{*}\ocp{(1)}$ need not be isomorphic to $\ocp{(1)}$. However, recall that the projective coordinates $t_0$ and $t_1$ on $\cp{1}$ form a basis of $H^0(\cp{1},{\mathcal O}_{\cp{1}}(1))$. Therefore, the action (\ref{i2}) of $\tau_{\cp{1}}$ on $t_0,t_1$ lifts to an involution on $H^0(\cp{1},{\mathcal O}_{\cp{1}}(1))$. One can show that this induces a bundle isomorphism 
\begin{equation}
\gamma : \tau_{\cp{1}}^{*}{\mathcal O}_{\cp{1}}(1) \to {\mathcal O}_{\cp{1}}(1),
\end{equation}
where $\gamma$ acts linearly on the fibers and trivially on $\cp{1}$. Combining this result with (\ref{t}), it follows that the map 
\begin{equation}\label{ib1}
\tau_{\cp{1}} :{\mathcal O}_{\cp{1}}(1) \to {\mathcal O}_{\cp{1}}(1),
\end{equation}
where $\tau_{\cp{1}}=t_{\cp{1}}\circ \gamma^{-1}$, is an involution on $\ocp{(1)}$. That is, $\tau_{\cp{1}}$ is the lift to $\ocp{(1)}$ of the involution $\tau_{\cp{1}}$ on $\cp{1}$. This can be extended to any bundle ${\mathcal O}_{\cp{1}}(k)$ with non-negative integer $k$ using the fact that
\begin{equation}\label{isk}
H^0(\cp{1},{\mathcal O}_{\cp{1}}(k))= S^k H^{0}(\cp{1},{\mathcal O}_{{\mathbb P}^{1}}(1)),
\end{equation}
where $S^k$ denotes the $k$-th symmetric product. As above, this implies an isomorphism between $\tau^{*}_{\cp{1}}{\mathcal O}_{\cp{1}}(k)$ and ${\mathcal O}_{\cp{1}}(k)$. In combination with the natural bundle map, this induces the involution
\begin{equation}\label{ibk}
\tau_{\cp{1}} :{\mathcal O}_{\cp{1}}(k) \to {\mathcal O}_{\cp{1}}(k),
\end{equation}
which is the lift of $\tau_{\cp{1}}$ to ${\mathcal O}_{\cp{1}}(k)$. Recalling from (\ref{P}) and (\ref{Cn}) that 
\begin{equation}\label{new}
P={\mathbb P}({\mathcal O}_{\cp{1}}\oplus{\mathcal O}_{\cp{1}}(2)\oplus{\mathcal O}_{\cp{1}}(3)),
\end{equation}
we see using (\ref{ibk}) that $\tau_{\cp{1}}$ induces an action on $P$, which we denote by $\tau_P$. That is,
\begin{equation}\label{iP}
\tau_P : P \to P.
\end{equation}
This involution lifts again to an involution
\begin{equation}\label{2.27}
\tau_{P} : {\mathcal O}_{P}(r) \to {\mathcal O}_{P}(r),
\end{equation}
where $r$ is a non-negative integer. The bundle ${\mathcal O}_{P}(1)$ was discussed in the previous section. Finally, note that the actions (\ref{ibk}) and (\ref{2.27}) induce a natural involution 
\begin{equation}\label{as}
\tau_P : H^0(P,{\mathcal O}_{P}(r)\otimes p^{*}{\mathcal O}_{\cp{1}}(s)) \to H^0(P,{\mathcal O}_{P}(r)\otimes p^{*}{\mathcal O}_{\cp{1}}(s)),  
\end{equation}
for any non-negative integers $r$ and $k$.

We can now give the action of $\alpha_B$ on $B$ in terms of the Weierstrass model $W_B$ and $\tau_P$. We refer the reader to \cite{dopw-i} for a complete discussion. Let us assume that $B$ admits an involution $\alpha_B$. Then, there must exist a unique involution
\begin{equation}\label{}
\alpha_{W_{B}} : W_{B} \to W_{B}
\end{equation}
satisfying
\begin{equation}\label{}
\alpha_{W_{B}}\circ \nu = \nu\circ\alpha_{B},
\end{equation}
where $\nu : B \to W_B$ is the natural map of $B$ into $P$. It was shown in  \cite{dopw-i} that $\alpha_{W_{B}}$ must  be either of the form
\begin{equation}\label{aw}
\alpha_{W_{B}}=\tau_{P}|_{W_{B}}
\end{equation}
or
\begin{equation}\label{aw1}
\alpha_{W_{B}}=\tau_{P}|_{W_{B}}\circ (-1)_{W_{B}}.
\end{equation}
An important implication of this result is that, for $\alpha_{W_B}$ and, hence, $\alpha_B$ to exist, the Weierstrass model $W_B \subset P$ must be stable under involution $\tau_P$, that is
\begin{equation}
\tau_P|_{W_B}: W_B\to W_B.
\end{equation}
This requirement puts significant restrictions on the allowed surfaces B, as we now show.

Recall from  (\ref{xyz}) that $z \in H^0({\mathcal O}_{P}(1))$. Since
\begin{equation}\label{}
H^0(P,{\mathcal O}_{P}(1))= H^0(\cp{1},{\mathcal O}_{\cp{1}}\oplus {\mathcal O}_{\cp{1}}(-2)\oplus {\mathcal O}_{\cp{1}}(-3)),
\end{equation}
we see that the section $z$ must be the constant section of ${\mathcal O}_{\cp{1}}$. Now, it follows from (\ref{as}) with $r=1, s=0$ that $\tau_P : H^0(P,{\mathcal O}_{P}(1)) \to H^0(P,{\mathcal O}_{P}(1))  $. Since $z$ is a constant section over $\cp{1}$, it must be invariant under $\tau_P$. That is
\begin{equation}\label{}
\tau_Pz=z.
\end{equation}
In a similar way, one can show that
\begin{equation}\label{}
\tau_Px=x,\;\;\; \tau_Py=y.
\end{equation}
Since $W_B$ is stable under $\tau_P$ and $x,y,z$ are invariant, it follows from ($\ref{We1}$) that $g_2$ and $g_3$ must satisfy 
\begin{equation}\label{fg}
\tau_{\cp{1}}g_2=g_2,\;\;\; \tau_{\cp{1}}g_3=g_3.
\end{equation}
Therefore, the existence of $\alpha_B$ on $B$ implies that the sections $g_2,g_3$ of the corresponding Weierstrass model must satisfy (\ref{fg}). The converse is also true, that is, if (\ref{fg}) is satisfied then there exists an associated involution $\alpha_B$ on $B$. Note that (\ref{fg}) is a strong restriction on the Weierstrass model $W_B$ and, hence, on the structure of $B$.

Expression (\ref{aw}) and (\ref{aw1}) allow us to establish a very important property of involutions $\alpha_B$ on $B$, namely that $\alpha_B$ must fix the fiber  $f_0=\beta^{-1}(0)$ point-wise and have four isolated fixed points on the fiber $f_\infty=\beta^{-1}(\infty)$. These four fixed points are $e,e{'},e{''}$ and $e{'''}$. To prove this, consider the Weierstrass representation $W_B$ and assume that  $\alpha_{W_{B}}=\tau_{P}|_{W_{B}}$. Let us first focus on $f_0$.

Recall that $p^{-1}(0)$ is the fiber in $P$ over $0$. It follows from (\ref{new}) that
\begin{equation}\label{}
P|_{p^{-1}(0)} = {\mathbb P}({\mathcal O}_{\cp{1}}|_{0}\oplus {\mathcal O}_{\cp{1}}(2)|_{0}
\oplus {\mathcal O}_{\cp{1}}(3)|_{0}).
\end{equation}
Note that $P|_{p^{-1}(0)}$ is isomorphic to $\cp{2}$. Since $t_1=0$ at $0=(1:0)$, we see that
\begin{equation}\label{}
1,\;\; t_0(0)^2,\;\;t_0(0)^3 
\end{equation}
are the projective coordinates on $P|_{p^{-1}(0)}$. However, it follows from (\ref{i2}) that $\tau_{\cp{1}} t_0(0)=t_0(0)$ and, hence, that 
\begin{equation}\label{}
\tau_{P} P|_{p^{-1}(0)}=P|_{p^{-1}(0)}
\end{equation}
point-wise. Therefore, $\alpha_{W_B}=\tau_P|_{W_B}$ leaves the fiber $f_0=W_B \cap   P|_{p^{-1}(0)}$ fixed point-wise, as claimed.

Now consider $f_\infty$. Again, $f_\infty$ is described by the vanishing locus of a homogeneous cubic polynomial, this time in $P|_{p^{-1}(\infty)}$ where 
\begin{equation}\label{}
P|_{p^{-1}(\infty)} = {\mathbb P}({\mathcal O}_{\cp{1}}|_{\infty}\oplus {\mathcal O}_{\cp{1}}(2)|_{\infty}
\oplus {\mathcal O}_{\cp{1}}(3)|_{\infty}).
\end{equation}
Since $t_0=0$ at $\infty=(0:1)$, we see that 
\begin{equation}\label{}
1,\;\; t_1(\infty)^2,\;\;t_1(\infty)^3 
\end{equation}
are the projective coordinates on $P|_{p^{-1}(\infty)}$. From (\ref{i2}), we know that $\tau_{\cp{1}} t_1(\infty)=-t_1(\infty)$ and, hence, that the invariant points of $P|_{p^{-1}(\infty)}$ must satisfy 
\begin{equation}\label{infty}
t_1(\infty)^3=0.
\end{equation}
Recalling from (\ref{tau}) that $z$ is a section of ${\mathcal O}_{\cp{1}}$ and, similarly, that $x$ and $y$ are sections of ${\mathcal O}_{\cp{1}}(2)$ and ${\mathcal O}_{\cp{1}}(3)$ respectively, we see that (\ref{infty}) is fulfilled when 
\begin{equation}\label{}
y|_{p^{-1}(\infty)} =0.
\end{equation}
This, however, when inserted into (\ref{We1}) leads precisely to the equation for the four fixed points $e,e{'},e{''},e{'''}$. We conclude, therefore, that $\alpha_{W_B}=\tau_P|_{W_B}$ leaves the fiber $f_{\infty}=W_B \cap   P|_{p^{-1}(\infty)}$ fixed on  the four isolated points $e,e{'},e{''}$ and $e{'''}$, as claimed. Note that if the involution $\alpha_{W_B}$ is of the form (\ref{aw1}), we get the same distribution of fixed points with $f_0$ and $f_{\infty}$ switched. That is, we get four isolated fixed points on $f_0$ and a trivial action on $f_{\infty}$. Therefore, after a possible relabeling of $f_0$ and $f_{\infty}$, we can assume that $\alpha_{W_B}$ is of the form (\ref{aw}), which we do henceforth.

Having established these results, we can now return to the question of the constraints on sections $\xi \in \Gamma(B)$ satisfying equation (\ref{xi}). It follows from the fact that $\alpha_B$ leaves fiber $f_0$ point-wise fixed that
\begin{equation}\label{2.41}
\alpha_B|_{f_0}=id.
\end{equation}
Hence, restricted to $f_0$, equation (\ref{xi}) becomes 
\begin{equation}\label{2.42}
\xi(0)=-\xi(0).
\end{equation}
Therefore, at $f_0$, any such section must pass through either the zero point $e$ or one of the remaining three invariant points of $(-1)_B$, namely, $e^{'},e^{''}$ and $e^{'''}$. Now note that 
\begin{equation}\label{2.43}
\alpha_B\circ (-1)_B=(-1)_B \circ \alpha_B.
\end{equation}
Therefore, if $\xi$ is a solution of (\ref{xi}), then so is $(-1)_B (\xi)$. We conclude that the solutions of equation (\ref{xi}) come in pairs
\begin{equation}\label{2.44}
(\xi,(-1)_B(\xi)),
\end{equation}
where
\begin{equation}\label{2.45}
\xi(0)=e,e^{'},e^{''}\text{or}\; e^{'''}.
\end{equation}
The only obvious solution is $\xi=e$, which corresponds to $\tau_B=\alpha_B$. However, it is not clear  that any other such sections $\xi$ exist. We will show later in the paper that any surface $B$ that admits an involution $\alpha_B$, also admits  sections $\xi \neq e$ which satisfy (\ref{xi}).

In principle, all three types of involutions on $B$, of the form (\ref{in1}), (\ref{in1a}) and (\ref{tau}) respectively, are of interest. However, in this paper we are interested in Calabi-Yau threefolds $X$ elliptically fibered over a base $B^{'}=d\mathbb{P}_9$. It will be shown in Section~{\ref{CY}} that such threefolds are always a fiber product over $\cp{1}$ of the base $B^{'}$ with second rational elliptic surface $B$. That is, $X=B\times_{\cp{1}} B^{'}$. Suffice it here to say that any involution on $X$ can always be constructed as the lift of involutions on $B$ and $B^{'}$. Furthermore, any involution on $X$ involving the lift of an involution on $B, \; B^{'}$ of the form (\ref{in1}) must necessarily have at least one fixed point. This will be discussed in \cite{dopr-ii}. Since, in this paper, we are interested in freely acting involutions on $X$  we, henceforth, do not consider involutions of type (\ref{in1}). Happily, involutions on $X$ constructed solely from the remaining two types (\ref{in1a}) or  (\ref{tau}) where $\xi\neq e$ can have fixed point free actions. Recall, however, that involutions of the form (\ref{in1a}) require a section which intersects each fiber at a point of order two. Such sections, as we will show, only exists in special circumstances and, hence, are of secondary importance. We turn, therefore, to the remaining involutions of the form (\ref{tau}).

\section{The Quotient $B/\alpha_{B}$}
In this section, we  analyze rational elliptic surfaces $B$ admitting an involution $\alpha_B$. Recall from Section~\ref{INV}, that the existence of $\alpha_B$ restricts the form of their Weierstrass model $W_B$. Note that such  models are only smooth for generic surfaces. Since, as  shown later,  we need to consider  surfaces with singular Weierstrass models, and since we must discuss the induced action of $\alpha_B$ on the cohomology $H_2(B,\mathbb{Z})$, we present, in this section, yet  another description of such surfaces $B$.

Let us assume the surface $B$ admits an involution $\alpha_B$. We now show that, in order to understand such surfaces $B$, it is sufficient to study the quotient $B/\alpha_B$.
First of all, since $\alpha_B$ maps fibers of $B$ to fibers, it is clear that $B/\alpha_{B}$ is a torus fibration over the base $\cp{1}/\tau_{\cp{1}}\cong \cp{1}$. We denote the projection map by $\beta_{\alpha_B} : B/\alpha_{B} \to \cp{1}  $ and get the following commuting diagram
\[
\xymatrix{
B \ar[r]^-{\kappa} \ar[d]_-{\beta} & B/\alpha_{B} \ar[d]^{\beta_{\alpha_B}}\ar@{}^-{,}[dr]&\\
{\mathbb P}^{1} \ar[r]_{\op{sq}}  & {\mathbb P}^{1}&
}
\] 
where $\kappa$ and sq are the  double covering maps associated with modding out $\alpha_B$ and $\tau_{\cp{1}}$ respectively.
Furthermore, since $\alpha_B$ preserves $e$, the curve $e/\alpha_B$ is a zero section of $B/\alpha_{B}$. Hence, $B/\alpha_{B}$ is an elliptic fibration. Note, that  we will call the image of $0$ and $\infty$ under $sq$ in $\cp{1}/\tau_{\cp{1}}$   $0$ and $\infty$  as well.

Secondly, since $\alpha_B$ is an involution of $B$ it must map singular fibers to singular fibers. Recall from Section~\ref{RES} that, generically, $B$ possesses twelve $I_1$ fibers. Therefore, $\alpha_B$ must map pairs of $I_1$ fibers into each other. It follows that $B/\alpha_{B}$ has six $I_1$ fibers, each the result of identifying a pair of $I_1$ fibers on $B$. 

\begin{figure}[!ht]
\begin{center}
\input{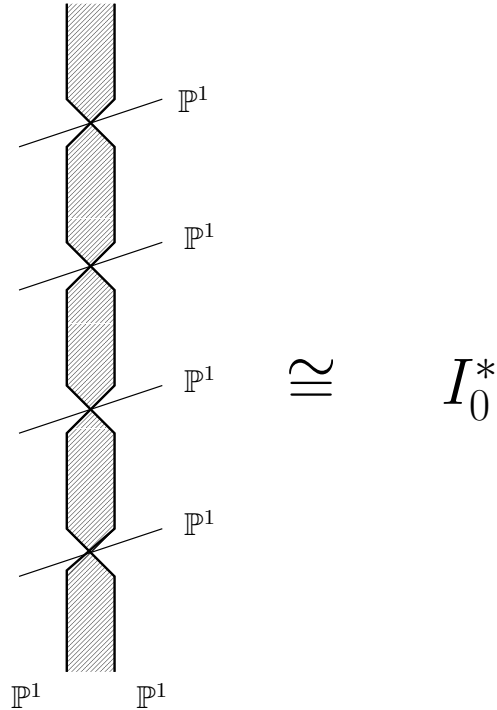} 
\end{center}
\caption{A double $\cp{1}$ line with four surface $A_1$ singularities. Its resolution is an $I_0^{*}$ Kodaira fiber. The cross-hatch indicates the identification of the two $\cp{1}$ lines.}
\label{fig2} 
\end{figure}

Are there other degenerate fibers? To answer this question, we must explore the singularity structure of $B/\alpha_{B}$. In the previous section, we showed that $\alpha_B$ leaves the fiber $f_0=\beta^{-1}(0)$ point-wise fixed. Furthermore, $\alpha_B$ leaves four isolated points $e,e{'},e{''},e{'''}$ fixed on $f_\infty=\beta^{-1}(\infty)$. Since the fixed point locus of $\alpha_B$ in $f_0$ is of codimension one in $B$, the associated fiber $f_0/\alpha_{B}$ in  $B/\alpha_{B}$ is a smooth elliptic fiber. For simplicity denote  $f_0/\alpha_{B}$ by $f_0$. Note that $f_0$ in $B/\alpha_{B}$ is given by $\beta_{\alpha_B}^{-1}(0)$.  However, $f_\infty/\alpha_B$, with its four fixed points each of codimension two in $B$, is more interesting. It is easy to see that the local neighborhood in $B/\alpha_{B}$ of any one such fixed point, call it $p$, is a complex surface in ${\mathbb C}^3$ with coordinates $u,v$ and $w$ where $p=(0,0,0)$. The precise embedding is described by the equation
\begin{equation}\label{}
w^2=uv.
\end{equation}
It is well known that this relation describes a surface with an isolated singular point of type $A_1$ at $p=(0,0,0)$. It follows that $f_\infty/\alpha_B$ in $B/\alpha_{B}$ is a degenerate fiber. Since $\alpha_B$ leaves $f_{\infty}$ stable, that is $\alpha_B: f_{\infty} \to f_{\infty}$, and has four fixed points, the map $f_{\infty} \to f_{\infty}/\alpha_B$ is a double cover branched over four points. As we show later in this section, this implies that $f_{\infty}/\alpha_B\cong \cp{1}$ is a line. Actually, as a fiber in $B/\alpha_B$ it is a double line passing through the four isolated $A_1$ singularities. This is illustrated in Figure~\ref{fig2}.
Henceforth, for simplicity, we will denote $f_{\infty}/\alpha_B$ by $f_{\infty}$. Note that  $f_{\infty}$ in $B/\alpha_{B}$ is given by $\beta_{\alpha_B}^{-1}(\infty)$. Hence, $B/\alpha_{B}$ is not a smooth surface. Rather, it has four isolated $A_1$ singularities on its $f_{\infty}$ fiber.

Although $B/\alpha_{B}$ is the surface of primary interest, it is useful to consider the related manifold $\widehat{B/\alpha_{B}}$, obtained from $B/\alpha_{B}$ by blowing-up each of these four $A_1$ singularities into a $\cp{1}$. The result is a smooth elliptic fibration which has six $I_1$ fibers. However, one can show that the double fiber with four isolated $A_1$ singularities will become a $I_0^{*}$ Kodaira fiber  once the  singularities are resolved. This is indicated in Figure~\ref{fig2}. Thus, $\widehat{B/\alpha_{B}}$ is a smooth elliptic fibration with seven degenerate fibers, six $I_1$ and one $I_0^{*}$. Recall from (\ref{I_1}) that $\chi(I_1)=1$. Furthermore, it is well-known that
\begin{equation}\label{77}
\chi(I_0^{*})=6.
\end{equation}
It follows that the Euler characteristic of $\widehat{B/\alpha_{B}}$ is
\begin{equation}\label{78}
\chi(\widehat{B/\alpha_{B}})=6\chi(I_1)+\chi(I_0^{*})=12
\end{equation}
and, hence, $\widehat{B/\alpha_{B}}$ is a rational elliptic surface. That is, $\widehat{B/\alpha_{B}}=d{\mathbb P}_9$. The singular fibration $\beta_{\alpha_B} : B/\alpha_{B} \to \cp{1}$ is its Weierstrass model. 

We can summarize  the relationship of $B/\alpha_{B}$ to $B$ as follows. The map
\begin{equation}\label{79}
\kappa : B \to B/\alpha_{B} 
\end{equation}
is a double cover of $B/\alpha_{B}$. However, since $f_0$ is point-wise fixed and $e,e{'},e{''},e{'''}$ on $f_\infty$ are fixed under $\alpha_B$, it follows that $\kappa$ has a branch locus in $B/\alpha_B$ consisting of all of $f_0$ and the four singular points in $f_\infty$. That is, the double cover of $B/\alpha_{B}$ is branched at all these points. It was shown in \cite{dopw-i} that for a  given rational elliptic surface $\widehat{B/\alpha_{B}}$ with six $I_1$ fibers and one $I_0^{*}$ fiber, $B$ is the unique double cover of $B/\alpha_{B}$ branched over the fiber $f_0$ and the four singular points in $f_{\infty}$. Therefore, a given $B/\alpha_{B}$ determines $B$ and  $\alpha_B$.

\begin{figure}[!ht]
\begin{center}
\input{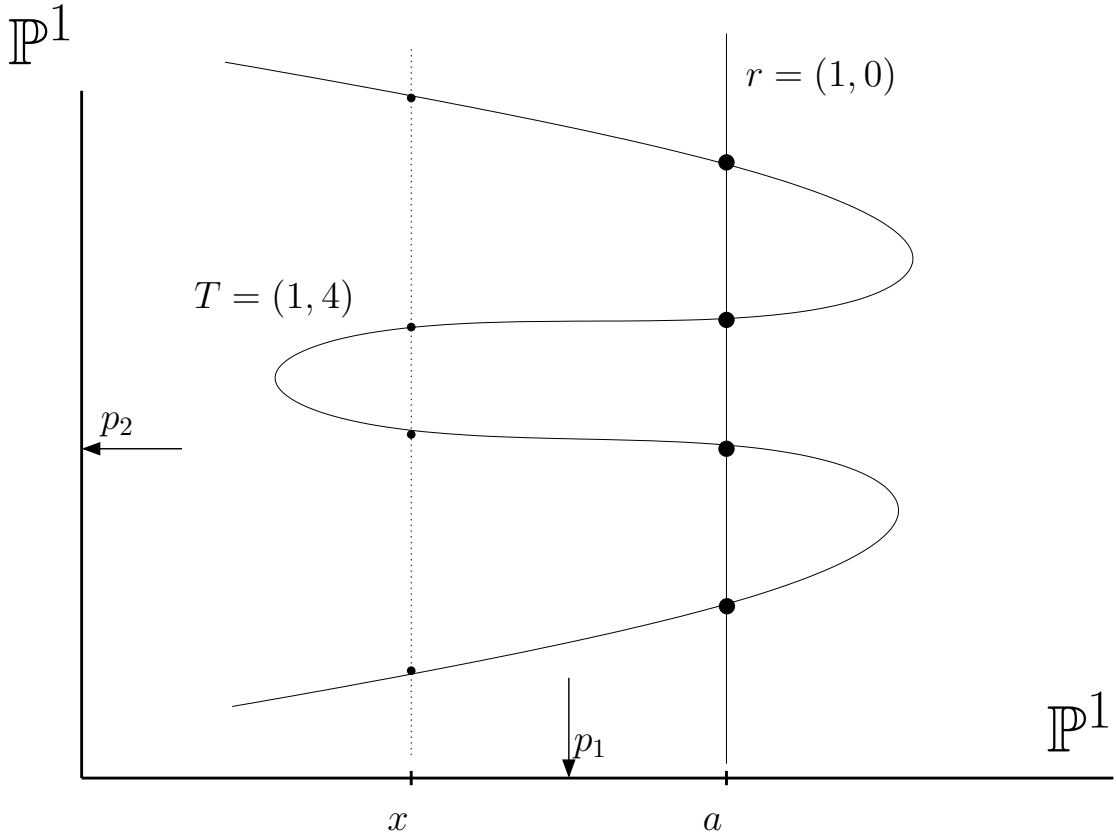} 
\end{center}
\caption{The bidegree $(2,4)$ curve $M=T\cup r$ in $Q=\cp{1}\times \cp{1}$. The projections $p_i: Q \to \mathbb{P}_i^1, i=1,2$ are explicitly shown.}
\label{fig3} 
\end{figure}

Thus far, we have discussed  generic properties of the quotient $B/\alpha_{B}$. To go further, we will give a complete geometric construction of $B/\alpha_{B}$. Specifically, every surface $B/\alpha_{B}$ can be obtained as a double cover of $Q=\cp{1}\times \cp{1}$ branched along a curve $M \subset Q$ of bidegree $(2,4)$ which splits as a union of two curves $T$ and $r$, of bidegree $(1,4)$ and $(1,0)$ respectively. That is
\begin{equation}\label{}
M= T \cup r.
\end{equation} 
To see this, first define the projection maps
\begin{equation}\label{pq}
p_i : Q \to \cp{1}_i
\end{equation}
for $i=1,2$ where $\cp{1}_i$ is the $i$-th $\cp{1}$ factor in $Q=\cp{1}\times \cp{1}$.
Let $z_i$ be the affine coordinates of $\mathbb{P}_i^1$ for $i=1,2$ respectively. In terms of these coordinates, the curve $r \subset Q$ of bidegree $(1,0)$ must satisfy the linear equation
\begin{equation}\label{ta}
-a +z_1 =0,
\end{equation}
where $a$ is a complex number. Hence, $z_1=a$. We refer the reader to Figure~\ref{fig3} for a pictorial representation of the curve $r$. The curve $T \subset Q$ of bidegree $(1,4)$, on the other hand, must satisfy an equation that is linear in $z_1$ and of degree four in $z_2$. A generic curve of this type is pictured in Figure~\ref{fig3} as well. 

Clearly, for the curve $r$ located at fixed $z_1=a$, $T$ will generically intersect $r$ at four distinct points. In this paper, we will assume that $T$ is smooth and intersects $r$ transversely. Having specified $Q$ and the curve $M$, let us now construct the double cover. Denote by $W_M$ the double cover of $Q$ branched along the curve $M$, and denote the double cover map by $\pi$. That is   
\begin{equation}\label{}
\pi : W_M \to Q .
\end{equation}
Combining this with (\ref{pq}), we see that there are two natural projection maps
\begin{equation}\label{pt}
\tilde{p}_i :  W_M  \to \cp{1}_i
\end{equation}
for $i=1,2$, where 
\begin{equation}\label{}
\tilde{p}_i = p_i \circ \pi.
\end{equation}
We will be primarily interested in $\tilde{p}_1 :  W_M  \to \cp{1}_1  $. We begin our analysis of $W_M$ by considering a generic point $x \in \cp{1}_1$, where $x \neq a$. Then $p^{-1}_1(x)$ is a $\cp{1}$, pictorially given by the dashed line in Figure~\ref{fig3}, that intersects the curve $T$ at four points. Its pre-image in $W_M$, $\tilde{p}^{-1}_1(x)$, is then the double cover of $p^{-1}_1(x)\cong \cp{1}$ branched over the four points where $p^{-1}_1(x)$ intersects $T$. We now show this is a torus $T^{2}$.

\begin{figure}[!ht]
\begin{center}
\input{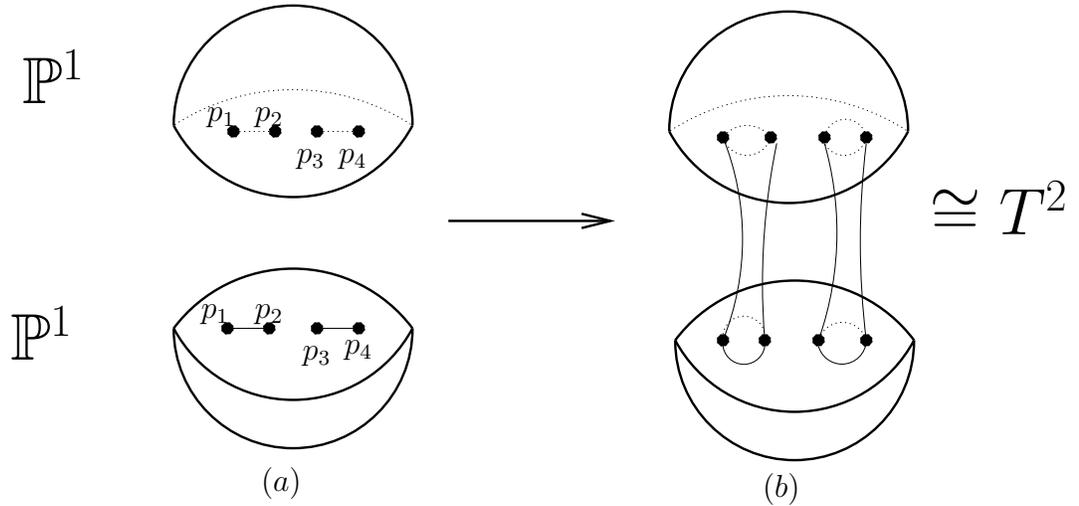} 
\end{center}
\caption{The double cover of $\cp{1}$ branched over four points.}
\label{fig4} 
\end{figure}

Having a double cover $\mathcal{C}$ of $\cp{1}$ which is branched over $4$ points means that  there is a map
\begin{equation}
c: \mathcal{C}\to \cp{1}
\end{equation}
such that the generic fiber of $c$ contains two points. However, there are four special points $b_i, i=1,...,4$ in $\cp{1}$ whose pre-images consist  of only a single point $R_i$. The  points $b_i$  are called branch points and the points $R_i$ in $\mathcal{C}$   are called ramification points. Here, since only two points get identified to a single point, they are called simple ramification points. Now $\mathcal{C}$ can be described as follows. Take two copies of $\cp{1}$ lines, we will call them the upper and lower sheets, and fix four points $p_1,...,p_4$ on each of them. Order the points pairwise and cut the line between the points of each pair, as shown in Figure~\ref{fig4}$a$. Open the cuts and identify the upper and lower sheets along them. The result of doing this is  a torus $T^2$. This is pictured in Figure~\ref{fig4}$b$.

\bigskip
\newpage
\noindent
It follows that
\begin{equation}\label{WM}
\tilde{p}_1 :  W_M  \to \cp{1}_1
\end{equation}
is a torus fibration, since the fiber $\tilde{p}_1^{-1}(x) $ over a generic point $x \in \cp{1}$ is a torus.

\begin{figure}[!ht]
\begin{center}
\input{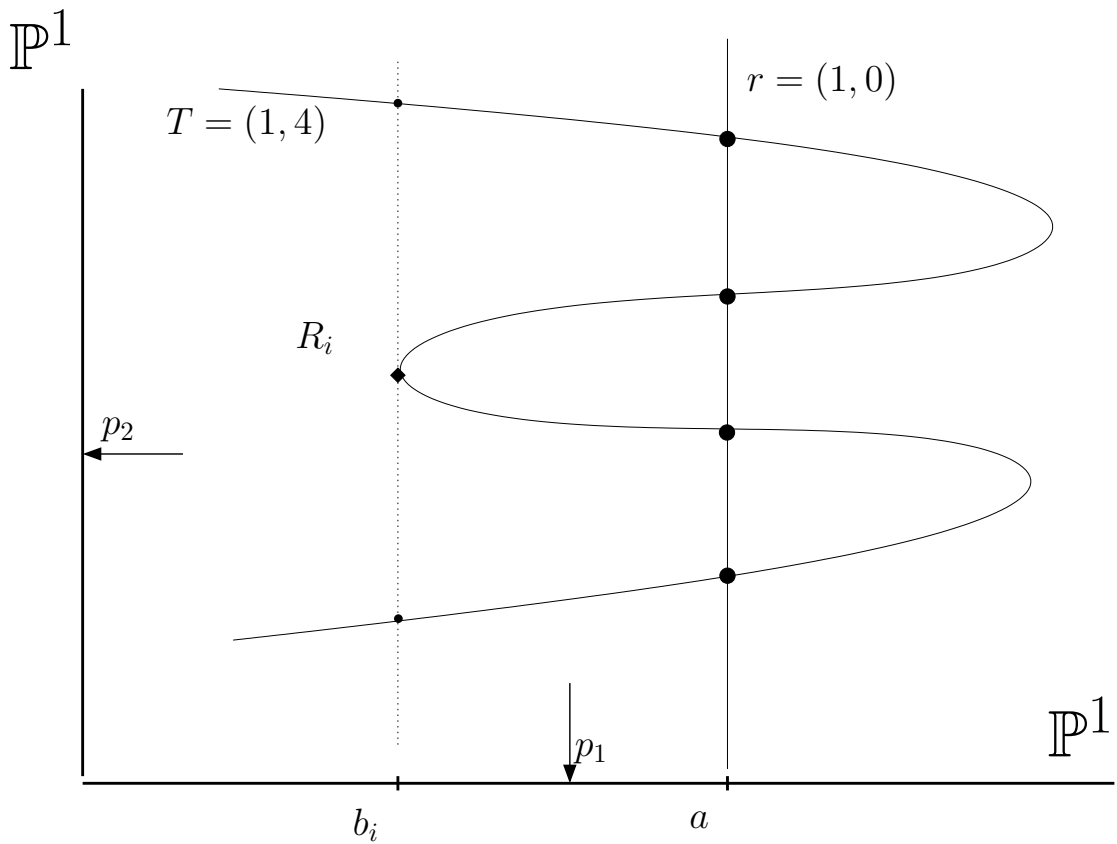} 
\end{center}
\caption{A schematic representation of the curve $T$ on $Q=\cp{1}\times \cp{1}$ indicating a simple ramification point $R_i$ over the branch point ${b_i}$ in $\cp{1}$.}
\label{fig5} 
\end{figure}

Are there any bad fibers in $W_M$ ? To explore this, we have to analyze more closely the structure of the curve $T \subset Q$. A naive form of $T$ was illustrated in Figure~\ref{fig3}. Consider, for a moment, the projection ${p}_2 :  Q  \to \cp{1}_2  $. Clearly, $p_2$ restricted to $T$ is a degree one cover of $\cp{1}$ and, hence,
\begin{equation}\label{ET}
\chi(T)=\chi(\cp{1})=2.
\end{equation}
Now consider ${p}_1 :  Q  \to \cp{1}_1  $.  Under $p_1$, $T$ is clearly a degree four cover of $\cp{1}$. Then the Riemann-Hurwitz formula states that for simple ramification points of $T$ we have
\begin{equation}\label{ET1}
\chi({T})=4\chi({\cp{1}\setminus \{b_i\})}+3A,
\end{equation}
where the $4$ arises since $T$ is a degree four cover and $\{b_i\}$ for $i=1,...,A$ are the branch points of the simple ramification points of $T$ over $\cp{1}_1$. For more information about the Riemann-Hurwitz formula, we refer the reader to \cite{gh}. By definition, above each such point  $T$  intersects $p^{-1}_1(b_i)$ in only three, not four, points. At  one of these points, the ramification point $R_i$, the fiber $p^{-1}_1(b_i)$ intersects $T$ tangentially. This is shown in Figure~\ref{fig5}.
Therefore, the branch locus for the double cover $\tilde{p}^{-1}_1(b_i)$ consists of three instead of four points. Looking back to Figure~\ref{fig4}, we see that over the points $b_i$ one of the pairs, for example $(p_1,p_2)$  degenerates to a single point. This implies that one of the $A$-cycles of the torus shrinks to zero. Therefore, as discussed earlier,  one obtains a $I_1$ fiber. Since the curve $M$ is smooth at this ramification point, the singularity is in the fiber only and not in $W_M$. Therefore, we need not blow-up this point. Combining  (\ref{ET}) and (\ref{ET1}), we can solve for the number $A$ of such branch points. We find that 
\begin{equation}\label{}
A=6.
\end{equation}
We conclude that there are six $I_1$ fibers in $W_M$ that arise in this manner.

Are there other degenerate fibers? Let us consider the specific point $a \in \cp{1}_1$. Then, $p^{-1}_1(a)=r$ is a $\cp{1}$ marked at the four points where it intersects $T$. Hence, one might imagine that $\tilde{p}_1^{-1}(a)$ is simply two copies of $\cp{1}$ identified at the four marked points and, hence, is a smooth torus. However, one must keep in mind that $p^{-1}_1(a)=r$ is itself in the branch locus and, hence, the two $\cp{1}$ lines must be identified as well. This produces a double line with four isolated singular points of type $A_1$. Unlike the $I_1$ fiber singularity at a ramification point, each of these four points is singular in $W_M$. This follows from the assumption that the curves $T$ and $r$ intersect, so the branch locus for $W_M$ is not smooth at these points. More precisely, since these curves intersect  transversally at these points, the singularity is an $A_1$ surface singularity. This has precisely the properties of the $f_\infty$  fiber in $B/\alpha_B$ discussed earlier that can be resolved into $I_0^{*}$.

To finish our analysis, let us chose any one of the four points of intersection of curves $T$ and $r$, and let $u$ be a curve of bidegree $(0,1)$ passing through that point. Then $\pi^{-1}(u)$ is two copies of $\cp{1}$ intersecting at that point. Clearly, each one of these $\cp{1}$ lines is a section of the fibration (\ref{WM}). Choose either copy as the zero section. Hence, the fibration (\ref{WM}) is an elliptic fibration.
Putting everything together, we see that $W_M$ in (\ref{WM}) is an elliptic fibration with six $I_1$ fibers and one fiber consisting of a double line with four isolated $A_1$ type surface singularities which can be blown-up to an $I_0^{*}$ Kodaira fiber. We conclude that
\begin{equation}\label{90}
W_M \cong B/\alpha_{B},
\end{equation}
as claimed above. Note, that this allows us to identify the point $0$ and the fiber $f_0$ in $W_M$ and, hence, in $Q$.

This characterization of $B/\alpha_{B}$ as the double cover of $Q$ branched over the curve $M$ allows us to analyze the properties of  $B/\alpha_{B}$ in detail. In particular, we can compute the dimension of the moduli space of $B/\alpha_{B}$. It is clear from the above construction that, up to some caveats we discuss below, the moduli of $W_M$ is the same as the moduli of the curve $M=T \cup r$ in $Q$. Consider some curve $C$ of bidegree $(n_1,n_2)$. Then, the number of parameters for $C$ is given by the dimension of the space 
\begin{equation}\label{3.16}
|{\mathcal O}_{Q}(n_1,n_2)|\equiv\mathbb{P}H^0(Q,{\mathcal O}_{Q}(n_1,n_2)),
\end{equation}
that is, the projectivization of the space of global sections of ${\mathcal O}_{Q}(n_1,n_2)$. These sections are specified by a polynomial of degree $n_1$ in the coordinate $z_1$ of $\cp{1}_1$ and of degree $n_2$ in the coordinate $z_2$ of $\cp{1}_2$, which have $n_1+1$ and $n_2+1$ parameters respectively. It follows that
\begin{equation}\label{91}
\dim |{\mathcal O}_{Q}(n_1,n_2)|=(n_1+1)(n_2+1)-1.
\end{equation}
Note that one must subtract $1$ in (\ref{91}) since one parameter can always be scaled away. We can now consider the component curves of $M$. The curve $r$ has bidegree $(1,0)$. Hence, using (\ref{91}), we see that 
\begin{equation}\label{r}
\dim|{\mathcal O}_{Q}(1,0)|=1.
\end{equation}
This is consistent with equation (\ref{ta}) and simply tells us that the complex constant $a$ is the modulus of $r$. The curve $T$ has bidegree $(1,4)$. It follows from (\ref{91}) that
\begin{equation}\label{T}
\dim |{\mathcal O}_{Q}(1,4)|=9.
\end{equation}
One might imagine that the total number of moduli of curve $M=T\cup r$
is simply the sum of (\ref{r}) and (\ref{T}). However, one must be careful to subtract out the automorphisms of $Q=\cp{1}\times \cp{1}$. Since Aut$(\cp{1})={\mathbb P}GL(2)$, it follows that $\dim$Aut$(\cp{1})=3$ and hence
\begin{equation}\label{94}
\dim \text{Aut} (\cp{1}\times \cp{1})=6.
\end{equation}
Furthermore, recall that in addition to $f_\infty$, which is specified by the location modulus $a$ of the curve $r$, there is the invariant fiber $f_0$ in $B/\alpha_B$.  The location of this zero point in $\cp{1}$ is yet another complex modulus. We conclude, therefore, that the moduli of $W_M \cong B/\alpha_{B}  $ are the moduli of $M$ reduced by the automorphisms of $\cp{1}\times \cp{1}$ but enhanced by the location modulus of fiber $f_0$. That is
\begin{equation}\label{}
\dim {\mathcal M}(B/\alpha_{B})=1+9-6+1=5,
\end{equation}
where ${\mathcal M}(B/\alpha_{B})$ stands for the moduli space of $B/\alpha_{B}$. Recall from (\ref{79}) that $B$ is the double cover of $B/\alpha_{B}$ branched over $f_0$ and the four singular points in $f_\infty$. Since this double cover is completely determined by the above quantities, it  has the same moduli space. Therefore, we have shown that
\begin{equation}\label{}
\dim {\mathcal M}(B)=5,
\end{equation}
where ${\mathcal M}(B)$ denotes the moduli space of $B$.

In this section, we have given a rather complete characterization of the rational elliptic surfaces $B$ that admit an involution  $ \alpha_B$ which leaves the zero section invariant and acts non-trivially on the base $\cp{1}$. The space of such surfaces is specified by five parameters. Recalling from (\ref{dim-8}) that generic rational elliptic surfaces are specified by eight parameters, we see that  the existence of such involution puts severe restrictions on $B$. From now on we will restrict ourselves to rational elliptic surfaces which are contained in this $5$ dimensional family and, hence,  an involution $\alpha_B$ will always exist.

\section{Surfaces with a ${\mathbb Z}_2$ Group of Automorphisms }
In the previous section, we discussed rational elliptic surfaces with an involution $\alpha_B$ leaving the zero section fixed. But, as stated at the end of Section~\ref{INV}, to construct freely acting involutions on the Calabi-Yau threefolds $X$ we need to find sections $\xi \neq e$ which fulfill condition (\ref{xi}). It was shown in \cite{dopw-i} that for each member of  the five dimensional family of rational elliptic surfaces just discussed, there is a rank four lattice of such sections. Since these are the surfaces which have a least one involution which can lead to a freely acting involution on the Calabi-Yau space  $X$, we will call them surfaces with a ${\mathbb Z}_2$  group of automorphisms.

We will further restrict ourselves to a four dimensional sub-family of surfaces with a ${\mathbb Z}_2$ group of automorphisms. We do so for the following two reasons.
First, the restriction to this sub-family will introduce new fiber classes. As shown in \cite{dopw-ii}, these classes are a necessary ingredient to construct stable, rank five vector bundles on $X$ which lead to standard models in string theory compactifications. Secondly, for the surfaces $B$ of this four dimensional family, we will be able to solve explicitly for some sections $\xi$.

Consider the five dimensional family of surfaces $B$ discussed so far. This family was obtained by constructing $B/\alpha_{B}\cong W_M$ as the double cover of $Q=\cp{1}\times \cp{1}$ branched over a curve $M \subset Q$ of bidegree $(2,4)$ which split as $M=T\cup r$, where $T$ is a smooth curve of bidegree $(1,4)$, $r$ has bidegree $(1,0)$ and we have assumed that $T$ and $r$ intersect transversely. In this section, we will further restrict $B$ by adding the requirement that
\begin{equation}\label{}
T=t\cup s,
\end{equation}
where the curve $t$ has bidegree $(1,3)$ and the curve $s$ has bidegree $(0,1)$. That is, we henceforth assume
\begin{equation}\label{M}
M=t\cup s \cup r.
\end{equation} 

\begin{figure}[!ht]
\begin{center}
\input{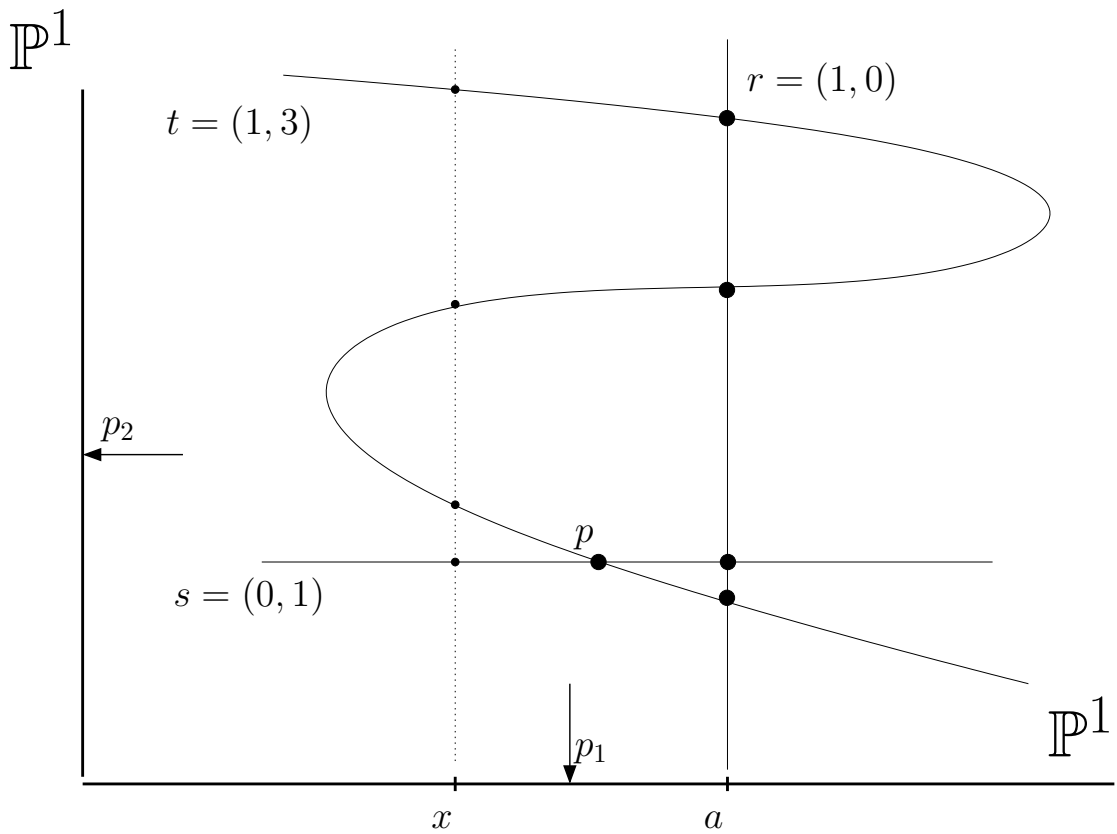} 
\end{center}
\caption{The bidegree (2,4) curve $M=t\cup s \cup r$ in $Q=\cp{1}\times \cp{1}$.}
\label{fig6} 
\end{figure}

In terms of the affine coordinates $z_i$ in  $\cp{1}_i$ for $i=1,2$,  the curve $s\subset Q$ of bidegree $(0,1)$ must satisfy the linear equation
\begin{equation}\label{99}
-b+z_2=0,
\end{equation}
where $b$ is a complex constant. Hence, $z_2=b$. We refer the reader to Figure~\ref{fig6} for a pictorial representation of the curve $s$, as well as the curve $r$ of bidegree $(1,0)$ given by (\ref{ta}).
 The curve $t$ of bidegree $(1,3)$, on the other hand, must satisfy an equation that is linear in $z_1$ and is of degree three in $z_2$. A generic such curve is described in Figure~\ref{fig6}. Note that the curves $t$ and $s$ intersect at a single point, which we will denote by $p$. Clearly, for the curve $r$ located at fixed $z_1=a$, $s$ will intersect $r$ at one point whereas the curve $t$ will generically intersect $r$ at three points. As before, we will assume that these intersections are transversal. Having specified the curve $M$ in (\ref{M}), we now analyze the double cover $W_M$ of $Q$ branched over $M$. As discussed above, $W_M$ is a fibration
\begin{equation}\label{100}
\tilde{p}_1 : W_M \to \cp{1}_1
\end{equation}
over $\cp{1}$. 

We begin by considering a generic point $x\in \cp{1}_1$ where $x\neq a$. Then $p_1^{-1}(x)$ is a $\cp{1}$  that intersects the curve $T=t\cup s$ at four marked points. Exactly as in the previous section, $\tilde{p}_{1}^{-1}(x)$ is a double cover of $\cp{1}$  branched at four points and, hence, is a torus. It follows that (\ref{100}) is a torus fibration, since the fiber $\tilde{p}_{1}^{-1}(x)$ over a generic point $x\in \cp{1}_1$ is a torus.

To explore possible degenerate fibers in $W_M$, we have to analyze more closely the structure of the curves $t$ and $s$ in $Q$. Consider, for a moment, the projection $p_2 : Q \to \cp{1}_2$. Clearly, $p_2 : t \to \cp{1}$ is a degree one cover and, hence, $t\cong \cp{1}$. Then
\begin{equation}\label{101}
\chi(t)=\chi(\cp{1})=2.
\end{equation}
Now consider $p_1: Q \to \cp{1}_1$. Under $p_1$, $t$ is clearly a three-fold cover of $\cp{1}$ with generically simple ramification points. Then the Riemann-Hurwitz formula states that
\begin{equation}\label{102}
\chi(t)=3\chi(\cp{1}\setminus \{b_i\})+2A,
\end{equation}
where the $3$ arises since $t$ is a three-fold cover and $\{b_i\}$ for $i=1,...,A$ is the set of branch points of $t$ in $\cp{1}_1$ corresponding to the simple ramification points. By definition, each such branch point is an isolated point in $\cp{1}$ above which $t$ intersects $p^{-1}_1(b_i)$ in only two, not three, points. Combining (\ref{101}) and (\ref{102}), we can solve for the number $A$ of such ramification points. We find that 
\begin{equation}\label{}
A=4.
\end{equation}
Let $b_i$ be such a point in $\cp{1}$. As discussed previously,  the fiber $\tilde{p}_1^{-1}(b_i)$ over such a point is branched over three instead of four points and, hence, $\tilde{p}_1^{-1}(b_i)$ is singular and of Kodaira type $I_1$. Since curve $M$ is smooth, this is not a singularity in $W_M$ and need not be blown-up. We conclude that there are four $I_1$ fibers that arise in this manner. 

\begin{figure}[!ht]
\begin{center}
\input{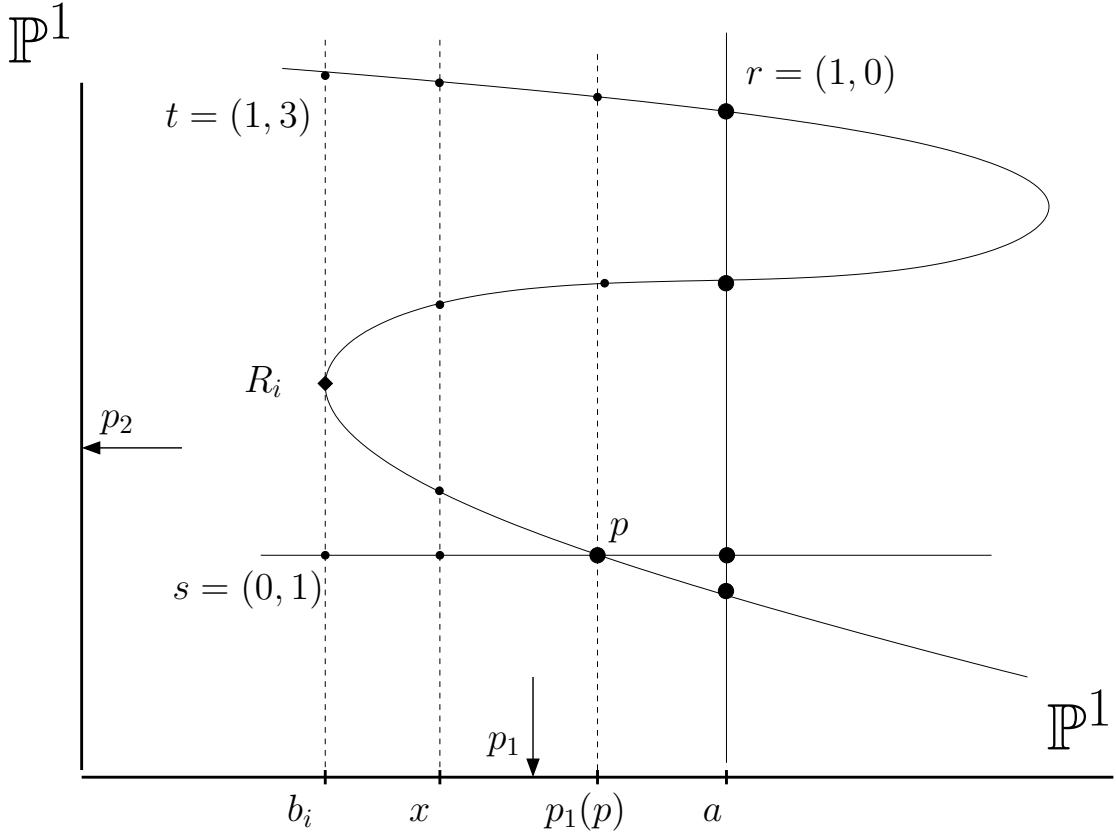} 
\end{center}
\caption{A schematic representation of curve $M=t\cup s \cup r$ showing a simple ramification point $R_i$ explicitly. We indicate the fibers over a generic point $x$, a branch point $b_i$, $p_1(p)$ and $a$.}
\label{fig7} 
\end{figure}

Are there other degenerate fibers? Let us consider the intersection point $p$ of the curves $t$ and $s$. This projects to a point $p_1(p)$ in $\cp{1}$ and lies in the fiber $p_1^{-1}(p_1(p))$ in $Q$. Note from Figure~\ref{fig7}
that $p_1^{-1}(p_1(p))$ intersects $t$ alone at two points and both $t$ and $s$ simultaneously at the point $p$. It follows that $\tilde{p}_1^{-1}(p_1(p))$ is a double cover of $\cp{1}$ in $W_M$ branched at three points. Furthermore, since both $t$ and $s$ are in the branch locus, the branch locus of $W_M$ has an ordinary nodal  singularity at $p$. Therefore, at this point $W_M$ is singular with a singularity of  type $A_1 $. One can show that blowing-up this singular point into a $\cp{1}$ produces a Kodaira fiber of type $I_2$. We conclude that there is an isolated fiber in $W_M$ that blows-up into an $I_2$ fiber.

Now consider the specific point $a\in \cp{1}$. Then $p^{-1}_{1}(a)=r$ is a $\cp{1} $ which intersects $t$ in three points and has a single point of intersection with $s$. Then, exactly as in the previous section, we see that the fiber $\tilde{p}_1^{-1}(a)$ in $W_M$ consists of two copies of $\cp{1}$ identified at the four intersection points. Furthermore, since $t,s$ and $r$ are all components of $M$ which intersect transversely, it follows that the two $\cp{1}$'s must be identified and that each of the intersection points is an isolated singularity of type $A_1$ in $W_M$. This is precisely the $f_\infty$ fiber discussed earlier. Recall that it  can be resolved into an $I^{*}_0$.

To finish the analysis of $W_M$, note that curve $s$ is a curve of bidegree $(0,1)$ which intersects $r$. It follows that $s$ can play the role of the curve $u$ in the previous section.  Note, however, that $s$ is a component of the branch curve $M=t\cup s \cup r$ and  $s$ intersects  $t$ and $r$. It follows that $\pi^{-1}(s)$ is a double  $\cp{1}$ which passes through  two isolated $A_1$ singularities, one at the intersection with $\pi^{-1}(r)$ and one at the intersection with $\pi^{-1}(t)$.  Be that as it may, $\pi^{-1}(s)$ is a section of the fibration (\ref{100}), which we will choose to be the zero section.

Putting everything together, we see that $W_M$ in (\ref{100}) is an elliptic fibration with four $I_1$ fibers. There are two additional degenerate fibers. The first consists of  two $\cp{1}$ lines identified at three points, one of which is an $A_1$ singularity in $W_M$, which blows-up to give a $I_2$ fiber. The second degenerate fiber is a double line with four isolated type $A_1$ singularities that can be blown-up to give a $I^{*}_0$ Kodaira fiber.

Having discussed $W_M$, we use the identification (\ref{90}) that $W_M\cong B/{\alpha_B}$ to construct the associated space $B$ and involutions $\alpha_B$. We will denote by $\widehat{W_M}$  the smooth surfaces $\widehat{{B/{\alpha_B}}}$, which are obtained by blowing up each isolated $A_1$ singularity in $B/{\alpha_B}$ with a $\cp{1}$. These surfaces are  elliptic fibration over $\cp{1}$ with four $I_1$ fibers, an $I_2$ fiber and one $I_0^{*}$ Kodaira fiber. Using  (\ref{18}) and (\ref{77}) we see that
\begin{equation}\label{}
\chi(\widehat{W_M})=12
\end{equation}
and, hence, $\widehat{W_M}\cong d{\mathbb P}_9$. The singular fibration $W_M\cong B/\alpha_B$ is its Weierstrass model.

\begin{figure}[!ht]
\begin{center}
\input{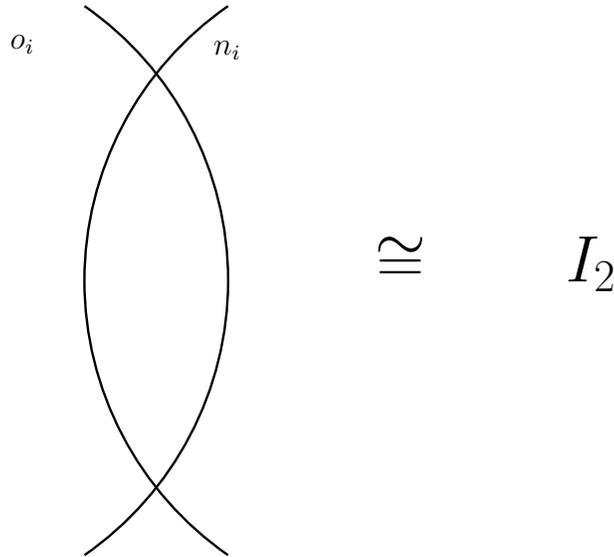} 
\end{center}
\caption{The form of the two reducible $I_2$ fibers consisting of $o_i \cup n_i$ for $i=1,2$.}
\label{fig8} 
\end{figure}

Now let us consider the space $B$. 
Recall from (\ref{79}) that $B$ is a double cover of $B/\alpha_B$ branched over the fiber $f_{0}$ and the four singular points $e,e{'},e{''}$ and $e{'''}$ on the fiber $f_{\infty}$. First consider the four $I_1$ fibers on $B/\alpha_B$. Since $\alpha_B$ must take fibers of the same Kodaira type to each other in order to be an involution, it follows that there must be eight $I_1$ fibers on $B$.
Next we discuss the fiber over $p_1(p)$, whose singular point is also  an $A_1$ singularity in $W_M$ and blows up to $I_2$ in $\widehat{W_M}$. Clearly, $\kappa^{-1}(\tilde{p}^{-1}(p_1(p)))$ are two disjoint copies of this fiber, each containing an $A_1$ singularity of $B$. Since $B$, unlike $B/\alpha_B$, must be smooth, it is essential that we blow-up each of the two singularities into a $\cp{1}$. This introduces two new classes of curves, which we denote by $n_1$ and $n_2$, beyond those associated with the generic surfaces $B$ discussed previously.
Be that as it may, we will continue to denote this blown-up surface  and its projection to $ B/\alpha_B$ by $B$ and $\kappa$ respectively. In addition, we will give names, $o_1$ and $o_2$, to the proper transforms of the old singular fibers. Recall that the proper transform of the singular curves are their pre-images in the blown-up surface $B$ reduced by the exceptional divisors $n_1$ and $n_2$. Both components, the proper transform and the new exceptional divisor, together form a new fiber of Kodaira  type $I_2$. These classes are illustrated in Figure~\ref{fig8}.
\noindent
Hence, there are two $I_2$ fibers on $B$. Now consider the fiber in $B/\alpha_B$ containing the original four isolated $A_1$ singularities of the surface $B$. Clearly, the pre-image of this fiber in $B$ must be the unique fiber $f_{\infty}$. The four singular points in $B/\alpha_B$ arise from the four invariant points $e,e{'},e{''},e{'''}$ on $f_{\infty}$ in $B$. That is, $f_{\infty}$ is smooth and needs no further resolution. 

We conclude that the surfaces $B$ constructed in this section have as degenerate fibers eight $ I_1$ and two $ I_2$ fibers. Note that this configuration satisfies equation (\ref{21}) for $n=4, m=2$. Therefore, these surfaces $B$ are, indeed, rational elliptic, that is $B=d{\mathbb P}_9$.

Finally, let us analyze the zero section of $B$. Recall that the section $\pi^{-1}(s)$ of $B/\alpha_B$ is composed of two identified $\cp{1}$ lines with two isolated $A_1$ singularities in $B/\alpha_B$, one at the intersection with $r$ and the other at the intersection with $t$. The pre-image of the curve $\pi^{-1}(r)$ under $\kappa$ is $f_{\infty}$. Hence, the $A_1$ singularity at the intersection with $r$ is simply the invariant point $e$ on $f_{\infty}$, which is smooth. However, the second $A_1$ singularity lives on $\pi^{-1}(t)$. Therefore, the pre-image of this point under $\kappa$ consists of the two isolated $A_1$ singularities.  These are the singularities that are blown-up by $n_1$ and $n_2$ to produce the non-singular surface $B$. Finally, note that, by construction, the zero section $e$ is stable under $\alpha_B$. It follows that $\kappa^{-1}(\pi^{-1}(s))$, after reducing it by the two exceptional divisors $n_1$  and $n_2$, is a single copy of $\cp{1}$ which intersects  $n_1$ and $ n_2$ and is a section of $B$. We choose the zero section $e$ of $B$ to be 
\begin{equation}\label{newzero}
e=\{\kappa^{-1}(\pi^{-1}(s))\setminus  n_1 \cup n_2\}.
\end{equation}

We can use the characterization of $B/\alpha_B$ as the double cover of $Q$ branched over the curve $M$ to compute the dimension of its moduli space, as well as the moduli space of $B$. Recall that $M=t\cup s \cup r$. The curve $r$ has bidegree $(1,0)$ and, hence, it follows from (\ref{91}) that
\begin{equation}\label{}
\dim|{\mathcal O}_Q(1,0)|=1.
\end{equation}
Similarly, since curves $s$ and $t$ have bidegree $(0,1)$ and $(1,3)$ respectively, we find from (\ref{91}) that 
\begin{equation}\label{106}
\dim|{\mathcal O}_Q(0,1)|=1
\end{equation}
and
\begin{equation}\label{}
\dim|{\mathcal O}_Q(1,3)|=7.
\end{equation}
Note that (\ref{106}) is consistent with equation (\ref{99}) and simply tells us that the complex constant $b$ is the modulus of $s$. As discussed previously, the moduli of $W_M\cong B/\alpha_{B}$ are the moduli of $M$ reduced by the number of automorphism of $\cp{1}\times \cp{1}$ given by (\ref{94}) plus the location of the fiber $f_0$. That is, 
\begin{equation}\label{}
\dim{\mathcal M}(B/\alpha_{B})=1+1+7-6+1=4,
\end{equation}
where ${\mathcal M}(B/\alpha_{B})$ is the moduli space of $B/\alpha_{B}$. Since surface $B$ is the unique double cover of $B/\alpha_{B}$ branched over $f_0$ and the four singular points on $f_{\infty}$, it follows that $B$ has the same number of moduli as $B/\alpha_{B}$, namely
\begin{equation}\label{}
\dim{\mathcal M}(B)=4.
\end{equation}
These results have a simple interpretation. Since the curve $t\cup s$ is a degeneration of the generic curve $T$, the rational elliptic surfaces $B$ obtained here are a subset of the generic surface $B$ constructed from $M=T\cup r$ in the previous section. Specifically, they are the $d{\mathbb P}_9$ surfaces where four of the twelve $I_1$ fibers have coalesce pairwise to form two $ I_2$ Kodaira fibers. These surfaces form a four parameter subspace of the full five parameter moduli space of surfaces $B$.

\begin{figure}[!ht]
\begin{center}
\input{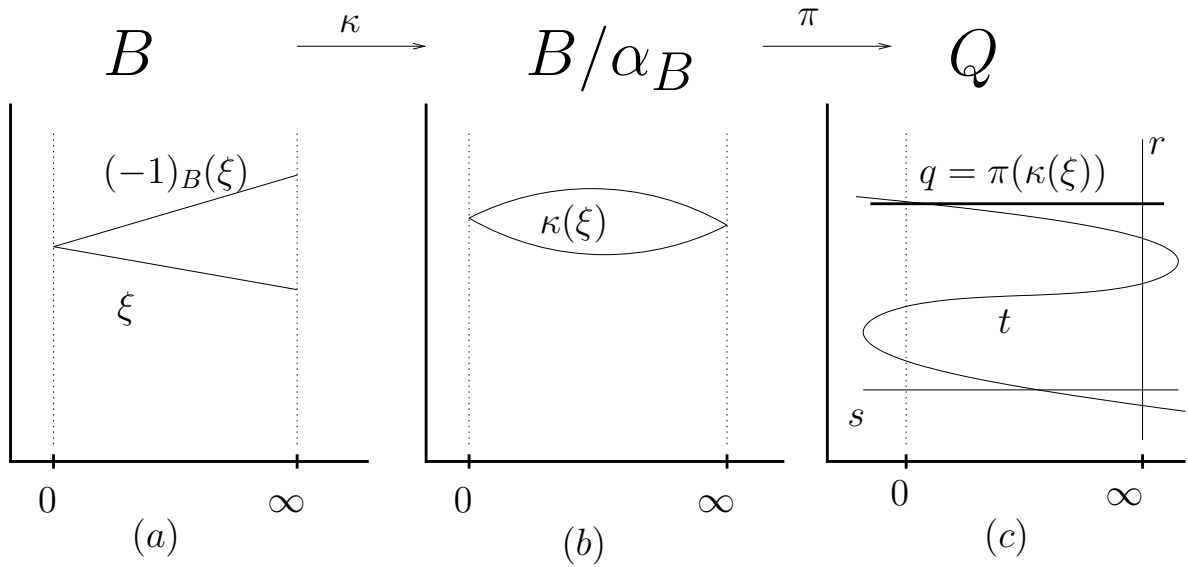} 
\end{center}
\caption{Successive images of a section $\xi$ under the various quotient maps.}
\label{fig9a} 
\end{figure}

Now, within this restricted four parameter class of surfaces, we will solve for  sections $\xi$ that satisfy equation (\ref{xi}), that is
\begin{equation}\label{110}
\alpha_B(\xi)=(-1)_B(\xi).
\end{equation}
To do so,  recall from (\ref{2.44}) and (\ref{2.45}) that generically solutions of (\ref{110}) come in pairs $(\xi,(-1)_B(\xi))$, with $\xi(0)=e, e{'},e{''}$ or $e{'''}$. See for example Figure~\ref{fig9a}$a$. The first thing to notice is that, since these sections map to each other under $\alpha_B$, it follows that their images after modding out by $\alpha_B$ are identical. That is 
\begin{equation}\label{}
\kappa(\xi)=\kappa((-1)_B(\xi)).
\end{equation}
Next, we are interested in the image of one of these sections, call it $\xi$, under the quotient maps $\kappa$ and $\pi$.

\bigskip
\noindent
Recall that $\xi \cong \cp{1}$. By construction  $\kappa(\xi\cup (-1)_B(\xi))$ is isomorphic to a single $\cp{1}$. However using the remark above that $\kappa(\xi)=\kappa((-1)_B(\xi))$, we see that  $\kappa(\xi)\cong \cp{1}$ as well. It is clear that $\kappa(\xi)$ intersects each fiber of $B/\alpha_B$ at least once. Noticing that  $\kappa$ identifies two different fibers of $B$ and that, generically, $\xi$ intersects these fibers at different points, we see that actually $\kappa(\xi)$ intersects each fiber of $B/\alpha_B$ twice. Hence $\kappa(\xi)$ is a double section of $B/\alpha_B\cong W_M$. Recalling that $\beta_{\alpha_B}: W_M\to \cp{1}$, it follows that the image ${\kappa(\xi)}$ is a double cover  of $\cp{1}$ induced by the restriction of the projection $\beta_{\alpha_B}$ to $\kappa(\xi)$. Naively, it therefore appears that  ${\kappa(\xi)}$ consists of two copies of $\cp{1}$. But we have just shown that $\kappa(\xi)\cong \cp{1}$. These two points of view  can be reconciled by noting that the double cover is branched over two points, $0$ and $\infty $, in the base $\cp{1}$. Therefore, one of the ramification points is in the fiber $f_0$ and the other in $f_{\infty}$ of $W_B$. This situation is indicated in   Figure~\ref{fig9a}$b$.
It can be shown that $\kappa(\xi)$ is stable under the covering involution for $\pi:W_M\to Q$. Since $\pi$ projects each fiber of $W_M$ to a unique fiber of $Q$, the two points of $\kappa(\xi)$ intersecting a fiber in $B/\alpha_ B$ get identified under $\pi(\kappa(\xi))$. Therefore, the image of $\xi$ in $Q$ is a section,  which we call $q$. Assuming that the double section $\kappa(\xi)$  is smooth, it can be shown that $q$ has bidegree $(0,1)$.  Henceforth, we will always make that assumption.
Furthermore, since the pre-image $\pi^{-1}(q)$ has only one point in the fibers $f_0$ and $f_{\infty}$ in $B/\alpha_B $, $q$ needs to intersect the branch locus of  $B/\alpha_B$ at the fibers in $Q$ over $0$ and $\infty$. One such curve $q$ is shown in Figure~\ref{fig9a}$c$.

\begin{figure}[!ht]
\begin{center}
\input{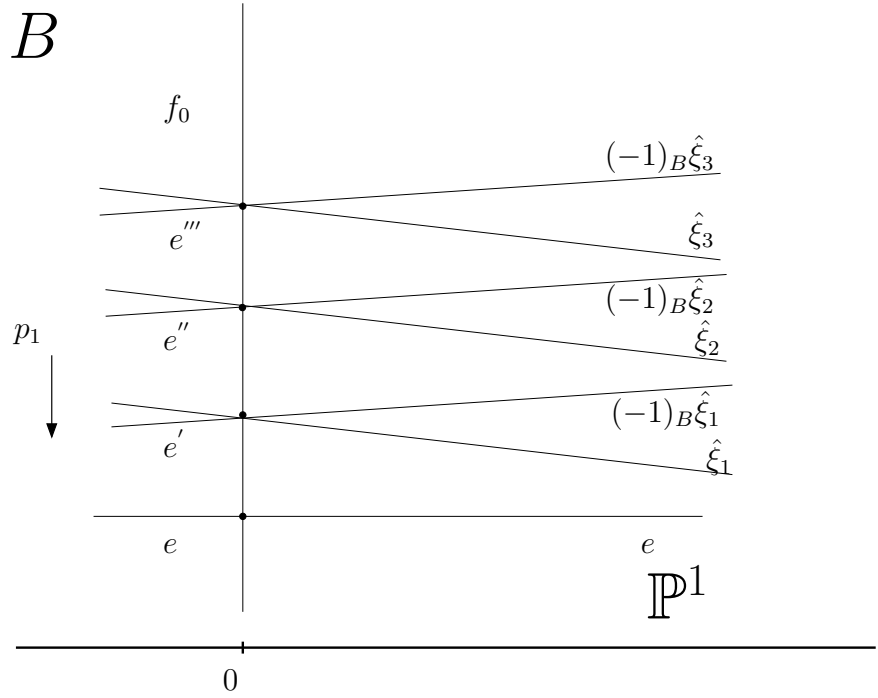} 
\end{center}
\caption{Diagram of $B$ showing the fiber $f_0$, the four points of order two on $f_0$ and the seven sections of $B$ satisfying $\alpha_B (\xi) =(-1)_B (\xi)$.}
\label{fig9} 
\end{figure}

Clearly, there are three distinct curves $q_i, i=1,2,3$ in $Q$, each of bidegree $(0,1)$, which satisfy these conditions. Their pre-image in $B$ are three pairs of sections $(\hat{\xi}_i,(-1)_B(\hat{\xi}_i))$ for $i=1,2,3$ passing through the invariant points $e{'},e{''},e{'''}$ of $f_0$ respectively. These sections, as well as the zero section, are shown in Figure~\ref{fig9}. We conclude that for restricted surfaces $B$, there are three pairs of  sections satisfying (\ref{xi}). 

\bigskip

\section {Surfaces with a ${{\mathbb Z}_2\times{\mathbb Z}_2}  $ and a ${{\mathbb Z}_2\times{\mathbb Z}_2\times{\mathbb Z}_2}  $ Group of Automorphisms }
Thus far, we have considered surfaces $B$ that admit involutions $\tau_B=t_{\xi}\circ \alpha_B$. Are there surfaces $B$ which admit two commuting involutions of this form? This can indeed be achieved. We will show that we have to further  restrict  to a three dimensional sub-family in the moduli space of $B$ in order to obtain two commuting involutions. Since these surfaces $B$ are the ingredients used to construct Calabi-Yau spaces $X$ with two commuting, freely acting involutions, we will call them surfaces with a ${\mathbb Z}_2\times{\mathbb Z}_2$ automorphism group.

To find the constraints on such surfaces, consider two involutions of the form 
\begin{equation}\label{}
\tau_{Ba}=t_{\xi_a}\circ \alpha_B
\end{equation}
for $a=1,2$, where each $\xi_a$ is one of the sections $\hat{\xi_1},\hat{\xi_2}$ or $\hat{\xi_3}$. Since $\tau_{B1}^2=\tau_{B2}^2=id$, it is straightforward to see that these two involutions will commute,
\begin{equation}\label{}
\tau_{B1}\circ\tau_{B2}=\tau_{B2}\circ\tau_{B1},
\end{equation}
if and only if
\begin{equation}\label{117}
(\tau_{B1}\circ\tau_{B2})^2=id.
\end{equation}
Let us see under what conditions this will be true. Note that
\begin{equation}\label{118}
\tau_{B1}\circ\tau_{B2}=t_{\xi_1-\xi_2},
\end{equation}
where we have used $\tau_{B2}=t_{\xi_2} \circ \alpha_B=\alpha_B \circ t_{-\xi_2}$. Hence, it follows from (\ref{117}) and (\ref{118}) that $\tau_{B1}$ and   $\tau_{B2}$ will commute if and only if 
\begin{equation}\label{119}
(t_{\xi_1-\xi_2})^2=id.
\end{equation}
Therefore, $\xi_1-\xi_2$ must be a section of $B$ whose intersection with any fiber $f$ is an invariant point of that fiber under $(-1)_B$. Those points are  the  points of order two.
One possible solution is $\xi_1-\xi_2=e$. But then $\xi_1=\xi_2$ and, hence, $\tau_{B1}=\tau_{B2}$. Therefore, we must find another section that is invariant under $(-1)_B$ which is not the zero section. For the surfaces $B$ studied thus far, there is no such section. However, as we will presently demonstrate, such a section will exist for surfaces $B$ whose quotient $B/\alpha_{B}$ is the double cover of $Q$ branched over curve $M$ of the form (\ref{M}) with the additional restriction that  
\begin{equation}\label{}
t=t^{'}\cup i,
\end{equation}
where the curve $t^{'}$ has bidegree $(1,2)$ and the curve $i$ bidegree $(0,1)$. That is, we henceforth assume
\begin{equation}\label{121}
M=t^{'}\cup i \cup s \cup r.
\end{equation}
As mentioned at the end of Section~\ref{INV}, a purely translational involution of the form (\ref{119}) can lift to freely acting involution on $X$.

\begin{figure}
\begin{center}
\input{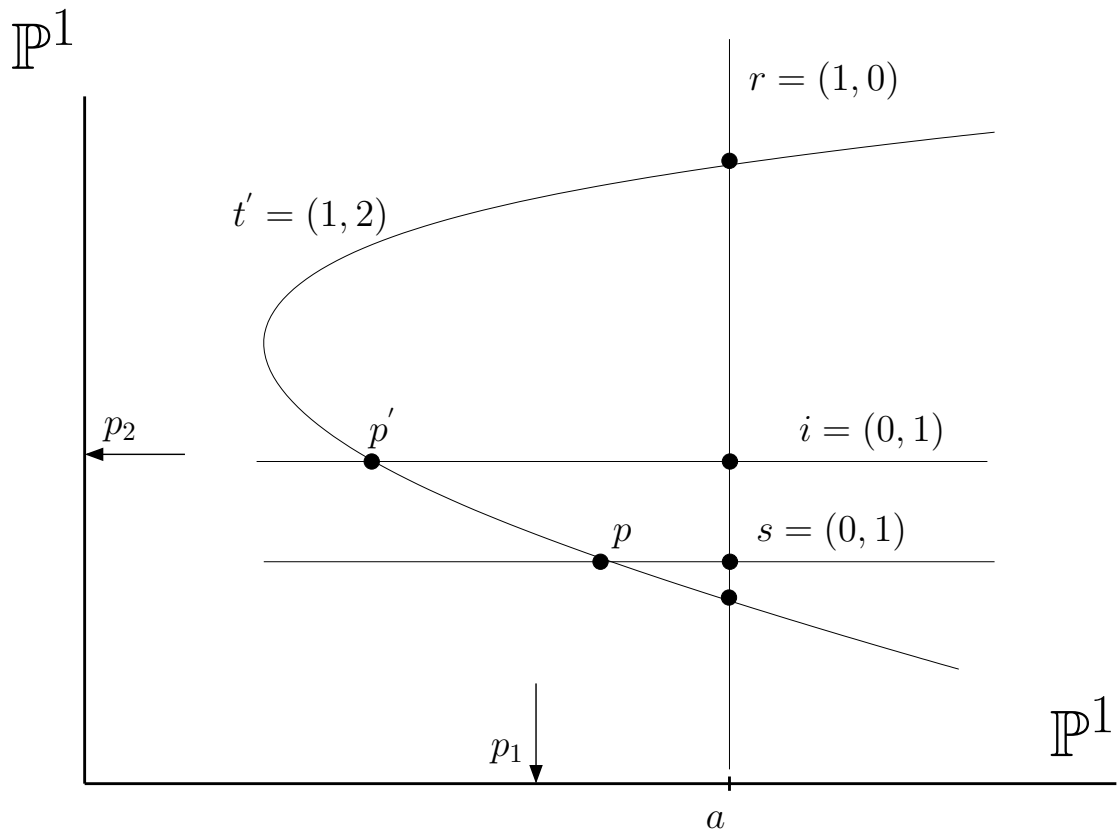} 
\end{center}
\caption{A schematic representation of curve $M=t^{'}\cup i \cup s \cup r$ in $Q=\cp{1}\times \cp{1}$.}
\label{fig10} 
\end{figure}

Recall that $z_i$ are the complex coordinates of $\cp{1}_i$ for $i=1,2$. In terms of these coordinates, the curve $i\subset Q$ of bidegree $(0,1)$ must satisfy the linear equation 
\begin{equation}\label{}
-c+z_2=0,
\end{equation}
where $c$ is a complex coefficient. Hence, $z_2=c$. We refer the reader to Figure~\ref{fig10}
\noindent
for a pictorial representation of curve $i$, as well as the curves $s$ and $r$ of bidegree $(0,1)$ and $(1,0)$ respectively given in (\ref{ta}) and (\ref{99}). The curve $t^{'}$, on the other hand, must satisfy an equation that is linear in $z_1$ and is of degree two in $z_2$. Note that $s$ and $i$ each intersect the curve $t^{'}$ at a single point, which we denote by $p$ and $p^{'}$ respectively. Clearly, the curve $r$ located at fixed $z_1=a$ will intersect each of $s$ and $i$ at a single point, whereas $r$ will generically intersect $t^{'}$ at two points. As before, we will assume that these intersections are transversal. Again, we refer the reader to Figure~\ref{fig10} for an illustration of these remarks.

Having specified the curve $M$ in (\ref{121}), we now analyze the double cover $W_M$ of $ Q$ branched over $M$. As discussed previously, $W_M$ is a fibration, $ \tilde{p}_1 : W_M \to \cp{1}_1$, over $\cp{1}$ where $\tilde{p}_1$ is defined by (\ref{pt}). We begin by considering a generic point $x \in \cp{1}$, where $x\neq a, p_1(p)$ or $ p_1(p^{'})$. Then $p^{-1}_1(x)$ is a $\cp{1}$  that intersects curve $T=t^{'}\cup i\cup s$ at four points. Therefore, $\tilde{p}_{1}^{-1}(x)$ is a double cover of  $\cp{1}$ branched at those four points. It follows that $\tilde{p}_{1}^{-1}(x)$  is a torus and, hence, the fibration (\ref{100}) is a torus fibration. 

To explore the possible degenerate fibers of $W_M$, we have to analyze more closely the structure of the curves $t{'},i$ and $s$ in $Q$. Consider, for a moment, the projection $p_2: Q \to \cp{1}_2$. Clearly $p_2: t^{'} \to \cp{1}_2$ is a degree one cover and, hence, $t^{'}\cong \cp{1}$.
Now consider $p_1: Q \to \cp{1}_1$. Under $p_1$, $t^{'}$ is clearly a two-fold cover of $\cp{1}$. As already mentioned several times in this paper, the condition for a double cover of $\cp{1}$ to be isomorphic to  $\cp{1}$ implies that the cover has two branch points. 
Let $b_i,\;i=1,2$ denote these  points in $\cp{1}$. Clearly, for each $b_i$ $\tilde{p}_1^{-1}(b_i)$ is a double cover of $\cp{1}$ branched at three, not four, points, where one of these points is in $t^{'}$, one in $i$ and one in $s$. Similarly to the discussion in the previous section, the fiber $\tilde{p}_1^{-1}(b_i)$ will be singular of Kodaira type $I_1$. Since the branch curve is smooth at this point, this is not a singularity in $W_M$ and need not be blown-up. We conclude that there are two $I_1$ fibers that arise in this manner.

Are there other degenerate fibers? Let us consider the intersection point $p^{'}$ of the curves $t^{'}$ and $i$. This projects to a point $p_1(p^{'})$ in $\cp{1}_1$ and lies in the fiber $p_1^{-1}(p_1(p^{'}))$ in $Q$. Note from Figure~\ref{fig10} that $p_1^{-1}(p_1(p^{'}))$ intersects each of $t^{'}$ and $s$ at a single point and both $t^{'}$ and $i$ simultaneously at $p^{'}$. It follows that $\tilde{p}_1^{-1}(p_1(p^{'}))$ is a fiber in $W_M$ which is a double cover of  $\cp{1}$ branched at three points. Furthermore, this fiber must be singular at the point $p^{'}$. In addition, since, by assumption, $t^{'}$ and $i$ intersect, it follows that the branch locus of $W_M$ is not smooth at  $p^{'}$ and, hence, the singularity is of type $A_1$ in the space $W_M$. Therefore, as discussed previously, blowing-up the singular point into a  $\cp{1}$ produces a Kodaira fiber of type $I_2$ in $\widehat{W_M}$.
Exactly the same comments can be made concerning the intersection point $p^{}$ of the curves $t^{'}$ and $s$. We conclude that there are two isolated fibers in $W_M$, each of which blows-up into an $I_2$ fiber.
Now consider the specific point $a \in \cp{1}_1$. Then $p_1^{-1}(a)=r$ is a $\cp{1}$   which intersects $t^{'}$ in two points and has a single intersection with each of $i$ and $s$. Then exactly as in the previous cases, we see that fiber $\tilde{p}_1^{-1}(a)$  in $W_M$ consists  of two copies of $\cp{1}$ identified at the four intersection points. Since $t^{'}, i, s$ and $r$ are all components of $M$ which intersect transversely, it follows that the two $\cp{1}$ copies  must be identified and that each of the intersection points is an isolated singularity of type $A_1$ in $W_M$. This is precisely the $f_{\infty}$ fiber discussed earlier that can be resolved into $I_0^{*}$.

To finish the analysis of $W_M$, note that curve $s$ has exactly the same properties as it had in the restricted $\mathbb{Z}_2$ case in the previous section. Therefore, $\pi^{-1}(s)$ is a section of the fibration (\ref{100}), which we choose to be the zero section. Now consider the curve $i$ of bidegree $(0,1)$, which is new to this analysis. It is not hard to see that it has exactly the same properties as the curve $s$. That is, $\pi^{-1}(i)$ is two identified  $\cp{1}$ lines with two isolated $A_1$ singularities, one at the intersection with $r$ and one at the intersection with $t^{'}$. Since both intersections are transverse, these are singularities in the space $W_M$. Be that as it may, $\pi^{-1}(i)$ is a section of the fibration (\ref{100}) which is distinct from $\pi^{-1}(s)$.

Putting everything together, we see that $W_M$ in (\ref{100}) is now an elliptic fibration with two $ I_1$ fibers and two fibers with an $A_1$ singularity in $W_M$, which can be resolved by blowing-up into $I_2$ fibers in $\widehat{W_M}$. In addition, there is another degenerate fiber which is a double line containing four isolated $A_1$ singularities in $W_M$, that can be blown-up into an $I_0^{*}$ Kodaira fiber. 
Note that the surface $\widehat{W_M}$, obtained by blowing up each isolated $A_1$ singularity in $W_M$ with a $\cp{1}$, is an elliptic fibration over $\cp{1}$ with two $ I_1$ fibers, two  $ I_2$ fibers and one $I_0^{*}$ Kodaira fiber. Using (\ref{18}) and (\ref{77}), we see that 
\begin{equation}\label{126}
\chi(\widehat{W_M})=2\chi(I_1)+2\chi(I_2)+\chi(I_0^{*})=12
\end{equation}
and, hence, $\widehat{W_M}=d{\mathbb{P}_9}$. The singular fibration $W_M$ is its Weierstrass model.

Having discussed $W_M$ and its resolution $\widehat{W_M}$, we use the identification (\ref{90}) that $W_M\cong B/\alpha_B$ to construct the associated spaces $B$ and involutions $\alpha_B$.  Recall from (\ref{79}) that $B$ is the double cover of ${B}/\alpha_B$ branched over the fiber $f_0$ and the four singular points $e, e^{'}, e^{''}$ and $e^{'''}$ on the fiber $f_{\infty}$. 
First, consider the two $I_1$ fibers in ${B}/\alpha_B$. Since $\alpha_B$ must map Kodaira fibers of the same type into each other, it follows that there must be four $I_1$ fibers in $B$. 
Next, we discuss the fiber over $p_1(p)$ which contains an $A_1$ singularity of $W_M$ and blows-up into a $I_2$ in $\widehat{W_M}$. As discussed above, $\kappa^{-1}(\tilde{p}^{-1}_1(p_1(p)))$ are two disjoint copies of this fiber, each containing an $A_1$ singularity of $B$. To make $B$ smooth, we must blow-up  each of these singularities into a $\cp{1}$, thus introducing two new classes of curves, which we denote by $n_1$ and $n_2$. We give the names $o_1$ and $o_2$ to the associated proper transforms. Each of the pairs $(n_i,o_i),\;i=1,2$ is an $I_2$ Kodaira fiber as illustrated in Figure~\ref{fig8}. 
Similarly, consider the fiber over $p_1(p^{'})$. Clearly, this behaves in exactly the same way as the pre-image of $p_1(p)$. That is, $\kappa^{-1}(\tilde{p}^{-1}_1(p_1(p)))$ is two disjoint fibers each with an $A_1$ singularity in $B$. To make $B$ smooth, we must also blow up these two singularities. This introduces two additional classes of curves, which we denote by $n_3$ and $n_4$. We also denote by $o_3$ and $o_4$ the associated proper transforms. Again, each of the pairs $(n_i,o_i),\;i=3,4$  is an $I_2$ Kodaira fiber. Therefore, there are, in total, four $I_2$ fibers in $B$. 
Now consider the fiber in $B/\alpha_B$ with four isolated $A_1$ singularities. Clearly, the pre-image of this fiber in $B$ must be the unique fiber $f_{\infty}$. The four singular points in the $B/\alpha_B$ fiber arise from the four invariant points $e, e^{'}, e^{''}$ and $e^{'''}$ on $f_{\infty}$ in $B$. That is, $f_{\infty}$ is smooth and needs  no further resolution. 
We conclude that the surfaces $B$ constructed in this section have as degenerate fibers four $ I_1$ and four $ I_2$ fibers. Note that this configuration satisfies equation (\ref{21}) for $n=2, m=4$. Therefore, these surfaces $B$ are, indeed, rational elliptic, that is, $B=d\mathbb{P}_9$.

Let us analyze $\kappa^{-1}\pi^{-1}(s)$ and $\kappa^{-1}\pi^{-1}(i)$ of $B$. First consider $\kappa^{-1}\pi^{-1}(s)$. Exactly as in the previous case, $\kappa^{-1}\pi^{-1}(s)$ reduced by the exceptional divisors $n_1$ and $n_2$ is a section of $B$, which is single copy of $\cp{1}$ which intersects $n_1$ and $n_2$. We define $\kappa^{-1}\pi^{-1}(s)$ minus the exceptional divisors to be the zero section $e$ of $B$.
In the present case, there is a second bidegree $(0,1)$ curve on $Q$, namely, $i$. Clearly, the properties of $\pi^{-1}(i)$ are identical to those of $\pi^{-1}(s)$ with the exception that the second $A_1$ singularity in $\pi^{-1}(i)$ lies at the point $p^{'}$ of the $\tilde{p}^{-1}_1(p_1(p^{'}))$ fiber instead of at $p$ in $\tilde{p}^{-1}_1(p_1(p))$. Note that the pre-image of $p^{'}$ under $\kappa$ consists of  the two isolated $A_1$ singularities that are blown-up by $n_3$ and $n_4$. A key property of the curve $i$ is that, by construction, it must coincide with one of the curves $q_i$ defined at the end of the previous section. The reason is that it is a bidegree $(0,1)$ curve distinct from $s$ which passes through an intersection point of $t=t^{'}\cup i$ with $f_0$. However, unlike the generic curves $q_i$, this specific curve, since it is a component of $T$, also passes through one of the invariant $A_1$ singular points on $r$. It is clear, therefore, that the proper transform of that curve in $B$, which we denote by $e_6$, is a section of $B$ satisfying
\begin{equation}\label{127}
e_6=(-1)_B (e_6).
\end{equation}
That is, $e_6$ is a section of $B$ with the property that $e_6|_f$ is an invariant point of each fiber $f$ under $(-1)_B$. To fix notation, we will label the points of order two so that 
\begin{equation}\label{128}
e_6|_f=e^{'}.
\end{equation}

The designation of the proper transform of $\kappa^{-1}\pi^{-1}(i)$ as $e_6$ arises as follows. Recall from Section~\ref{RES} that there are nine sections $e_i, i=1,...,9$ of $B$ with fibration (\ref{100}), corresponding to the nine $\cp{1}$ blow-ups of the $d{\mathbb{P}_9}$. We have already designated $e_9$  corresponding to the proper transform of $\kappa^{-1}\pi^{-1}(s)$ as the zero section in (\ref{11}). In \cite{opr}, we will show that $e_6$ can indeed be identified with the section associated to $\kappa^{-1}\pi^{-1}(i)$. In anticipation of that result, we here denote $e_6$ as the proper transform of $\kappa^{-1}\pi^{-1}(i)$. To conclude, we see that $\kappa^{-1}\pi^{-1}(i)$ minus the exceptional divisors $n_3$ and $n_4$ is a section of $B$, which is a single copy of $\cp{1}$ intersecting the exceptional divisors  $n_3$ and $n_4$. 
Having identified $i$ with one of the curves $q_i$, let us arbitrary set $i=q_3$. It follows that, in addition  to $e$ and $e_6$, there are two pairs of sections $(\hat{\xi}_i,(-1)_B(\hat{\xi}_i))$ for $i=1,2$ that satisfy condition (\ref{110}). We can choose them so that $\hat{\xi}_1(0)=e^{''}$ and $\hat{\xi}_2(0)=e^{'''}$ respectively.

It is now clear that these restricted $B$ surfaces admit a $\mathbb{Z}_2\times\mathbb{Z}_2$ action. To see this, consider
\begin{equation}\label{129}
\tau_{B1}=t_{\xi}\circ \alpha_B,
\end{equation}
where section $\xi$ is chosen to be either $\hat{\xi}_1$ or $\hat{\xi}_2$ and 
\begin{equation}\label{130}
\tau_{B2}=t_{\xi-e_6}\circ \alpha_B.
\end{equation}
Note that both $\tau_{B1}$ and $\tau_{B2}$ are involutions since $\xi$ and $\xi-e_6$ satisfy condition (\ref{110}). Furthermore,
\begin{equation}\label{131}
\tau_{B1}\circ \tau_{B2}=t_{e_6}.
\end{equation}
Since $e_6$ is a point of order two on each fiber, it follows that condition (\ref{117}) is satisfied and, hence, by the previous discussion
\begin{equation}\label{132}
\tau_{B1}\circ \tau_{B2}=\tau_{B2}\circ \tau_{B1}.
\end{equation}
Therefore, $B$ admits a $\mathbb{Z}_2\times\mathbb{Z}_2$ action.

We can use the characterization of $B/\alpha_B$ as the double cover of $Q$ branched over curve $M$ to compute the dimension of its moduli space, as well as the moduli space of $B$. Recall that for these restricted surfaces $M=t^{'}\cup i\cup s\cup r$. The curve $r$ has bidegree $(1,0)$ and, hence, it follows from (\ref{91}) that
\begin{equation}\label{133}
\dim|\mathcal{O}_Q(1,0)|=1.
\end{equation}
Similarly, the curves $s$ and $i$ have bidegree $(0,1)$. We then find, using (\ref{91}), that
\begin{equation}\label{134}
\dim|\mathcal{O}_Q(0,1)|=1
\end{equation}
for each curve. Finally, since the curve $t^{'}$ has bidegree $(1,2)$, equation (\ref{91}) implies that
\begin{equation}\label{135}
\dim|\mathcal{O}_Q(1,2)|=5.
\end{equation}
As discussed previously, the number of the moduli of $W_M\cong B/\alpha_B$ is the number of  the moduli of $M$ reduced by the dimension of the automorphism group of $\cp{1}\times \cp{1}$ given in (\ref{94}) and enhanced by  the location parameter of fiber $f_0$. That is,
\begin{equation}\label{136}
\dim \mathcal{M}(B/\alpha_B)=1+1+1+5-6+1=3.
\end{equation}
Since surface $B$ is the double cover of $B/\alpha_B$ branched over $f_0$ and the four singular points in $f_{\infty}$, it follows that $B$ has the same number of moduli as $B/\alpha_B$, namely
\begin{equation}\label{137}
\dim \mathcal{M}(B)=3.
\end{equation}
These results have a simple interpretation. Since the curve $t^{'}\cup i \cup s$ is a degeneration of the generic curve $T$, the rational elliptic surfaces $B$ obtained here are a subset of the surfaces $B$ constructed from $M=t\cup s\cup r$ and $M=T\cup r$. Specifically, they are $d\mathbb{P}_9$ surfaces where eight of the twelve $I_1$ fibers coalesce together in pairs to form four $I_2$ Kodaira fibers. These surfaces form a three parameter subspace of the full five parameter moduli space of surfaces $B$ admitting an involution $\alpha_B$. Finally, we note that, in addition to the $\mathbb{Z}_2\times\mathbb{Z}_2$ action, the restricted surfaces discussed  in this section contain four classes of curves, namely $n_1, n_2, n_3$ and $n_4$, that are not present for a generic surface. 

However, we are not quite finished. 
Our aim is to construct rational elliptic surfaces giving rise to Calabi-Yau threefolds of the form $X=B\times_{\cp{1}}B^{'}$ admitting two freely acting commuting involutions. Furthermore, $X$ must admit stable, holomorphic vector bundles which can be used to obtain standard-like models in string theory compactifications. In order to obtain standard-like models, as we will show in \cite{opr}, it is necessary to have invariant classes in $H_2(B,\mathbb{Z})$ which consist of fiber components, but do not contain a whole fiber. To find such invariant classes, we find it necessary to restrict the surfaces $B$ once more. As we will see, surfaces of this highly restricted type admit three commuting involutions $\tau_{B1}, \tau_{B2}$ and $\tau_{B3}$, so that we call them surfaces with a $\mathbb{Z}_2\times\mathbb{Z}_2\times\mathbb{Z}_2$ automorphism group.

\begin{figure}[!ht]
\begin{center}
\input{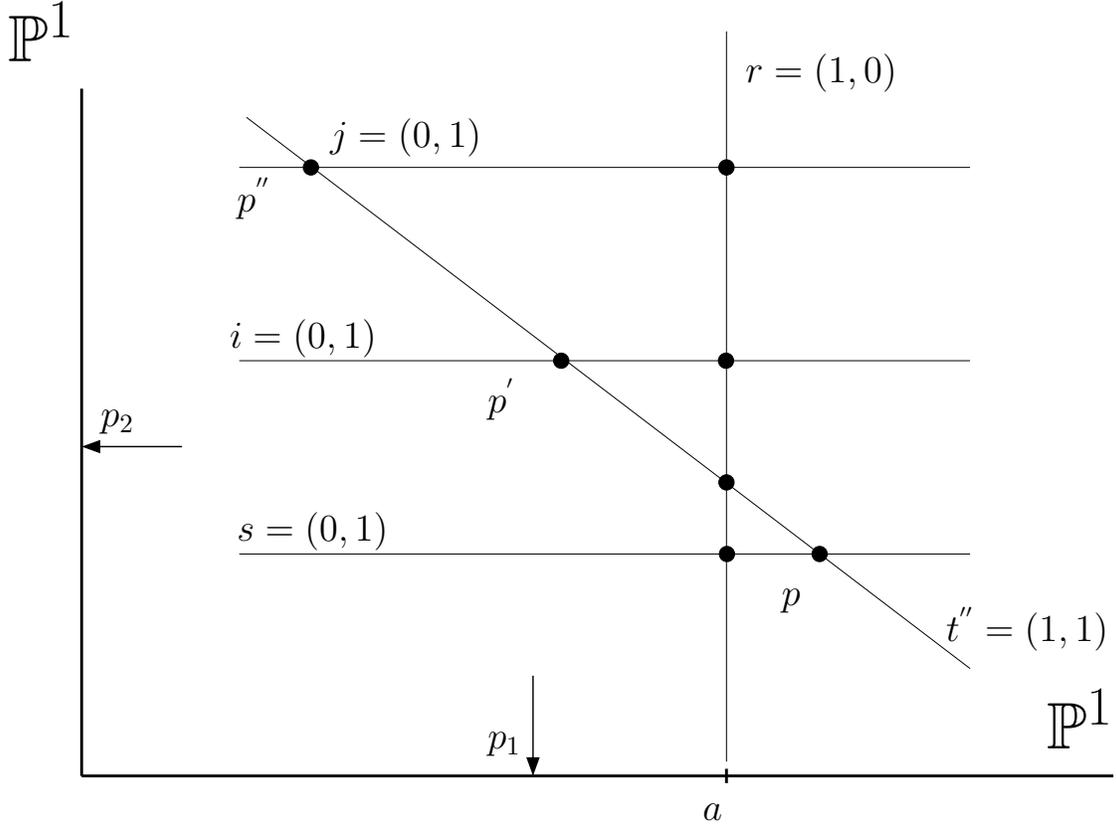} 
\end{center}
\caption{A schematic representation of curve $M=t^{''}\cup j\cup i \cup s \cup r$ in $Q=\cp{1}\times \cp{1}$.}
\label{fig11} 
\end{figure}

Since the techniques involved have all been discussed above, we will only present the results. Let us split the curve $t^{'}$ as
\begin{equation}\label{138}
t^{'}=t^{''}\cup j,
\end{equation}
where curve $t^{''}$ has bidegree $(1,1)$ and curve $j$ has bidegree $(0,1)$. That is, we henceforth assume
\begin{equation}\label{139}
M=t^{''}\cup j\cup i \cup s\cup r.
\end{equation}
We refer the reader to Figure~\ref{fig11} 
for a pictorial representation of the curve $M$ in $Q$. The points of intersection of $t^{''}$ with $s, i$ and $j$ are denoted by $p,p^{'}$ and $p^{''}$ respectively.
\noindent
Now consider the double cover $W_M$ of $Q$ branched over $M$. This is a fibration over $\cp{1}$ with projection map $\tilde{p}_1$ specified in (\ref{100}). For $x \in \cp{1}$ where $x\neq a, p_1(p), p_1(p^{'})$ or $p_1(p^{''})$ one can show, exactly as in the previous cases, that
\begin{equation}\label{140}
\tilde{p}^{-1}_1(x)=T^2,
\end{equation}
where $T^2$ is a smooth torus. Hence, (\ref{100}) is a torus fibration. However, as above, we expect there to be degenerate fibers. Using the fact that $t^{''}\cong \cp{1}$ and $p_1|_{t^{''}}: t^{''}\to \cp{1}$ is an isomorphism,
it follows that there are no $I_1$ fibers on $W_M$. Now, exactly as above, we see that $\tilde{p}_1^{-1}(p_1(p))$ is a fiber of $W_M$ which is a double cover of  $\cp{1}$ branched at three points. One of these points, $p$, is an $A_1$ singularity of $W_M$. This fiber can be blown-up at $p$ to produce an $I_2$ Kodaira fiber. The fibers $\tilde{p}_1^{-1}(p_1(p^{'}))$ and $\tilde{p}_1^{-1}(p_1(p^{''}))$ each have the same properties. That is, there are three isolated singular fibers which blow-up to $I_2$. Finally, there is the fiber $\tilde{p}_1^{-1}(a)$ with four $A_1$ singularities in $W_M$. This is the $f_{\infty}$ fiber, which can be resolved into $I_0^{*}$.

Note that $\widehat{W_B}$, obtained by blowing-up all $A_1$ singularities in $W_M$, is an elliptic fibration over $\cp{1}$ with no $I_1$ fibers, three $ I_2$ fibers and one $I_0^{*}$ fiber. Using (\ref{18}) and (\ref{77}), we see that $\chi(\widehat{W_M})=12$ and, hence, $\widehat{W_M}\cong d\mathbb{P}_9$. The singular fibration $W_M$ is its Weierstrass model.
As previously, $\pi^{-1}(s)$ is the zero section of the fibration (\ref{100}). Clearly, each of $\pi^{-1}(i)$ and $\pi^{-1}(j)$ have the same properties. That is, each consists of a double $\cp{1}$ containing two isolated $A_1$ singularities of $W_M$, one at the intersection of $\pi^{-1}(r)$ and one at the intersection with $\pi^{-1}(t^{''})$. Both are sections of $W_M$, distinct  from each other and from the zero section.

We now identify $W_M \cong B/\alpha_B$ and consider space $B$. Recall from (\ref{79}) that $B$ is the double cover of ${B}/\alpha_B$ branched over the fiber $f_0$ and the four singular points $e, e^{'}, e^{''}$ and $e^{'''}$ on the fiber $f_0$.
Since there are no $I_1$ fibers on ${B}/\alpha_B$, there will be none in $B$. Recall that $\kappa^{-1}(\tilde{p}^{-1}(p_1(p)))$ are two separated copies of this fiber, each with an $A_1$ singularity in $B$. These must be blown-up, thus introducing two new classes $n_1$ and $n_2$. The associated proper transforms are denoted $o_1$ and $o_2$. Each of these blown-up fibers is of Kodaira type $I_2$. Clearly, $\kappa^{-1}(\tilde{p}^{-1}(p_1(p^{'})))$ and  $\kappa^{-1}(\tilde{p}^{-1}(p_1(p^{''})))$ have the same properties. That is, each must have its pair of singularities blown-up, thus introducing additional curve classes $n_3,n_4$ and $n_5,n_6$, with $o_3,o_4$ and $o_5,o_6$ denoting the proper transforms. This yields four  more $I_2$ fibers in $B$, for a total of six. 

Finally, consider the fiber  in ${B}/\alpha_B$ with four isolated $A_1$ singularities. Clearly, the pre-image of this fiber in $B$ is the non-singular fiber $f_{\infty}$. We conclude that the surfaces $B$, so constructed, have six $I_2$ degenerate fibers. Note that this configuration satisfies equation (\ref{21}) for $n=0, m=6$. Therefore, these surfaces are, indeed, rational elliptic, that is $B=d\mathbb{P}_9$.

Let us analyze the sections associated with   $\kappa^{-1}\pi^{-1}(s),\; \kappa^{-1}\pi^{-1}(i)$ and $\kappa^{-1}\pi^{-1}(j)$. As above, $\kappa^{-1}\pi^{-1}(s)$ reduced by the exceptional divisors $n_1$ and $n_2$ is a section of $B$, which is a $\cp{1} $ which intersects  $n_1$ and $n_2$. We define as in (\ref{newzero})
\begin{equation}\label{144}
e=\{\kappa^{-1}\pi^{-1}(s)\setminus n_1 \cup n_2 \}
\end{equation}
to be the zero section of $B$.
Similarly, $\kappa^{-1}\pi^{-1}(i)$ reduced by the exceptional divisors is a $\cp{1}$  intersecting  the pair of $I_2$ fibers $\kappa^{-1}(\tilde{p}^{-1}(p_1(p^{'})))$ in $n_3$ and $n_4$. As discussed above,
\begin{equation}\label{145}
e_6=\{\kappa^{-1}\pi^{-1}(i)\setminus n_3 \cup n_4\}
\end{equation}
is the section of $B$ which passes through the invariant point $e^{'}$ on each fiber. As will be shown elsewhere \cite{opr}, this section can be identified with  $e_6$ defined in Section~\ref{RES}. 
Clearly, by construction, $\kappa^{-1}\pi^{-1}(j)$ has similar properties. That is, $\kappa^{-1}\pi^{-1}(j)$ reduced by the exceptional divisors is a $\cp{1}$ section of $B$  intersecting  the pair of $I_2$ fibers on $\kappa^{-1}(\tilde{p}^{-1}(p_1(p^{''})))$ in $n_5$ and $n_6$. This section has the property that it passes through a  point of order two distinct from $e$ and $e^{'}$ on each fiber. To fix notation we denote this to be $e^{''}$. It will be shown in \cite{opr} that this section can be identified with the section $e_4$ defined in Section~\ref{RES}. In anticipation of this, we denote
\begin{equation}\label{146}
e_4=\{\kappa^{-1}\pi^{-1}(j)\setminus n_5 \cup n_6 \}.
\end{equation}

As discussed above, curve $i$, by construction, must be identified with one of the curves $q_i$. As before, we choose $i=q_3$. Since $i$ is in the branch locus of $M$, its pre-image in $B$ determines a unique section, $e_6$. This is consistent with the fact (\ref{127}) that $e_6=(-1)_B (e_6)$. In exactly the same way, the curve $j$ can be identified with one of the two remaining curves $q_i$, which we arbitrarily choose as $j=q_2$. Since $j$ is a component of $M$, its pre-image in $B$ determines a unique section $e_4$. This leaves one more curve, $q_1$. Hence, there exists exactly one pair of sections $(\xi,(-1)_B (\xi))$ fulfilling condition (\ref{xi}). However, in this restricted case, there is an important caveat. 
Note that the section $e_4+e_6$ is also a section of $B$, which intersects each fiber at a point of order two. It can be shown that this section is determined by the pre-image of $t^{''}$ in $B$ and  intersects all  of the six new classes $n_i, i=1,...,6$. However, even through both sections $\xi$ and $e_4+e_6$ pass through the invariant point $e^{'''}$ at the fiber $f_0$, they are not identical. Therefore, the section $e_4+e_6$ is not of the type discussed previously in this paper. It arises in this very restricted surface $B$, due to the existence of the $t^{''}$ curve in the branch locus of bidegree $(1,1)$.

It is now straightforward to show that these restricted surfaces $B$ admit a $\mathbb{Z}_2\times\mathbb{Z}_2\times\mathbb{Z}_2$ action. To see this, define
\begin{equation}\label{148}
\tau_{B1}=t_{\xi}\circ \alpha_B, \;\;\;\tau_{B2}=t_{\xi-e_4}\circ \alpha_B, \;\;\;\tau_{B3}=t_{\xi-e_6}\circ \alpha_B.
\end{equation}
Clearly $\tau_{B1}^2=\tau_{B2}^2=\tau_{B3}^2=id$, since $\xi, \xi-e_4$ and $\xi-e_6$ fulfill condition (\ref{xi}). Furthermore,
\begin{equation}\label{149}
\tau_{B1}\circ\tau_{B2}=t_{e_6},\;\;\;\tau_{B1}\circ\tau_{B3}=t_{e_4},\;\;\;\tau_{B2}\circ\tau_{B3}=t_{e_4-e_6}.
\end{equation}
Note that since $e_6$ is a section of order two, $e_4-e_6$ is equal to the section $e_4+e_6$ discussed above.
Since each of these quantities squares to the identity map, it follows from condition (\ref{117}) that
\begin{equation}\label{150}
\tau_{Bi}\circ\tau_{Bj}=\tau_{Bj}\circ\tau_{Bi}
\end{equation}
for $i,j=1,2,3$. Therefore, $B$ admits a $\mathbb{Z}_2\times\mathbb{Z}_2\times\mathbb{Z}_2$ action.

Notice that for these surfaces $B$ one could take the $\mathbb{Z}_2\times\mathbb{Z}_2$ action generated by the translations $t_{e_6}$ and $t_{e_4}$. Recalling the remarks of the end of Section~\ref{INV}, these involutions could be lifted to freely acting involutions on the Calabi-Yau threefold $X$. However, it would not be possible to obtain invariant classes consisting of fiber components only. Hence, they are very unlikely to admit vector bundles which allow standard model-like string compactifications and, hence,  we will not consider them further.

Finally, recall that for these restricted surfaces $M=t^{''}\cup j\cup i\cup s\cup r$, where $r$ has bidegree $(1,0)$, the curves $s,i$ and $j$ have bidegree $(0,1)$ and $t^{''}$ has bidegree $(1,1)$. Then, using (\ref{133}), (\ref{134}) and the fact that (\ref{91}) implies
\begin{equation}\label{151}
\dim|\mathcal{O}_Q(1,1)|=3,
\end{equation}
we find that
\begin{equation}\label{152}
\dim \mathcal{M}(B/\alpha_B)=1+1+1+1+3-6+1=2,
\end{equation}
where we have subtracted off the number of automorphisms (\ref{94}) of $\cp{1}\times \cp{1}$ and added the location modulus of fiber $f_0$. As discussed above, it follows that
\begin{equation}\label{153}
\dim \mathcal{M}(B)=2.
\end{equation}
Since $t^{''}\cup j\cup i\cup s\cup r$ is a degeneration of the generic curve $T$, the rational elliptic surfaces $B$ obtained here are a subset of the surfaces  previously constructed. Specifically, they are $d\mathbb{P}_9$ surfaces where all twelve $I_1$ fibers have coalesced in  pairs to form six  $ I_2$ Kodaira fibers. These surfaces form a two parameter subspace of the full five parameter moduli space of surfaces $B$ that admit involutions $\alpha_B$. We note that, in addition to the $\mathbb{Z}_2\times\mathbb{Z}_2\times\mathbb{Z}_2$ action, these restricted surfaces contain six classes of curves, namely $n_i$ for $i=1,\dots ,6$, that are not present in the generic fibration.

\section{The Calabi-Yau Threefolds $X$ and $Z$}\label{CY}

Having constructed the rational elliptic surfaces $B$ that admit several commuting involutions, we are finally in a position to construct torus-fibered Calabi-Yau threefolds $Z$ with fundamental group 
\begin{equation}\label{154}
\pi_1(Z)=\mathbb{Z}_2\times\mathbb{Z}_2.
\end{equation}
To do so, we construct elliptically fibered Calabi-Yau spaces $X$ as the fiber product over $\cp{1}$ of two $d\mathbb{P}_9$ surfaces, $B$ and $B^{'}$ \cite{schoenCY}. Choosing each surface to admit several commuting involutions, we will lift these to two, commuting, freely acting involutions on $X$. These generate the involution group $\mathbb{Z}_2\times\mathbb{Z}_2$. The quotient 
\begin{equation}
Z=X/(\mathbb{Z}_2\times\mathbb{Z}_2)
\end{equation}
must be a smooth manifold  with fundamental group $\pi_1(Z)=\mathbb{Z}_2\times\mathbb{Z}_2$. We will then prove that, in addition, $Z$ is a torus-fibered Calabi-Yau space. Notice, however, that the quotient $Z$ does not admit  any section, and, hence it is not an elliptic fibration.

The fiber product of two rational elliptic surfaces $B$ and $B^{'}$
\begin{equation}\label{155}
\beta: B \to \cp{1}, \;\;\;\;\;\beta^{'}: B^{'} \to \mathbb{P}^{1'}
\end{equation}
is defined as
\begin{equation}\label{156}
B \times_{\cp{1}}B^{'}=\{(b,b^{'})\in B \times B^{'}|\beta(b)=\beta^{'}(b^{'})\},
\end{equation}
where we need to make an identification of $\cp{1}$ and $\mathbb{P}^{1'}$ discussed in more detail below. The manifold $B\times B^{'}$ has complex dimension $4$. Clearly, however, the constraint $\beta(b)=\beta^{'}(b^{'})$ reduces the dimension by one. That is, 
\begin{equation}\label{157}
\dim_{\mathbb{C}} (B \times_{\cp{1}}B^{'})=3.
\end{equation}
Now let $\pi:B \times_{\cp{1}}B^{'}\to B^{'} $ and $\pi^{'}:B \times_{\cp{1}}B^{'}\to B $ be the natural mappings defined by
\begin{equation}\label{158}
\pi(b,b^{'})=b^{'},\;\;\;\;\pi^{'}(b,b^{'})=b.
\end{equation}
Then, we obtain the commutative diagram
\begin{equation} 
\xymatrix{
& X \ar[dl]_-{\pi'} \ar[dr]^-{\pi} & \\
B \ar[dr]_-{\beta} & & B' \ar[dl]^-{\beta'} \\
& \cp{1} &
}
\end{equation}
and, hence, one can define a  map $\rho :B \times_{\cp{1}}B^{'} \to \cp{1} $ as
\begin{equation}\label{159}
\rho=\beta^{'}\circ \pi=\beta\circ \pi^{'}.
\end{equation}
Let us focus on $\pi:B \times_{\cp{1}}B^{'}\to B^{'} $. Choose $b^{'}\in B^{'}$ and denote $\beta^{'}(b^{'})=p$. Then the fiber $\pi^{-1}(b^{'})$ in $X$ given by
\begin{equation}\label{160}
\pi^{-1}(b^{'})=\{(b,b^{'})\in B \times_{\cp{1}} B^{'}|b \in \beta^{-1}(p)\},
\end{equation} 
is isomorphic to the fiber in $B$. That is, $\pi^{-1}(b^{'})$ is either a torus or a degeneration thereof. It follows that
\begin{equation}\label{161}
\pi:B \times_{\cp{1}}B^{'}\to B^{'}
\end{equation}
is a torus fibration over base $B^{'}$. Of course, the same is true for $\pi^{'}:B \times_{\cp{1}}B^{'}\to B$. However, following the conventions in previous literature, we will consider $B^{'}$ as the base of the $B \times_{\cp{1}}B^{'}$ fibration. This explains the, somewhat perverse, notation $\pi$ as the projection map in (\ref{161}). Now, recall that $e: \cp{1}\to B$ is the zero section of the fibration $\beta: B \to \cp{1}$. It follows from this and (\ref{160}) that mapping $\sigma: B^{'} \to B \times_{\cp{1}}B^{'}$, defined by its image
\begin{equation}\label{162}
\sigma=e\times_{\cp{1}}B^{'},
\end{equation} 
is a section of fibration (\ref{161}). We will choose $\sigma$ as the zero section. Therefore, (\ref{161}) is an elliptic fibration. We conclude that $B \times_{\cp{1}}B^{'}$ is a threefold that is elliptically fibered over base surface $B^{'}$ with projection map $\pi$.

The definition of $X$ seems, at first, rather contrived. However, as we now show, every elliptic  Calabi-Yau threefold $X$ fibered over a $d\mathbb{P}_9$ base is, indeed, a fiber product over $\cp{1}$ of two rational elliptic surfaces. Let $X$ be an elliptic fibration
\begin{equation}
\pi: X \to B^{'},
\end{equation}
where $B^{'}$ is isomorphic to $d\mathbb{P}_9$. By definition, a generic fiber of $\pi$ is a torus and there must exist a zero section $\sigma: B^{'}\to X$. Recall from Section~\ref{RES} that $B^{'}$ itself is an elliptic fibration, that is, $\beta^{'}:B^{'}\to \cp{1}$.

Let us describe the Weierstrass model $W_X$ of $X$. To do so, recall the construction of the Weierstrass model $W_B$ of an elliptic fibered surface $B$ given in Section~\ref{RES}. $W_B$  is cut out of $P={\mathbb{P}}(\mathcal{O}_{\cp{1}}\oplus\mathcal{O}_{e}(-2e)\oplus \mathcal{O}_e(-3e) )$ by a global section of $\mathcal{O}_P(3)\otimes p^{*}\mathcal{O}_e(-e)^6$. That is
\begin{equation}
\xymatrix{
         W_B \ar[d] \ar@{}[r]|-\subset  &   {\mathbb{P}}(\mathcal{O}_{\cp{1}}\oplus\mathcal{O}_{e}(-2e)\oplus \mathcal{O}_e(-3e) ) =P\ar[d]^p  \ar@{}[drr]|-{.} &   &  \\ 
     \cp{1}\ar@{}[r]|{=}   &     \cp{1}     &  &  }
\end{equation}
Note that $\mathcal{O}_e(-ne)\cong \ocp(n)$ for any integer $n$.  We argued that, for generic rational elliptic surfaces B, $W_B\cong B$.
In analogy to this, the Weierstrass model $W_X$ of $X$ is cut out of $P^{'}={\mathbb{P}}(\mathcal{O}_{B^{'}}\oplus\mathcal{O}_{\sigma}(-2\sigma)\oplus \mathcal{O}_{\sigma}(-3 \sigma) )$ by a global section of $\mathcal{O}_{P^{'}}(3)\otimes p^{'*}\mathcal{O}_{\sigma}(-\sigma)^6$. The associated diagram is
\begin{equation}\label{W_X}
\xymatrix{
        W_X \ar[d] \ar@{}[r]|-\subset  &    {\mathbb{P}}(\mathcal{O}_{B^{'}}\oplus\mathcal{O}_{\sigma}(-2\sigma)\oplus \mathcal{O}_{\sigma}(-3 \sigma) )=P^{'}\ar[d]^{p^{'}}  \ar@{}[drr]|-{.} &   &  \\ 
        B^{'}\ar@{}[r]|{=}   &      B^{'}     & &    }
\end{equation}
Again, for generic Calabi-Yau threefolds $X$, $W_X\cong X$.

Since for Calabi-Yau spaces the canonical bundle is trivial, the adjunction formula on $X$ allows us to re-express $P^{'}$ in terms of a projective bundle over $\cp{1}$. To see this, recall that the adjunction formula states
\begin{equation}
K_X|_{D}=K_{D}\otimes \mathcal{O}_{D}(D),
\end{equation}
where $\mathcal{O}_{D}(D)$ is the normal bundle for a divisor $D\subset X$,$\; K_{D}$ the canonical bundle of $D$ and $K_X|_{D}$ the canonical bundle of $X$ restricted to $D$. Since the canonical bundle of $X$ must be trivial, it follows that $K_X|_{D}=\mathcal{O}_D$. Furthermore, choosing $D=\sigma$  we find
\begin{equation}
\mathcal{O}_{\sigma}(\sigma)=K_{\sigma}.
\end{equation}
Using the fact that the section $\sigma$ is isomorphic to the base $B^{'}$, it follows that $K_{\sigma}\cong K_{B^{'}}$. Therefore, $\mathcal{O}_{\sigma}(-n\sigma)=K_{B^{'}}^{-n}$ for any integer $n$. Furthermore, we also have shown in Section~\ref{RES} that, for a rational elliptic surface $B^{'}$, the canonical bundle is given as $K_{B^{'}}=\mathcal{O}_{B^{'}}(-f^{'})=\beta^{'*}\ocp{(-1)}$, where $f^{'}$ denotes the fiber class of $B^{'}$ and $\beta^{'}$ the projection. Summarizing, we find that
\begin{equation}
\mathcal{O}_{\sigma}(-n\sigma)=\beta^{'*}\ocp{(n)}.
\end{equation}
Using this result, we see that $P^{'}$ can be written as $P^{'}={\mathbb{P}}(\beta^{'*}\mathcal{O}_{\cp{1}}\oplus \beta^{'*}{\ocp(2)}\oplus \beta^{'*}{\ocp(3)})$. That is, 
\begin{equation}
\xymatrix{
        W_X \ar[d] \ar@{}[r]|-\subset  &    {\mathbb{P}}(\beta^{'*}\mathcal{O}_{\cp{1}}\oplus \beta^{'*}{\ocp(2)}\oplus \beta^{'*}{\ocp(3)})=P^{'}\ar[d]^{p^{'}} \\ 
        B^{'}\ar@{}[r]|{=}   &      B^{'}       }
\end{equation}
and  the Weierstrass model $W_X$ of $X$ is cut out of $P^{'}$ by a global section of $\mathcal{O}_{P^{'}}(3)\otimes p^{'*}\circ \beta^{'*}\ocp(6)$.

Observe that $P^{'}$ is the projectivization of a rank three vector bundle on $B^{'}$, which is a pull-back of a vector bundle on $\cp{1}$. For such bundles, there is a natural projection from the pulled back bundle to the original bundle. We will call this projection $\lambda$ and obtain
\begin{equation}
\xymatrix{ P^{'}= \mathbb{P}(\beta^{'*}\mathcal{O}_{\cp{1}}\oplus \beta^{'*}{\ocp(2)}\oplus \beta^{'*}{\ocp(3)}) \ar[r]^-{\lambda} \ar[d]^{p^{'}}  &   {\mathbb{P}}(\mathcal{O}_{\cp{1}}\oplus {\ocp(2)}\oplus \ocp{(3)})\ar[d]^p \ar@{}[drr]|-{.} &   &  \\
B^{'}\ar[r]^{\beta^{'}} & {\cp{1}} & &}
\end{equation}
Recall that $W_X$ is the zero locus  of  a global section of $\mathcal{O}_{P'}(3)\otimes p^{'*}\circ \beta^{'*}\ocp(6)$ in  $P^{'}$. What is the  image of $W_B$ under $\lambda$ ? Clearly, it is the zero locus of a  section of the bundle $\lambda_{*}(\mathcal{O}_P(3)\otimes p^{'*}\circ \beta^{'*}\ocp(6))$ on ${\mathbb{P}}(\mathcal{O}_{\cp{1}}\oplus {\ocp(2)}\oplus \ocp{(3)})$. It can be shown that this push-forward bundle is $\mathcal{O}_{{\mathbb{P}}(\mathcal{O}_{\cp{1}}\oplus {\ocp(2)}\oplus \ocp{(3)})}(3)\otimes p^{*}\ocp(6)$. Hence, the image of $W_X$ under $\lambda$ is a divisor given by the zero locus of  a section of $\mathcal{O}_{{\mathbb{P}}(\mathcal{O}_{\cp{1}}\oplus {\ocp(2)}\oplus \ocp{(3)})}(3)\otimes p^{*}\ocp(6)$. But as discussed in Section~\ref{RES}, such a  divisor corresponds to a Weierstrass model $W_B$ of a rational elliptic surface. Since $W_B$ is generically isomorphic to the surface itself, we will denote it by $B$. In conclusion, we obtain the commuting diagram
\begin{equation} \label{eq-main-diagram}
\xymatrix{
 W_X \ar[r]^-{\pi'}\ar[d]^-{\pi} & B \ar[d]^-{\beta}   \ar@{}[dr]|-{.} &    \\
B^{'} \ar[r]^-{\beta^{'}} &  \cp{1} &  } 
\end{equation}
where we  denote the restriction $\lambda|_{W_X}$ as $\pi^{'}$, $p^{'}|_{W_X}=\pi$ and $p|_{W_B}=\beta $. But, generically, $W_X\cong X$. Therefore, we have proven that each generic elliptic Calabi-Yau threefold $X$ fibered over a rational elliptic surface $B^{'}$ is a fiber product $B\times_{\cp{1}}B^{'}$of two rational elliptic surfaces $B$ and $B^{'}$.

We now compute  the Chern classes for the tangent bundle $T_{B \times_{\cp{1}}B^{'}}$ of $B \times_{\cp{1}}B^{'}$. The results of this calculation will be used to  prove that $B \times_{\cp{1}}B^{'}$ is indeed a Calabi-Yau threefold.  We begin by considering the exact sequence
\begin{equation}\label{163}
\ses{T_D}{T_X|_D}{N_D},
\end{equation}  
where $X$ is a complex manifold, $D \subset X$ is a divisor and $T$ and $N$ denote the tangent and normal bundles respectively. It is straightforward to see that $B \times_{\cp{1}}B^{'}$ is a divisor of $B\times B^{'}$. For these manifolds, (\ref{163}) becomes
\begin{equation}\label{164}
\ses{T_{B \times_{\cp{1}}B^{'}}}{T_{B \times B^{'}}|_{B \times_{\cp{1}}B^{'}}}{N_{B \times_{\cp{1}}B^{'}}}.
\end{equation}  
Since $B \times_{\cp{1}}B^{'}$ is a divisor, it is the zero locus of a section of the line bundle $\mathcal{O}_{B \times B^{'}}(B \times_{\cp{1}}B^{'})$. Its normal bundle is given by 
\begin{equation}\label{165}
N_{B \times_{\cp{1}}B^{'}}=\mathcal{O}_{B \times B^{'}}(B \times_{\cp{1}}B^{'})|_{B \times_{\cp{1}}B^{'}}.
\end{equation} 
Now consider the Chern characters. From (\ref{164}), we see that
\begin{equation}\label{166}
ch(T_{B \times_{\cp{1}}B^{'}})=ch(T_{B \times B^{'}}|_{B \times_{\cp{1}}B^{'}})-ch(N_{B \times_{\cp{1}}B^{'}}).
\end{equation} 
Using the relation
\begin{equation}\label{167}
T_{B\times B^{'}}=\pi^{'*}T_B\oplus \pi^{*}T_{B^{'}}
\end{equation} 
and the fact that for any vector bundle $V$ over a threefold
\begin{equation}\label{168}
ch(V)=rank V +c_1+\frac{1}{2}(c_1^2-2c_2)+\frac{1}{6}(c_1^3-3c_1c_2+3c_3),
\end{equation} 
where $c_i$ is the $i$-th Chern class, we can compute $ch(T_{B \times_{\cp{1}}B^{'}})$ in (\ref{166}). First, consider $ch(T_{B \times B^{'}}|_{B \times_{\cp{1}}B^{'}})$. It follows from (\ref{26}) that
\begin{equation}\label{169}
c_1(T_B)=f,\;\;\;\;c_1(T_{B^{'}})=f^{'}
\end{equation} 
and, hence, using (\ref{167})
\begin{equation}\label{170}
c_1(T_{B\times B^{'}})=f\times B^{'}+B\times f^{'}.
\end{equation} 
Similarly,
\begin{equation}\label{171}
c_2(T_{B\times B^{'}})=12pt\times B^{'}+B\times 12pt^{'}
\end{equation} 
where we used the fact  that $c_2(T_B)=\chi(B)=12$ and similarly for $B^{'}$.
Setting $V=T_{B\times B^{'}}|_{B \times_{\cp{1}}B^{'}}$ in (\ref{168}), and restricting the the classes  (\ref{170}) and (\ref{171}) to $B \times_{\cp{1}}B^{'} $, we find that 
\begin{equation}\label{173}
ch(T_{B\times B^{'}}|_{B\times_{\cp{1}} B^{'}})=4+f\times f^{'}+f\times f^{'}-12pt\times f^{'}+f\times 12pt^{'}.
\end{equation} 
Now, consider $ch(N_{B\times_{\cp{1}} B^{'}})$. Using expressions (\ref{165}), we see
\begin{equation}\label{174}
ch(N_{B\times_{\cp{1}} B^{'}})=ch(\mathcal{O}_{B\times_{\cp{1}} B^{'}}(2(f\times f^{'}))).
\end{equation} 
It follows from (\ref{168}) that
\begin{equation}\label{175}
ch(N_{B\times_{\cp{1}} B^{'}})=1+2(f\times f^{'}).
\end{equation} 
Plugging expressions (\ref{173}) and (\ref{175}) into (\ref{166}), one obtains
\begin{equation}\label{176}
ch(T_{B\times_{\cp{1}} B^{'}})=3+0+\frac{1}{2}(0-2(12pt\times f^{'}+f\times 12 pt^{'}))
\end{equation}
from which, using (\ref{168}), one can read off the Chern classes of $T_{B\times_{\cp{1}} B^{'}}$. They are given by
\begin{equation}\label{177}
c_1(T_{B\times_{\cp{1}} B^{'}})=0,\;\;\;\;\;\;c_2(T_{B\times_{\cp{1}} B^{'}})=12(pt\times f^{'}+f\times pt^{'}),\;\;\;\;\;\;c_3(T_{B\times_{\cp{1}} B^{'}})=0.
\end{equation}
Hence, a necessary condition for $B\times_{\cp{1}} B^{'}$ to be a Calabi-Yau manifold is satisfied, namely that $c_1(T_{B\times_{\cp{1}} B^{'}})=0$.

Recall that the canonical bundle of $B\times_{\cp{1}} B^{'}$ is given by 
\begin{equation}\label{178}
K_{B\times_{\cp{1}} B^{'}}=\wedge^3 T^{*}_{B\times_{\cp{1}} B^{'}},
\end{equation}
where $T^{*}_{B\times_{\cp{1}} B^{'}}$ is the cotangent bundle. It follows from (\ref{177}) and (\ref{178}) that
\begin{equation}\label{179}
c_1(K_{B\times_{\cp{1}} B^{'}})=0.
\end{equation}
This, plus the fact that $B\times_{\cp{1}} B^{'}$ is simply connected, then implies that $K_{B\times_{\cp{1}} B^{'}}$ is the trivial bundle. Using (\ref{178}), we conclude that
\begin{equation}\label{180}
\wedge^3 T^{*}_{B\times_{\cp{1}} B^{'}}=\mathcal{O}_{B\times_{\cp{1}} B^{'}}
\end{equation}
and, hence, $B\times_{\cp{1}} B^{'}$ admits a global, holomorphic three-form. Therefore,
\begin{equation}\label{181}
X=B\times_{\cp{1}} B^{'}
\end{equation}
is an elliptically fibered Calabi-Yau threefold over base $B^{'}$ with projection map $\pi$.

The above remarks apply to a manifold of the form $B\times_{\cp{1}} B^{'}$ for any rational elliptic surfaces $B$ and $B^{'}$ discussed in this paper. We now restrict those surfaces to be such that they both admit a $\mathbb{Z}_2\times\mathbb{Z}_2$ or a $\mathbb{Z}_2\times\mathbb{Z}_2\times \mathbb{Z}_2$ action, as presented in the previous section. In either case, $B$ will have at least two commuting involutions $\tau_{Bi}:B \to B$ for $i=1,2$, where $\tau_{Bi}$ is of the form $\tau_{Bi}=t_{\xi_i}\circ \alpha_B$. The involution $\alpha_B: B\to B$ induces, through the projection map in (\ref{155}), the involution $\tau_{\cp{1}}:\cp{1}\to \cp{1}$ on the base space. Recall that $\tau_{\cp{1}}$ has two fixed points at $0$ and $\infty$. Furthermore, $\alpha_B$ leaves the fiber $f_0=\beta^{-1}(0)$ point-wise fixed and has four isolated fixed points $e, e^{'}, e^{''}, e^{'''}$ in $f_{\infty}=\beta^{-1}(\infty)$. Exactly the same statements are true for $B^{'}$. That is, $B^{'}$ admits at least two commuting involutions $\tau_{B^{'}i}:B^{'} \to B^{'}$ for $i=1,2$, where $\tau_{B^{'}i}$ is of the form $\tau_{B^{'}i}=t_{\xi^{'}_i }\circ \alpha_{B^{'}}$. The involution $\alpha_{B^{'}}$ induces the involution $\tau_{\mathbb{P}^{1'}}:\mathbb{P}^{1'}\to \mathbb{P}^{1'}$, which has two fixed points at $0^{'}$ and $\infty^{'}$. Again, $\alpha_{B^{'}}$ leaves $f^{'}_{0^{'}}=\beta^{'-1}(0^{'})$ point-wise fixed and has four isolated fixed points on $f_{\infty^{'}}^{'}=\beta^{'-1}(\infty^{'})$. 

Now, as mentioned above, to construct $B\times_{\cp{1}} B^{'}$ in (\ref{156}) it is necessary to make an identification 
\begin{equation}\label{182}
i: \cp{1}\to \mathbb{P}^{1'}
\end{equation}
of the bases of $B$ and $B^{'}$ respectively. We, henceforth, will choose the diffeormorphism  $i$ so that it identifies the involutions $\tau_{\cp{1}}$ and $\tau_{\mathbb{P}^{1'}}$, but satisfies
\begin{equation}\label{183}
i(0)=\infty^{'},\;\;\;\;i(\infty)=0^{'}.
\end{equation}
This identification is shown in Figure~\ref{fig11}.

\begin{figure}[!ht]
\begin{center}
\input{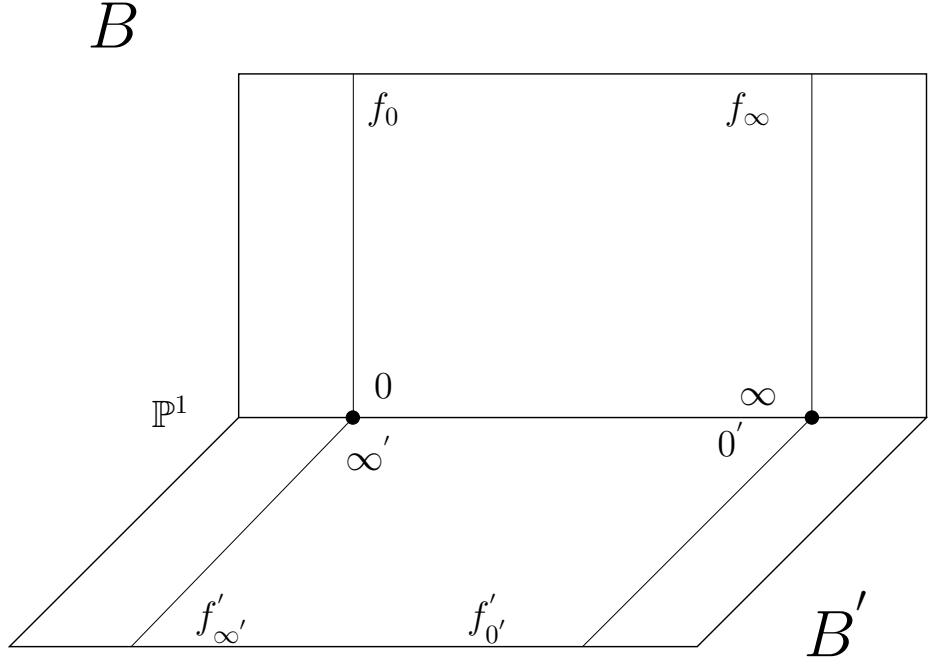} 
\end{center}
\caption{A diagrammatic representation of $B\times_{\cp{1}}B^{'}$ showing the identification of the base of $B$ with the base of $B^{'}$.}
\label{fig12} 
\end{figure}

It follows from the definition (\ref{156}) and the identification (\ref{182}), (\ref{183}) that the involutions on $B$ and $B^{'}$ naturally induce the maps $\tau_{Xi}: X \to X $ defined by
\begin{equation}\label{184}
\tau_{Xi}=\tau_{Bi}\times_{\cp{1}}\tau_{B^{'}i}
\end{equation}
for $i=1,2$. Of course, maps of the form $\tau_{Bi}\times_{\cp{1}}\tau_{B^{'}j}$ for $i\neq j $ are also induced but, as discussed below, these will not be of interest. Clearly
\begin{equation}\label{185}
\tau_{Xi}^2=\tau_{Bi}^2\times_{\cp{1}}\tau_{B^{'}i}^2
\end{equation}
and, hence, $\tau_{Xi}^2=id$ for $i=1,2$. That is $\tau_{Xi}, i=1,2$ are involutions on $X$. Do these involutions commute? To answer this, note that
\begin{equation}\label{186}
\tau_{X1}\circ\tau_{X2} =(\tau_{B1}\circ\tau_{B2})\times_{\cp{1}}(\tau_{B^{'}1}\circ\tau_{B^{'}2}).
\end{equation}
At this point, let us first examine the case defined in (\ref{121}) where both $B$ and $B^{'}$ admit a $\mathbb{Z}_2\times\mathbb{Z}_2 $ action. Then we see from (\ref{129}) and (\ref{130}) that 
\begin{equation}\label{187}
\tau_{B1}=t_{\xi}\circ\alpha_{B},\;\;\;\;\tau_{B2}=t_{\xi-e_6}\circ\alpha_{B}
\end{equation}
and
\begin{equation}\label{188}
\tau_{B^{'}1}=t_{\xi^{'}}\circ\alpha_{B^{'}},\;\;\;\;\tau_{B^{'}2}=t_{\xi^{'}-e_6^{'}}\circ\alpha_{B^{'}}.
\end{equation}
Furthermore, we showed in (\ref{132}) that $\tau_{B1}\circ\tau_{B2}=\tau_{B2}\circ\tau_{B1}$ and $\tau_{B^{'}1}\circ\tau_{B^{'}2}=\tau_{B^{'}2}\circ\tau_{B^{'}1}$. It follows from (\ref{186}) that
\begin{equation}\label{189}
\tau_{X1}\circ\tau_{X2} =\tau_{X2}\circ\tau_{X1} 
\end{equation}
and, hence, $X$ admits a $\mathbb{Z}_2\times\mathbb{Z}_2$ action generated by the involutions $\tau_{Xi},i=1,2$.

However, unlike the case of the rational elliptic surfaces $B$ and $B^{'}$, it is now essential to prove that the $\mathbb{Z}_2\times\mathbb{Z}_2$ action on manifold $X$ is fixed point free. To show that this is indeed the case, recall that $\alpha_B$ leaves the fiber $f_0=\beta(0)$ point-wise fixed. Since $t_{\xi}$ is a pure translation by $\xi\neq  e$, it follows the $\tau_{B1}=t_{\xi}\circ\alpha_{B}$ acts without fixed points on $f_0$. However, $\alpha_B$ leaves four isolated fixed points $e, e^{'}, e^{''}, e^{'''}$ on $f_{\infty}=\beta^{-1}(\infty)$. That is, $\alpha_B$ acts like $(-1)_B$ on $f_{\infty}$. Pick one of these fixed points, say $e^{'}$. Let $\tilde{\xi}$ be the point in $f_{\infty}$ with the property that $\tilde{\xi}+\tilde{\xi}={\xi}|_{f_{\infty}}$. Then 
\begin{equation}\label{190}
t_{\xi}\circ\alpha_{B}(e^{'}+\tilde{\xi})=t_{\xi}(-e^{'}-\tilde{\xi})=e^{'}+\tilde{\xi},
\end{equation}
where we have used the fact that $-e^{'}=e^{'}$. It follows that
\begin{equation}\label{191}
\tau_{B1}(e^{'}+\tilde{\xi})=e^{'}+\tilde{\xi}
\end{equation}
and, hence, $e^{'}+\tilde{\xi}$ in $f_{\infty}$ is an isolated fixed point of $\tau_{B1}$. Clearly, $\tilde{\xi},\; e^{''}+\tilde{\xi}$ and $e^{'''}+\tilde{\xi}$ in $f_{\infty}$ are also isolated fixed points of $\tau_{B1}$. Note that since $\alpha_B$ exchanges all other fibers pairwise, these are the only fixed points of  $\tau_{B1}$. Clearly,  $\tau_{B1'}$ has exactly the same properties. That is, the only fixed points of  $\tau_{B1'}$, are the four isolated points $\tilde{\xi^{'}},(e^{'})^{'}+\tilde{\xi}^{'}, (e^{''})^{'}+\tilde{\xi}^{'} $ and $(e^{'''})^{'}+\tilde{\xi}^{'}$ in $f_{\infty^{'}}^{'}$. Now consider $\tau_{X1}=\tau_{B1}\times_{\cp{1}}\tau_{B^{'}1}$. Here, the wisdom of identifying the bases $\cp{1}$ of $B$ and $\mathbb{P}^{1'}$ of $B^{'}$ with the ``twist'' given in (\ref{183}) becomes apparent. Note that, with this identification, when $\tau_{B1}$ is acting on $f_{\infty}$ with four fixed points, $\tau_{B^{'}1}$ is acting on $f_{0^{'}}^{'}$, with none. Similarly, when  $\tau_{B^{'}1}$ is acting on $f_{\infty ^{'}}^{'}$ with four fixed points, $\tau_{B1}$, is acting on $f_{0}$ with none. Hence, $\tau_{X1}$ acts without fixed points on $X$. It is straightforward to show, in exactly the same way, that $\tau_{X2}$ acts without fixed points on $X$. However, we are not yet finished. To show that the  $\mathbb{Z}_2\times\mathbb{Z}_2 $ action is fixed point free, we must also demonstrate that $\tau_{X1}\circ \tau_{X2}$ does  not have invariant points. First note, using (\ref{186}), (\ref{186}), (\ref{188}) and (\ref{47}), that $\tau_{X1}\circ \tau_{X2}$ is the pure translation
\begin{equation}\label{192}
\tau_{X1}\circ \tau_{X2}=t_{e_6}\times_{\cp{1}}t_{e_6^{'}}.
\end{equation}
Therefore, acting on non-singular fibers, this action is fixed point free. However, $t_{e_6}$ and $t_{e^{'}_6}$ can have fixed points on singular fibers, specifically on each of the four $I_1$ and four $I_2$ fibers on $B$ and $B^{'}$. Happily, under a generic identification (\ref{182}), none of the singular fibers of $B$ is paired with a singular fiber of $B^{'}$. If follows that $\tau_{X1}\circ \tau_{X2}$ and, therefore, the  $\mathbb{Z}_2\times\mathbb{Z}_2 $ action on $X$ generated by $\tau_{X1}$ and $\tau_{X2}$ is fixed point free, as required.

Finally, recall that in addition to $\tau_{X1}$ and $\tau_{X2}$ defined in (\ref{184}), maps of the form
\begin{equation}\label{193}
\tau_{X3}=\tau_{B1}\times_{\cp{1}}\tau_{B^{'}2},\;\;\;\;\tau_{X4}=\tau_{B2}\times_{\cp{1}}\tau_{B^{'}1}
\end{equation}
are also induced on $X$. These are clearly involutions and they also commute with each other and with $\tau_{X1}$ and $\tau_{X2}$. It would appear, therefore, that we actually have a $(\mathbb{Z}_2)^4$ action on $X$. However, as we now demonstrate, this action is not fixed point free. It suffices to consider $\tau_{X1}\circ \tau_{X3}$. Using (\ref{186}), (\ref{187}), (\ref{188}) and (\ref{47}) we see
\begin{equation}\label{193a}
\tau_{X1}\circ \tau_{X3}=id\times_{\cp{1}}t_{e_6^{'}}
\end{equation}
which has fixed points  $(b,b^{'})$ whenever $b^{'}$  is a fixed point of an $I_1$ or $I_2$ fiber in $B^{'}$. Similarly, $\tau_{X2}\circ \tau_{X4}$ has fixed points. We conclude, that only the $\mathbb{Z}_2\times \mathbb{Z}_2$  action generated by $\tau_{X1}$ and $\tau_{X2}$ (or equivalently, the $\mathbb{Z}_2\times \mathbb{Z}_2$ action generated by $\tau_{X3}$ and $\tau_{X4}$ ) acts freely on $X$. The larger group $(\mathbb{Z}_2)^4$ generated by $\tau_{X1},\; \tau_{X2},\;\tau_{X3}$ and $\tau_{X4}$ acts with fixed points.

We now turn to the case of $X=B\times_{\cp{1}}B^{'}$ where both $B^{}$ and $B^{'}$ are defined by (\ref{139}) and admit a $\mathbb{Z}_2\times \mathbb{Z}_2\times \mathbb{Z}_2$ action. The situation is almost identical to the case just discussed, so we will simply present the result. Recall from (\ref{148}) that
\begin{equation}\label{194}
\tau_{B1}=t_{\xi}\circ \alpha_{B^{'}},\;\;\;\tau_{B2}=t_{\xi-e_4}\circ \alpha_{B^{'}},\;\;\;\tau_{B3}=t_{\xi-e_6}\circ \alpha_{B}
\end{equation}
and
\begin{equation}\label{195}
\tau_{B^{'}1}=t_{\xi^{'}}\circ \alpha_{B^{'}},\;\;\;\tau_{B^{'}2}=t_{\xi^{'}-e_4^{'}}\circ \alpha_{B^{'}},\;\;\;\tau_{B^{'}3}=t_{\xi^{'}-e_6^{'}}\circ \alpha_{B^{'}}.
\end{equation}
These naturally induce the maps $\tau_{X_{i}}:X\to X$ for $i=1,2,3$ defined by
\begin{equation}\label{196}
\tau_{X1}=\tau_{B1}\times_{\cp{1}}\tau_{B^{'}1}\;\;\;\;\tau_{X2}=\tau_{B3}\times_{\cp{1}}\tau_{B^{'}3} 
\end{equation}
and
\begin{equation}\label{197}
\tau_{X3}=\tau_{B2}\times_{\cp{1}}\tau_{B^{'}2}.
\end{equation}
Clearly all three of these mappings are involutions on $X$ and it follows from (\ref{150}) that they mutually commute. Hence, $\tau_{X_i}$ for $i=1,2,3$ provide a $\mathbb{Z}_2\times\mathbb{Z}_2\times\mathbb{Z}_2 $ action on $X$. However, this action is not entirely fixed point free. Proceeding exactly as above, it is clear that the action of  a $\mathbb{Z}_2\times\mathbb{Z}_2$ subgroup generated by, say, $\tau_{X_1}$ and $\tau_{X_2}$ is indeed fixed point free. Note, however, that
\begin{equation}\label{198}
\tau_{X1}\circ\tau_{X2}\circ\tau_{X3}=t_{\xi+e_4+e_6}\circ\alpha_B\times_{\cp{1}}t_{\xi^{'}+e_4^{'}+e_6^{'}}\circ\alpha_{B^{'}}.
\end{equation}
But recall that  $(\xi+e_4+e_6)|_{f_0}=e|_{f_0}$. Clearly, $\tau_{X1}\circ\tau_{X2}\circ\tau_{X3}$ has many fixed points. We conclude that even for surfaces with involution $\mathbb{Z}_2\times\mathbb{Z}_2\times\mathbb{Z}_2$, the fixed point free action induced on $X$ is only $\mathbb{Z}_2\times\mathbb{Z}_2$.
 
As in the previous case, there are several different $\mathbb{Z}_2\times\mathbb{Z}_2$ actions that one could construct, for example, generated by $\tau_{B1}\times_{\cp{1}}\tau_{B^{'}2}$ and $\tau_{B2}\times_{\cp{1}}\tau_{B^{'}1}$ rather then the $\tau_{X1},\; \tau_{X2}$ defined in (\ref{196}). However, in all cases, the maximal fixed point free group action on $X$ in $\mathbb{Z}_2\times\mathbb{Z}_2$.

We are now, at long last, in a position to construct torus-fibered Calabi-Yau threefolds with $\mathbb{Z}_2\times\mathbb{Z}_2$ homotopy. Let $X$ be either of the two types of Calabi-Yau threefolds discussed  in this section admitting a fixed point free $\mathbb{Z}_2\times\mathbb{Z}_2$ action. Denote the generators of this action by $\tau_{X1}$ and $\tau_{X2}$. Now consider the quotient space
\begin{equation}\label{199}
Z=X/(\mathbb{Z}_2\times\mathbb{Z}_2).
\end{equation}
Since $\mathbb{Z}_2\times\mathbb{Z}_2$ acts freely on $X$, $Z$ is a smooth threefold. In addition, the first Chern class of $Z$ is given by
\begin{equation}\label{200}
c_1(Z)=\frac{1}{4}c_1(X),
\end{equation}
since $rank \;(\mathbb{Z}_2\times\mathbb{Z}_2)=4  $. Then, $c_1(X)=0$ implies that
\begin{equation}\label{201}
c_1(Z)=0,
\end{equation}
a necessary condition for $Z$ to be a Calabi-Yau threefold. It remains to show that $Z$ admits a holomorphic three-form. To see this, note that $\tau_{Xi},\;i=1,2$ act on $H^0(X,\Omega_X^3)$ as multiplication by elements $g_i$ in $\mathbb{C}^{*}=\mathbb{C}\setminus 0$. Since $\tau_{Xi}$ are involutions, it follows that each $g_i=\pm 1$. It is straightforward to show that, in fact, $g_i=+1$ for both $i=1,2$. We refer the reader to \cite{dopr-ii} for a proof. That is, both $\tau_{X1}$ and $\tau_{X2}$ leave the global holomorphic three-form on $X$, $\omega_X$, invariant. It follows that, under the quotient by $\mathbb{Z}_2\times\mathbb{Z}_2$, this descends to a global holomorphic three-form $\omega_Z$ on $Z$. Property (\ref{201}) and the existence of $\omega_Z$ then guarantee that $Z$ is a Calabi-Yau threefold. However, it can be shown that the global sections of $X$ do not descend to sections on $Z$. In fact, $Z$ has no global sections. Therefore $Z$ is only torus fibered and not elliptically fibered. Since it has been constructed as a $\mathbb{Z}_2\times\mathbb{Z}_2$ quotient of $X$, $Z$ has the non-trivial first homotopy group
\begin{equation}\label{203}
\pi_1(Z)=\mathbb{Z}_2\times\mathbb{Z}_2,
\end{equation}
as desired.

\section*{Acknowledgments}

We would like to thank Ron Donagi and Madeeha Khalid for helpful conversations. Ren\'e Reinbacher and Burt Ovrut are supported in part by DOE under contract No. DE-AC02-76-ER-03071 and the NSF Focused Research Grant DMS 0139799. In addition, Ren\'e Reinbacher wishes to acknowledge partial support from an SAS Dissertation Fellowship and a Daimler-Benz Fellowship. Tony Pantev is supported in part by NSF grants DMS 0099715 and DMS 0139799 and the A.P. Sloan Research Fellowship.


\begin{thebibliography}{OPR}

\bibitem{hw1} 
P.~Ho\v rava and E.~Witten,
\newblock  Heterotic and Type I String Dynamics from Eleven Dimensions,
\newblock  {\em Nucl. Phys.} {B460} (1996) 506.

\bibitem{hw2}
P.~Ho\v rava and E.~Witten,
\newblock  Eleven-Dimensional Supergravity on a Manifold with Boundary,
\newblock {\em Nucl. Phys.} {B475} (1996) 94.

\bibitem{w1}
E.~Witten,
\newblock Strong Coupling Expansion Of Calabi-Yau Compactification,
\newblock {\em Nucl. Phys.} {B471} (1996) 135.



\bibitem{losw1}
A.~Lukas, B.~A. Ovrut, K.S.~Stelle and D.~Waldram,
\newblock   The Universe as a Domain Wall,
\newblock {\em Phys.Rev.} {D59} (1999) 086001.

\bibitem{losw2}
A.~Lukas, B.~A. Ovrut, K.S. Stelle and D. Waldram,
\newblock  Heterotic M--theory in Five Dimensions,
\newblock  {\em Nucl.Phys.} {B552} (1999) 246-290.

\bibitem{low1}
A.~ Lukas, B.~ Ovrut, and D.~Waldram,
\newblock Non-standard embedding and five-branes in heterotic M-Theory,
\newblock {\em Nucl.Phys.}, {B552} (1999) 246-290.

\bibitem{low2}
R.~Donagi, A.~ Lukas, B.~ Ovrut, and D.~Waldram,
\newblock Non-Perturbative Vacua and Particle Physics in M-Theory,
\newblock {\em JHEP}, {9905} (1999) 018.


\bibitem{dlow}
R.~Donagi, A.~ Lukas, B.~ Ovrut, and D.~ Waldram,
\newblock Holomorphic Vector Bundles and Non-Perturbative Vacua in M-Theory,
\newblock {\em  JHEP}, {9906} (1999) 034.

\bibitem{schoenCY}
C.~Schoen,
\newblock On fiber products of rational elliptic surfaces with section,
\newblock {\em Math.Z.}, {197(2)} (1988) 177--199.


\bibitem{dopw-i}
R.~Donagi, B.~Ovrut, T.~Pantev, and D.~Waldram,
\newblock Spectral involutions on rational elliptic surfaces,
\newblock {\em Adv.Theor.Math.Phys.} {5} (2002) 499-561,  math.AG/0008011.

\bibitem{dopw-ii}
R.~Donagi, B.~Ovrut, T.~Pantev, and D.~Waldram,
\newblock Standard-{M}odel bundles,
\newblock {\em  Adv.Theor.Math.Phys.} {5} (2002) 563-615, math.AG/0008010.

\bibitem{dopw-iii}
R.~Donagi, B.~Ovrut, T.~Pantev, and D.~Waldram,
\newblock Standard-{M}odel bundles on non-simply connected {C}alabi-{Y}au threefolds,
\newblock {\em JHEP}, {0108} (2001) 053,   hep-th/0008008.

\bibitem{dopw-iv}
R.~Donagi, B.~Ovrut, T.~Pantev, and D.~Waldram,
\newblock Standard Models from Heterotic M-theory,
\newblock {\em Adv.Theor.Math.Phys.}, {5} (2002) 93-137.

\bibitem{w2}
E.~Witten,
\newblock Symmetry Breaking Patterns in Superstring Theory,
\newblock {\em Nucl. Physics} {B258} (1985) 75-100.

\bibitem{opr}
B.~Ovrut, T.~Pantev, R.~Reinbacher,
\newblock Invariant Homology on Standard Model Manifolds,
\newblock UPR 1016-T.


\bibitem{dopr-i}
R.~Donagi, B.~Ovrut, T.~Pantev, R.~Reinbacher,
\newblock Holomorphic Vector Bundles, Calabi-Yau Spaces with Non-Trivial Homotopy and the Standard Model,
\newblock {in preparation}.

\bibitem{dopr-ii}
R.~Donagi, B.~Ovrut, T.~Pantev, R.~Reinbacher,
\newblock Stable Vector Bundles on Non-Simply Connected Calabi-Yau Spaces,
\newblock{UPR 1018-T}.

\bibitem{dopr-iii}
R.~Donagi, B.~Ovrut, T.~Pantev, R.~Reinbacher.
\newblock Standard-like Models, Nucleon Decay and SU(4) Instantons,
\newblock {in preparation}.

\bibitem{gh}
Griffiths \& Harris,
\newblock Principles of Algebraic Geometry,
\newblock {\em Wiley Classics Library Edition}, Published 1994.

\bibitem{kodaira-casIII}
K.~Kodaira,
\newblock On compact analytic surfaces,
\newblock {\em Ann. of Math.}, 78:1--40, 1963.
















\end{thebibliography}

\end{document}